\documentclass[a4paper,11pt]{article}
\usepackage{jinstpub} 

\usepackage{duckuments}

\usepackage{graphicx}
\usepackage{multirow}

\usepackage{amsmath,amssymb,amsfonts}
\usepackage{amsthm}
\usepackage{mathrsfs}
\usepackage[title]{appendix}
\usepackage{xcolor}
\usepackage{textcomp}
\usepackage{manyfoot}
\usepackage{booktabs}
\usepackage{algorithm}
\usepackage{algorithmicx}
\usepackage{algpseudocode}
\usepackage{listings}
\usepackage{xspace}
\usepackage{tabularx}
\usepackage{makecell}
\usepackage{caption}
\usepackage{subcaption}
\usepackage{placeins}
\usepackage{bookmark}
\usepackage[shortlabels]{enumitem}
\usepackage[capitalise]{cleveref}
\usepackage{orcidlink}
\usepackage[absolute]{textpos}
\usepackage{siunitx}
\usepackage{circledsteps}

\usepackage{pgfplots}
\pgfplotsset{compat=1.18}
\definecolor{kit-green}     {RGB}{0,150,130}
\definecolor{kit-blue}      {RGB}{70,100,170}
\definecolor{kit-black}     {RGB}{0,0,0}
\definecolor{kit-gray}      {gray}{0.3}
\definecolor{kit-lightgray} {gray}{0.84}
\definecolor{kit-yellow}    {RGB}{252,229,0}
\definecolor{kit-orange}    {RGB}{223,155,27}
\definecolor{kit-lightgreen}{RGB}{140,182,60}
\definecolor{kit-red}       {RGB}{162,34,35}
\definecolor{kit-purple}    {RGB}{163,16,124}
\definecolor{kit-brown}     {RGB}{167,130,46}
\definecolor{kit-cyan}      {RGB}{35,161,224}

\Crefname{figure}{figure\,}{figures.\,}
\Crefname{table}{table\,}{tables\,}
\Crefname{equation}{eq.\,}{eqs.\,}

\newcommand\thirdwidth{0.325}
\newcommand\halfwidth{0.475}
\newcommand\fullwidth{0.975}

\newcommand{\superkekb}{SuperKEKB\xspace}
\newcommand{\belletwo}{Belle~II\xspace}

\newcommand{\trg}{L1~trigger\xspace}
\newcommand{\trgs}{L1~triggers\xspace}

\newcommand{\gev}{\ensuremath{\mathrm{\,Ge\kern -0.1em V}}\xspace}
\newcommand{\mev}{\ensuremath{\mathrm{\,Me\kern -0.1em V}}\xspace}
\newcommand{\kev}{\ensuremath{\mathrm{\,ke\kern -0.1em V}}\xspace}

\newcommand{\icncluster}{\ensuremath{\text{\textit{ICN cluster}}}\xspace}
\newcommand{\icnhits}{\ensuremath{\text{\textit{ICN hits}}}\xspace}
\newcommand{\icnhit}{\ensuremath{\text{\textit{ICN hit}}}\xspace}
\newcommand{\datawindow}{ECL-TRG data window\xspace}
\newcommand{\datawindows}{ECL-TRG data windows\xspace}
\newcommand{\triggerwindow}{ECL-TRG trigger window\xspace}
\newcommand{\triggerwindows}{ECL-TRG trigger windows\xspace}

\newcommand{\trgeff}{\ensuremath{\varepsilon_{\text{trg}}}\xspace}
\newcommand{\trgpur}{\ensuremath{\mathfrak{p}_{\text{trg}}}\xspace}

\newcommand{\eres}{\ensuremath{\eta(E_{\text{trg}})}\xspace}
\newcommand{\eresuncorr}{\ensuremath{\eta(E_{\text{trg}})'}\xspace}
\newcommand{\gnnetm}{\texttt{GNN-ETM}\xspace}
\newcommand{\cgnnetm}{\ensuremath{\texttt{GNN-ETM}_{{97.5}}}\xspace}
\newcommand{\icnetm}{\texttt{ICN-ETM}\xspace}
\newcommand{\baseline}{\texttt{ICN-ETM}\xspace}
\newcommand{\keras}{\texttt{KERAS}\xspace}
\newcommand{\qkeras}{\texttt{QKERAS}\xspace}

\newcommand{\offlinecluster}{offline reconstructed cluster\xspace}
\newcommand{\offlineclusters}{offline reconstructed clusters\xspace}
\newcommand{\triggercluster}{\trg cluster\xspace}
\newcommand{\triggerclusters}{\trg clusters\xspace}
\newcommand{\gnncluster}{\gnnetm cluster\xspace}
\newcommand{\gnnclusters}{\gnnetm clusters\xspace}
\newcommand{\icnetmcluster}{\icnetm cluster\xspace}
\newcommand{\icnetmclusters}{\icnetm clusters\xspace}

\newcommand{\eebkg}{\hbox{$e^+e^- \rightarrow e^+e^-$}}
\newcommand{\mumubkg}{\hbox{$e^+e^- \rightarrow \mu^+\mu^-$}}
\usepackage{etoolbox}   

\begin{document}
\setcounter{tocdepth}{2}

\title{Real-time graph neural networks on FPGAs\\ for the Belle II electromagnetic calorimeter}

\author[a]{I.~Haide\,\orcidlink{0000-0003-0962-6344}}
\author[a]{M.~Neu\,\orcidlink{0000-0002-4564-8009}}
\author[a]{Y.~Unno\,\orcidlink{0000-0003-3355-765X}}
\author[a]{T.~Justinger\,\orcidlink{0009-0008-8739-9400}}
\author[a]{V.~Dajaku\,\orcidlink{0009-0008-2629-5069}}
\author[a]{F.~Baptist\,\orcidlink{0009-0004-5007-5729}}
\author[a]{T.~Lobmaier\,\orcidlink{0009-0000-3104-4688}}
\author[a]{J.~Becker\,\orcidlink{0000-0002-5082-5487}}
\author[[a,*]{T.~Ferber\note{corresponding author}\,\orcidlink{0000-0002-6849-0427}}

\author[a]{H.~Bae\,\orcidlink{0000-0003-1393-8631}}
\author[a]{A.~Beaubien\,\orcidlink{0000-0001-9438-089X}}
\author[a]{J.~Eppelt\,\orcidlink{0000-0001-8368-3721}}
\author[a]{R.~Giordano\,\orcidlink{0000-0002-5496-7247}}
\author[a]{G.~Heine\,\orcidlink{0009-0009-1827-2008}}
\author[a]{T.~Koga\,\orcidlink{0000-0002-1644-2001}}
\author[a]{Y.-T.~Lai\,\orcidlink{0000-0001-9553-3421}}
\author[a]{K.~Miyabayashi\,\orcidlink{0000-0003-4352-734X}}
\author[a]{H.~Nakazawa\,\orcidlink{0000-0003-1684-6628}}
\author[a]{M.~Remnev\,\orcidlink{0000-0001-6975-1724}}
\author[a]{L.~Reuter\,\orcidlink{0000-0002-5930-6237}}
\author[a]{K.~Unger\,\orcidlink{0000-0001-7378-6671}}
\author[a]{R.~van~Tonder\,\orcidlink{0000-0002-7448-4816}}

\affiliation[a]{\url{https://www.belle2.org}}

\emailAdd{torben.ferber@kit.edu}

\abstract{
We present the development and evaluation of a real-time Graph Neural Network-based trigger module for the electromagnetic calorimeter of the Belle\,II experiment at the SuperKEKB collider.
The algorithm processes calorimeter trigger cells as graph nodes to perform clustering, feature extraction, and per-cluster signal classification with deterministic latency.
The model predicts cluster positions and energies and provides a signal classification score, enabling a more flexible clustering strategy than the baseline trigger algorithm.
Implemented on an FPGA and integrated into the Belle\,II trigger readout infrastructure for synchronous operation, the system sustains the \SI{8}{\mega\hertz} trigger throughput with an end-to-end latency of $3.168\,\mu$s.
The performance is evaluated on simulated events and collision data.
The energy resolution is comparable to the baseline trigger, while the position resolution for high-energy clusters improves by up to 18\,\% in the central detector region.
Cluster purity increases by up to 20\,\% at low energies for isolated clusters, and cluster efficiency improves by up to 20\,\% for overlapping clusters.
The signal classifier enables additional background suppression at fixed signal retention.
These results demonstrate a first step towards GNN-based real-time reconstruction on FPGAs in a collider trigger.
While the end-to-end latency exceeds the trigger decision budget, the system already sustains full operational conditions with $100\,\%$ uptime.}


\maketitle

\section{Introduction}
\label{sec:introduction}
Calorimeters are a main component of modern high-energy physics experiments, designed to measure the energy of particles through their interactions with dense absorber materials. 
In collider experiments, electromagnetic calorimeters are specifically optimized to detect electrons and photons by capturing their energy via electromagnetic showers. 
Due to the high event rates and data volumes in such environments, trigger systems are used to perform very fast event selection during data acquisition. 
These systems identify physics events of interest among a large number of background events before persistent data storage.

The \belletwo experiment~\cite{Belle-II:2010dht} is located at the SuperKEKB collider~\cite{Akai:2018mbz} in Tsukuba, Japan, an asymmetric-energy electron–positron collider primarily designed for high-precision flavour physics including rare decays and missing energy searches. 
\superkekb collides 4\,GeV positrons with 7\,GeV electrons at a center-of-mass energy near the $\Upsilon(4S)$ resonance at approximately 10.58\,GeV. 
Compared to its predecessor KEKB, \superkekb aims to increase the instantaneous luminosity by an order of magnitude to boost sensitivity to rare processes. 
However, this increase in luminosity leads to a corresponding increase in beam-induced backgrounds, resulting in a high rate of spurious detector hits and a large number of particles not originating from the interaction point~\cite{natochii_measured_2023}. 
Typical \hbox{$\Upsilon(4S)\to B\bar{B}$} decays produce on average 10 charged particles with momenta from a few tens of MeV up to several GeV, along with approximately 10 photons, mostly originating from neutral pion decays with energies of up to a few hundred MeV. 
In contrast, hypothetical particles such as dark photons~\cite{Fabbrichesi:2020wbt,Ferber:2022ewf,Jaeckel:2023huy}, inelastic Dark Matter~\cite{Duerr:2019dmv,Belle-II:2025bhd}, or axionlike particles dominantly coupled to photons~\cite{Dolan:2017osp, Belle-II:2020jti} typically produce only a few low-energy charged particles or photons, making them appear similar to background events

At bunch crossing frequencies set by the SuperKEKB bucket filling pattern, reaching up to 254.4~MHz, it is infeasible to read out and process the full detector information for every crossing due to bandwidth and storage limitations.
Instead, \belletwo employs a two-stage trigger system to identify potentially interesting events during data acquisition using a reduced subset of detector signals.
It consists of a low-latency first-level~(L1) trigger implemented in custom hardware operating under hard latency constraints, followed by a high-level trigger~(HLT) executed on a CPU farm where soft constraints dominate.
Hard latency constraints refer to strict upper bounds imposed by detector readout buffer limits, whereas soft constraints refer to throughput-driven requirements, such as a maximum average rate of 30,000~events per second, where rate variations are tolerable.

To make real-time decisions about which events to retain, the \trg system evaluates simplified logic conditions based on the available detector data. 
These logic conditions are encoded as so-called trigger bits. 
Trigger bits are binary signals that represent whether certain criteria are met, such as energy sum thresholds, cluster counts, or track multiplicities.
Most \trg input bits are computed on the Global Reconstruction Logic (GRL) \cite{Lai:2025gac}, as they require input from multiple subdetectors. 
These bits are then sent to the Global Decision Logic (GDL), where they are combined into \trg output bits used to issue the final \trg decision.
The current \belletwo electromagnetic calorimeter~(ECL) \trg system is a multi-stage pipeline implemented on FPGAs, responsible for identifying energy depositions in real-time~\cite{Kim:2017uee}. 
It employs a clustering algorithm originally developed for the Belle experiment, which has proven effective under its original operating conditions.
The current implementation is referred to as \icnetm in this work.
However, the system was designed with fixed logic and limited resources, introducing constraints in scalability, clustering granularity, and adaptability to changing conditions. 
For instance, only a maximum of six \triggerclusters can be processed due to FPGA resource limitations, duplicated \triggerclusters hits are not filtered, two close-by photons can not be resolved, and high-energy photon cluster position resolution is limited.
Such limitations have a disproportionately large impact on events with only a few low-energy particles. 
The increase in detector hits originating from beam-induced backgrounds further reduces the performance of the current system.
These limitations motivate the development of a more flexible and accurate approach.

\begin{figure}
    \centering
    \includegraphics[width=\linewidth]{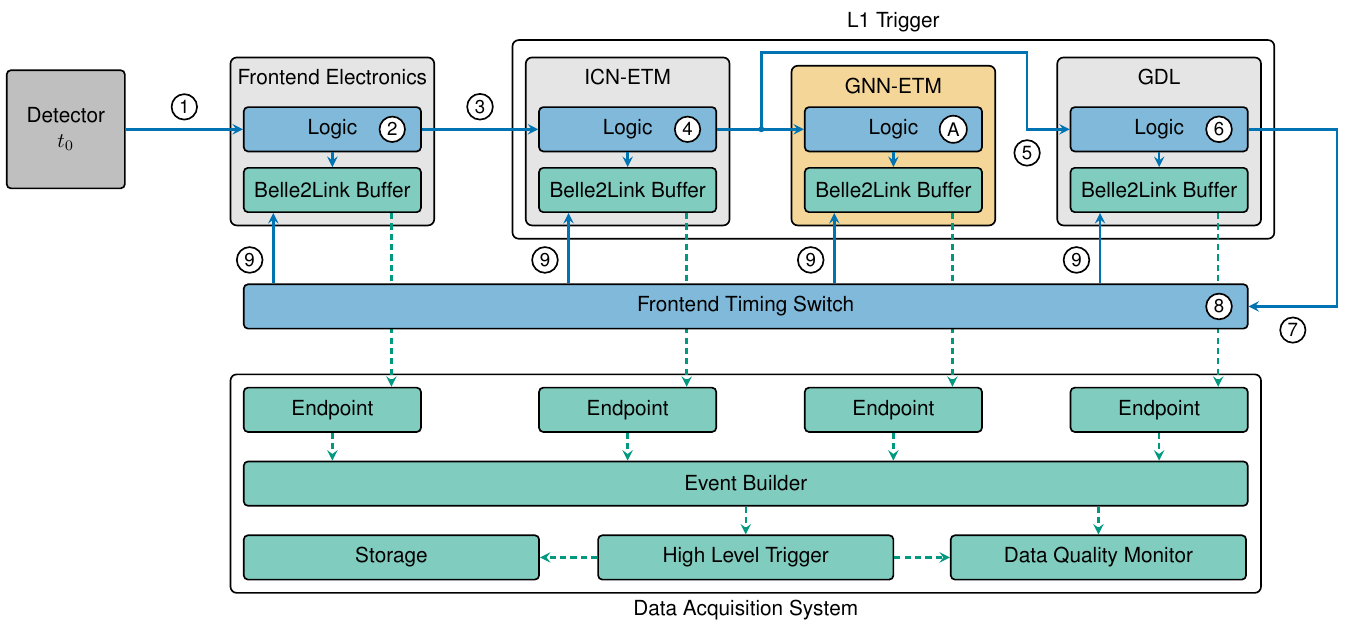}
    \caption{Simplified schematic overview of the data acquisition system and the \trg system at \belletwo. 
    Components of the \trg system, highlighted in blue, must satisfy hard real-time constraints. 
    Components of the data acquisition system, highlighted in green, must satisfy soft real-time constraints. 
    The \gnnetm module developed in this work is highlighted in orange.
    Numbered circles label selected components and connections referenced in the text.
    $t_0$ denotes the time of a detected bunch crossing.
    Belle2Link is described in \cref{sec:gnnetm_hardware:belle2link}.}
    \label{fig:introduction:dataflow}
\end{figure}

A simplified overview of the \belletwo ECL \trg system is shown in \cref{fig:introduction:dataflow}.
For a subsystem trigger to participate in the global trigger decision, four requirements must be met:
\begin{enumerate}
    \item The system must exhibit deterministic latency to satisfy hard real-time deadlines.
    \item The critical-path latency \Circled{1} $\rightarrow$ \Circled{9} must not exceed $R_\mathrm{L} = 5.0\,\mu$s, a constraint imposed by the finite depth of the data buffers on the frontend electronics modules.
    In practice, only a small fraction of this latency is available for a subsystem trigger module.
    For the \belletwo ECL \trg system, the latency \Circled{4} of the module replacing the current \icnetm must not exceed $1.221\,\mu$s.
    \item The system must sustain the full input rate of the respective subdetector, which for the \belletwo ECL \trg amounts to $R_\mathrm{th} = \SI{8}{\mega\hertz}$.
    \item The system must operate with $100\,\%$ uptime, since any disruption halts the entire experiment for the duration of the fault.
\end{enumerate}

To address the limitations of the current algorithm, we have developed and deployed \gnnetm, a real-time Graph Neural Network (GNN)-based ECL \trg module implemented on an FPGA. 
By treating calorimeter trigger cells as graph nodes, \gnnetm exploits spatial correlations in energy deposits to perform clustering and cluster parameter inference with improved position resolution and photon separation, maintaining performance under high beam-induced background.
Although its latency \Circled{A} currently exceeds the \trg constraint, \gnnetm records reduced-precision monitoring data that can be matched with full-precision offline data for performance evaluation.
Deployed in late 2024, \gnnetm has since been operated in parallel with the existing \trg system to validate its real-time performance under realistic experimental conditions.
In this work, we present the design, training, and hardware implementation of \gnnetm for the \belletwo ECL system.
We compare its performance to the existing \trg logic and demonstrate its potential for improving physics event selection at high luminosity.

The remainder of this paper is organized as follows: Section~\ref{sec:related_work} provides an overview of related work on machine learning (ML) for calorimeter, and ML-based \trgs and real-time inference in high-energy physics. 
Section~\ref{sec:belle2_ecl} introduces the Belle II ECL, including its structure, readout segmentation, and relevance for low-latency triggering. 
The architecture and logic of the existing ECL \trg system are described in Section~\ref{sec:belle2_icn}, highlighting key components, clustering algorithms, and hardware limitations. 
Section~\ref{sec:gnnetm_software} details the design and training of \gnnetm, including graph construction, model architecture, and training datasets. 
Section~\ref{sec:gnnetm_hardware} presents the hardware implementation on FPGA and analyzes latency, resource usage, and throughput. 
Section~\ref{sec:results} evaluates the physics performance of \gnnetm using simulation and collision data. 
Finally, Section~\ref{sec:summary} summarizes the results and outlines future directions.

\section{Related work}
\label{sec:related_work}

Machine learning~(ML) techniques are widely used for offline calorimeter-based reconstruction in HEP experiments, without hard latency constraints, for clustering\,\cite{canudas2022graph, Valsecchi_2023, Belle-II:2023cal}, energy regression\,\cite{ereg, Belayneh2019CalorimetryWD}, and particle identification\,\cite{Boldyrev_2020, Novosel:2023cki}.

Calorimeter systems, due to their irregular geometry especially in the detector endcaps and spatially correlated energy deposits, benefit from the relational structure modeled by GNNs\,\cite{Shlomi:2020gdn, Duarte:2020ngm, DeZoort:2023vrm}. 
Dynamic GNNs that construct edge connections based on learned spatial relations rather than fixed geometry for calorimeter clustering and energy regression have shown improved performance over traditional algorithms\,\cite{Wang:2018nkf,Qasim:2019otl}. 
GNN architectures have been evaluated for applications at highly granular calorimeters\,\cite{CMSHGCAL:2024esz} and specifically for photon reconstruction in the \belletwo electromagnetic calorimeter\,\cite{Belle-II:2023cal}. 
This contrasts with static GNNs\,\cite{ExaTrkX:2020nyf}, where edge connectivity is fixed by detector geometry or spatial proximity thresholds.

Methods that support inference over a variable number of entities are designed to identify and reconstruct distinct object instances directly from raw inputs such as detector hits. 
Instead of relying on fixed output structures or predefined object counts, these models learn to associate inputs with dynamically determined object representations. 
This enables end-to-end reconstruction, where the model directly maps detector inputs to identified particles and their properties within a single inference process.
While many such techniques, such as those based on bounding boxes\,\cite{7780460}, operate on grid-like data, object condensation\,\cite{Kieseler:2020wcq} is particularly well-suited for irregular and sparse data structures such as graphs, and has recently been applied in the context of calorimeter clustering\,\cite{Qasim:2022rww}.

The use of ML inference on FPGA hardware has become a viable option for low-latency inference in trigger applications. 
Tools such as hls4ml\,\cite{Duarte:2018ite,fastml_hls4ml} or FINN~\cite{blott:2018} allow for translating trained neural networks into high-level synthesis~(HLS) descriptions suitable for FPGA deployment, with deterministic latency and resource utilization. 
However, these frameworks are currently limited to relatively simple model architectures due to constraints in logic resources, memory bandwidth, and timing closure. 
Additionally, more complex network architectures are often not supported within these frameworks.
To date, deployed models under strict resource constraints with an end-to-end inference latency below 20\,$\mu$s have been standard feed-forward neural networks for regression\,\cite{Kohne:1997ph, Afanasyev:2002zd, Bahr:2024dzg} or autoencoder-based architectures for anomaly detection\,\cite{Zipper:2023ybp}.

Building on these hardware-oriented efforts, recent studies have investigated the feasibility of deploying GNNs under hard latency constraints with inference times on the order of microseconds, relevant for \trg systems\,\cite{Iiyama:2020wap,Que:2022kmo,Neu:2023sfh,Kvapil:2025ukm}.
To our knowledge, GNN-based approaches have so far been validated only on dedicated hardware test benches or standalone evaluation platforms and have not yet been integrated into the data acquisition pipelines of collider experiments.

\section{The Belle II electromagnetic calorimeter}
\label{sec:belle2_ecl}

The \belletwo detector is composed of several subdetectors arranged cylindrically around the beam pipe. 
A detailed description is available in Refs.~\cite{Kou:2018nap, Belle-II:2010dht}. 
The symmetry axis of these subdetectors is defined as the $z$-axis which is pointing approximately in the direction of the electron beam. 
The $x$-axis is horizontal and oriented away from the accelerator center, and the $y$-axis is vertical, pointing upward. 
The longitudinal and transverse directions, as well as the azimuthal angle $\phi$ and polar angle $\theta$, are defined with respect to this coordinate system.

The \belletwo ECL consists of 8736~thallium-doped cesium iodide (CsI(Tl)) crystals, divided into three regions: the forward endcap ($12.4^{\circ} < \theta < 31.4^{\circ}$), the barrel ($32.2^{\circ} < \theta < 128.7^{\circ}$), and the backward endcap ($130.7^{\circ} < \theta < 155.1^{\circ}$). 
Each crystal has a trapezoidal shape with a nominal cross-section of approximately $6 \times 6$~cm$^2$ and a length of 30~cm, corresponding to 16.1 radiation lengths. 
Barrel crystals are mostly uniform in geometry, whereas endcap crystals vary in shape and mass, ranging from 4.03~kg to 5.94~kg~\cite{ikeda:1999}. 
Additionally, the endcaps include more upstream passive material than the barrel.
All crystals are oriented toward the interaction point with small tilts in $\theta$ to minimize efficiency losses due to gaps between adjacent crystals. 
In the barrel, an additional small tilt is applied in the $\phi$ direction. 
Light produced by scintillation in the CsI(Tl) material is collected by two photodiodes attached to the rear face of each crystal. 
Signals from all 8736 calorimeter crystals are read out and sent to 576 ShaperDSP modules, where digital signal processing~(DSP) extracts the pulse amplitude and timing information for each channel.
Each ShaperDSP module generates two versions of the shaped signal: one integrating the signal over 1\,$\mu$s for offline processing, and another using a shorter shaping time of 0.2\,$\mu$s for the use in the \trg~\cite{Aulchenko:2017lmh}.

For the High Level Trigger~(HLT) and offline data processing, the energy and the time of each crystal signal relative to the event \trg time are stored.
In offline reconstruction, photon interactions typically result in energy deposits spanning up to $5 \times 5$ crystals. 
The clustering algorithm aims to associate all energy from a given photon while excluding contributions from other particles and beam background. 
In low-background conditions, around 17\% of crystals register energy above 1~MeV, increasing to about 30\% in recent data-taking periods with high beam background conditions, which complicates clustering. 
A detailed description of the baseline offline reconstruction algorithm can be found in \cite{Kou:2018nap,Belle-II:2023cal}.

\begin{figure*}[ht!]
    \centering
    \includegraphics[width=\textwidth]{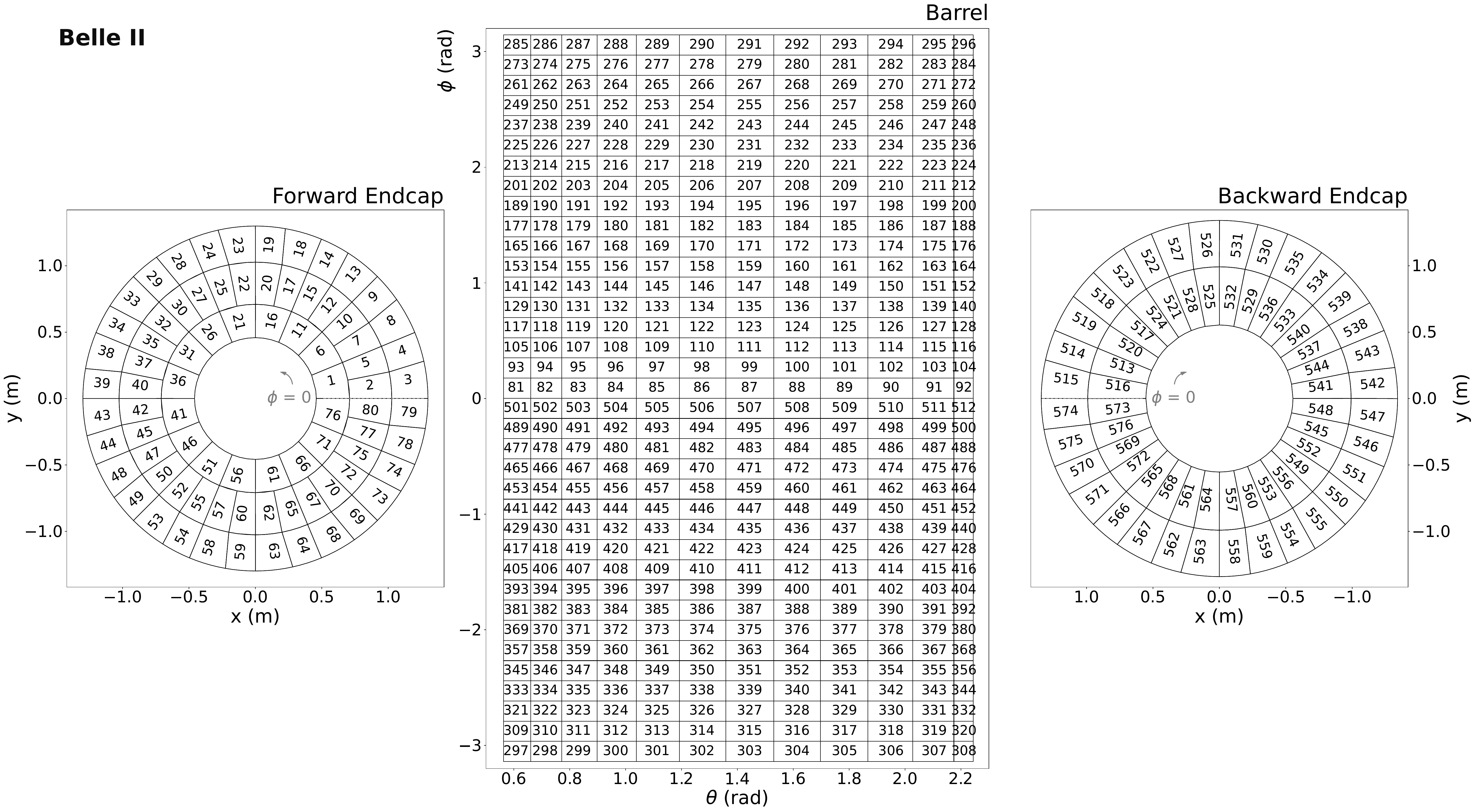}
    \caption{Placement of the 576~TCs with their corresponding TC-ID.}
    \label{fig:ecl_tc}
\end{figure*}

\begin{figure*}[ht!]
    \centering
    \includegraphics[width=\textwidth]{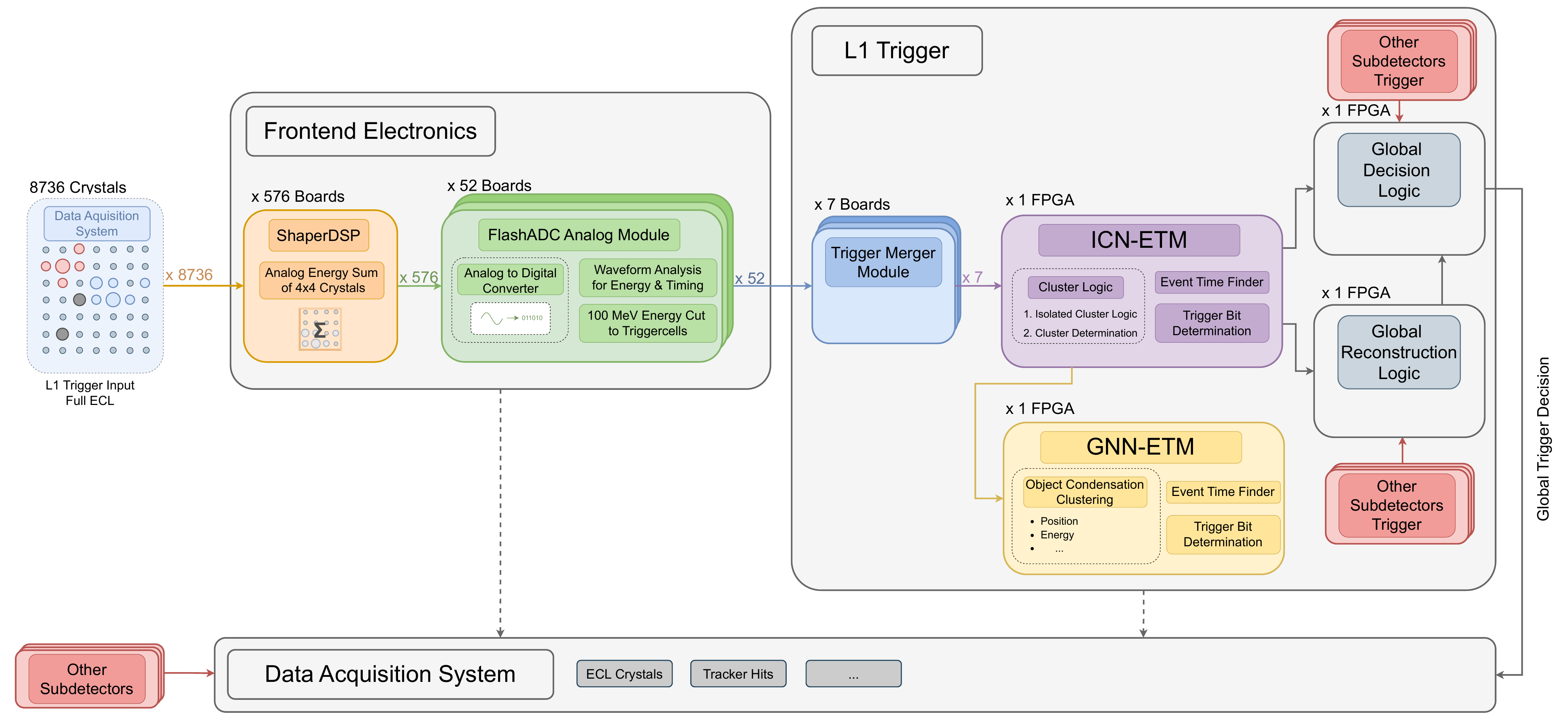}
    \caption{An overview of the modules in the ECL readout chain. In comparison to \cref{fig:introduction:dataflow} some details on the trigger signal distribution are omitted for clarity.}
    \label{fig:icn_chain}
\end{figure*}

For the \trg, up to 16 adjacent crystals are grouped and analogously summed into a single trigger cell~(TC) and forwarded to a front end analysis module~(FAM), as described in Ref.~\cite{Aulchenko:2017lmh}. 
The placement of the TCs is shown in \cref{fig:ecl_tc}.
Owing to the long scintillation decay time of the CsI(Tl) crystals of a few $\mu$s~\cite{Beylin:2004in} and the response of the shaper electronics, a sampling rate of 8~MHz is sufficient for this system~\cite{Aulchenko:2017lmh}.
The full \trg processing chain operates synchronously at this frequency to meet the \belletwo \trg timing requirements.
The analog signals are digitized by a fast analog-to-digital converter~(FADC) and processed by an FPGA, which performs waveform analysis to extract energy and timing information. 
Each FAM receives 12 TC inputs. 
An energy threshold of 100~MeV is applied to each input to reduce beam-related background and suppress electronic noise. 
The resulting data are passed to the Trigger Merger Module~(TMM), which aggregates information from multiple FAMs and forwards it to the ECL Trigger Module~(ETM). 
The ETM performs the final clustering and generates the \trg decision. 
The output data of the ETM consists of general event information, clustering information, and \trg bits derived solely from ECL information.
The ETM sends information both to the GDL and a subset of this information to the GRL.
An overview of the different modules and their functions in the ECL \trg with the number of boards per module is shown in Fig.~\ref{fig:icn_chain}.
 
To distinguish the current ETM implementation from the GNN-based algorithm described in Section~\ref{sec:gnnetm_software}, we refer to the current ETM implementation as \icnetm and describe it in Section~\ref{sec:belle2_icn}.

\section{Existing level~1 calorimeter trigger}
\label{sec:belle2_icn}

\begin{figure*}[ht!]
    \centering

    \begin{subfigure}{0.44\textwidth}
        \centering
        \includegraphics[width=\linewidth]{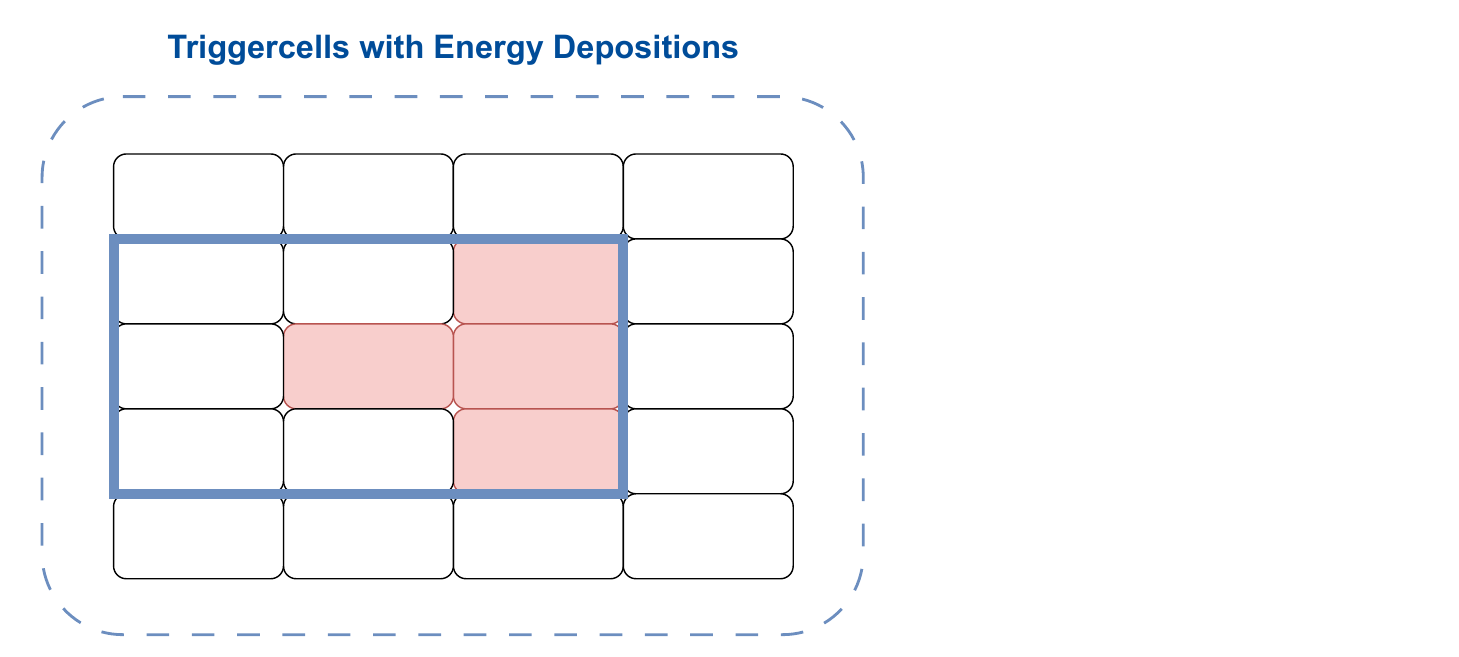}
        \caption{Move $3\times3$ window.}
        \label{fig:icn_logic_1}
    \end{subfigure}
    \hspace{0.9cm}
    \begin{subfigure}{0.44\textwidth}
        \centering
        \includegraphics[width=\linewidth]{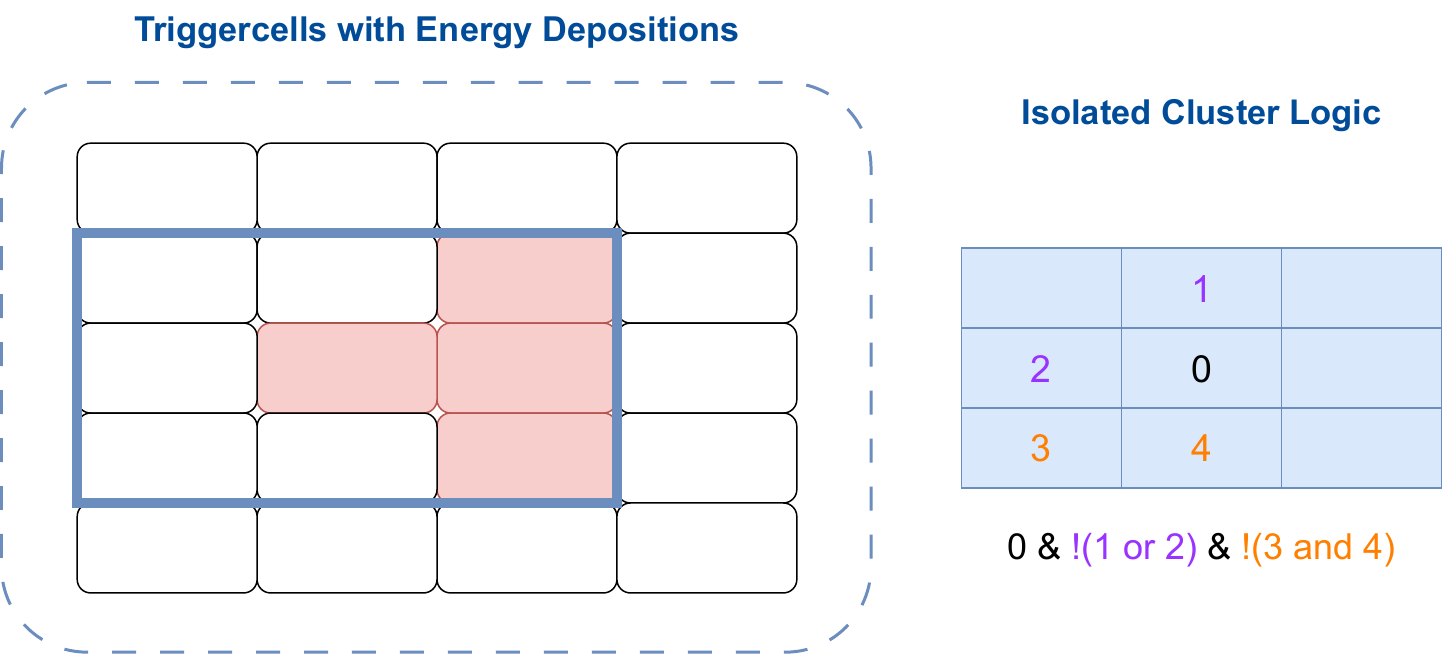}
        \caption{Check ICN hit condition in $3\times3$ window.}
        \label{fig:icn_logic_2}
    \end{subfigure}

    \vspace{0.2cm}

    \begin{subfigure}{0.44\textwidth}
        \centering
        \includegraphics[width=\linewidth]{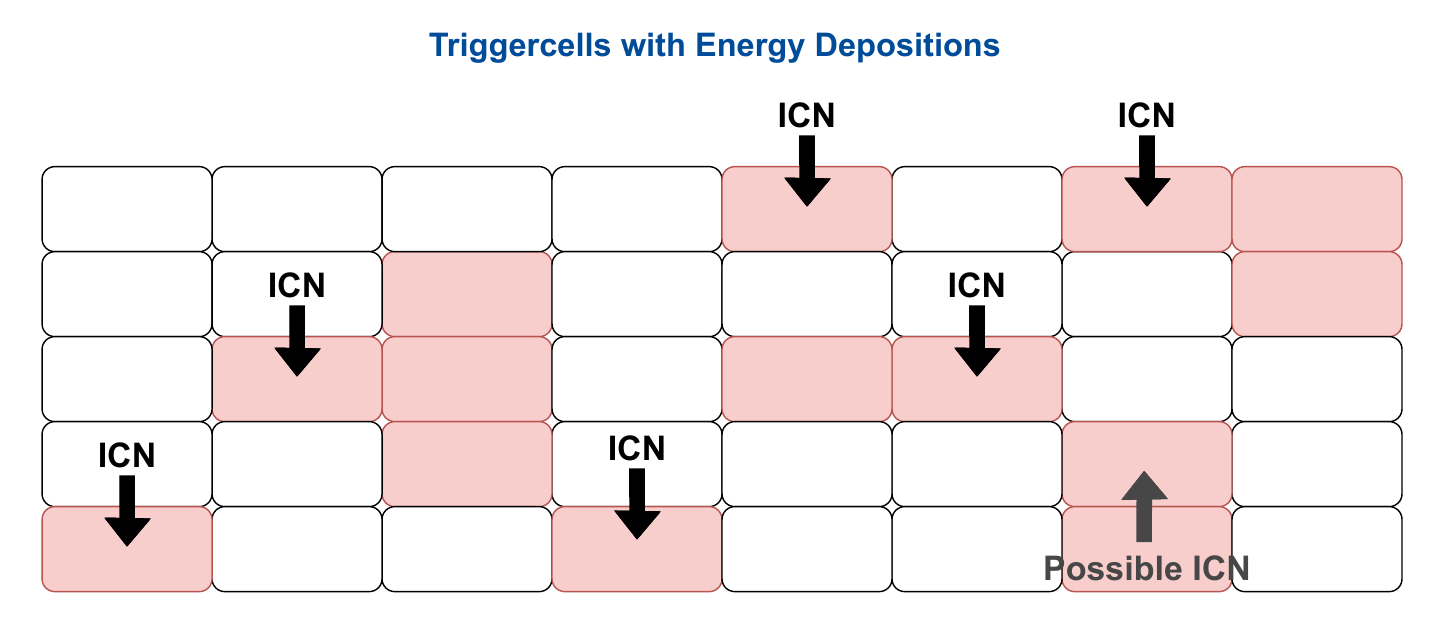}
        \caption{Select up to six ICN hits.}
        \label{fig:icn_logic_3}
    \end{subfigure}
    \hspace{0.9cm}
    \begin{subfigure}{0.44\textwidth}
        \centering
        \includegraphics[width=\linewidth]{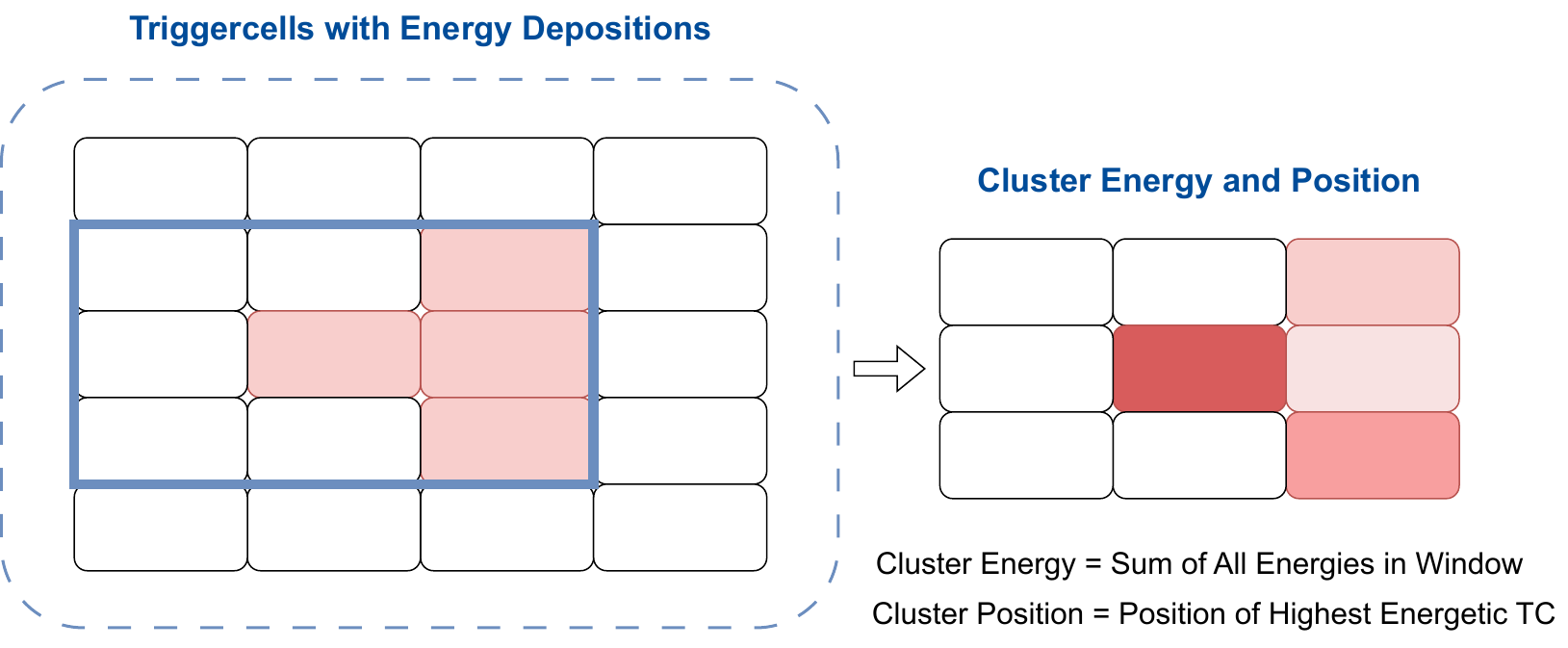}
        \caption{Calculate cluster properties if TC0 is the highest-energy TC within the evaluation window.}
        \label{fig:icn_logic_4}
    \end{subfigure}

    \vspace{0.2cm}

    \begin{subfigure}{0.44\textwidth}
        \centering
        \includegraphics[width=\linewidth]{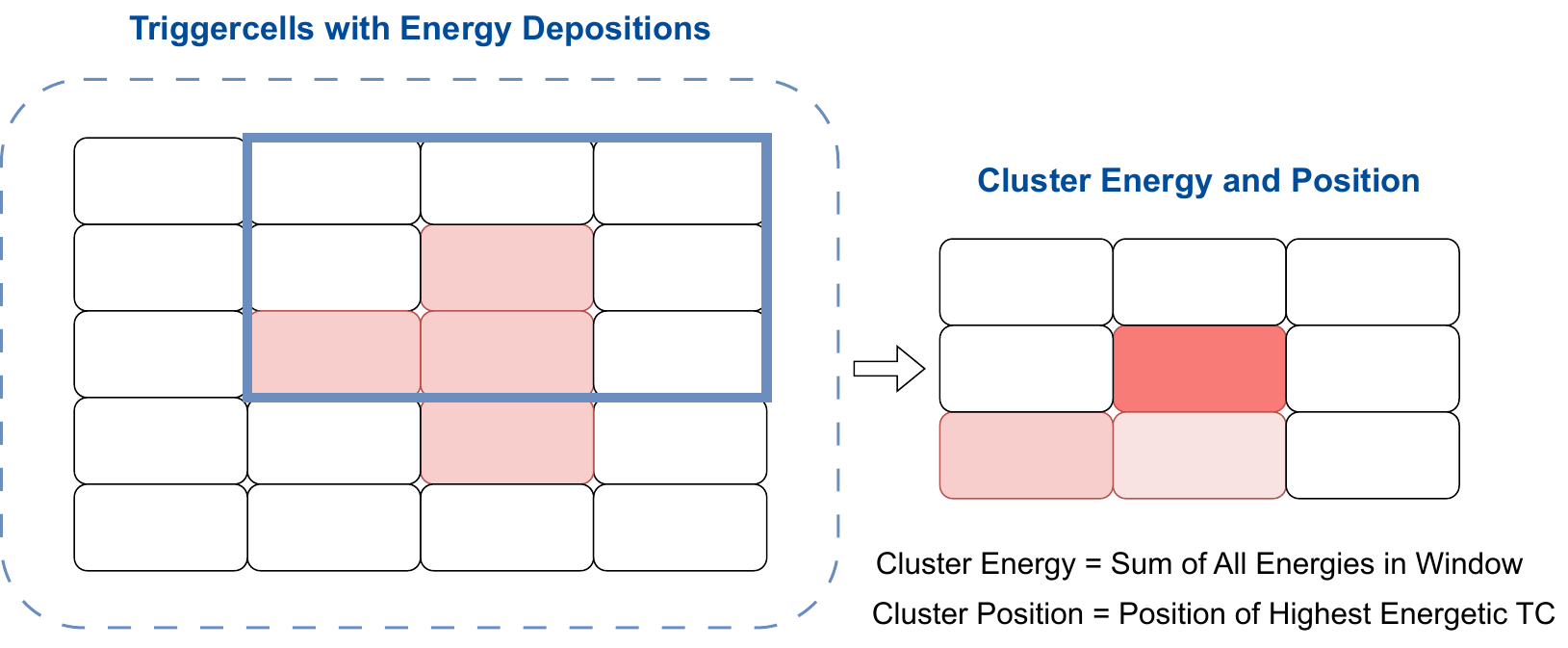}
        \caption{If TC0 is not highest-energy TC, shift window and recalculate cluster properties.}
        \label{fig:icn_logic_5}
    \end{subfigure}

    \caption{Illustration of the isolated cluster logic: (\subref{fig:icn_logic_1}) Move window, (\subref{fig:icn_logic_2}) check ICN hit condition, (\subref{fig:icn_logic_3}) select ICN hits, (\subref{fig:icn_logic_4}) calculate cluster properties, (\subref{fig:icn_logic_5}) calculate cluster properties after shifting to highest-energy TC.}
    \label{fig:icn_logic}
\end{figure*}

\begin{figure}[ht!]
    \centering
    \includegraphics[width=0.6\linewidth]{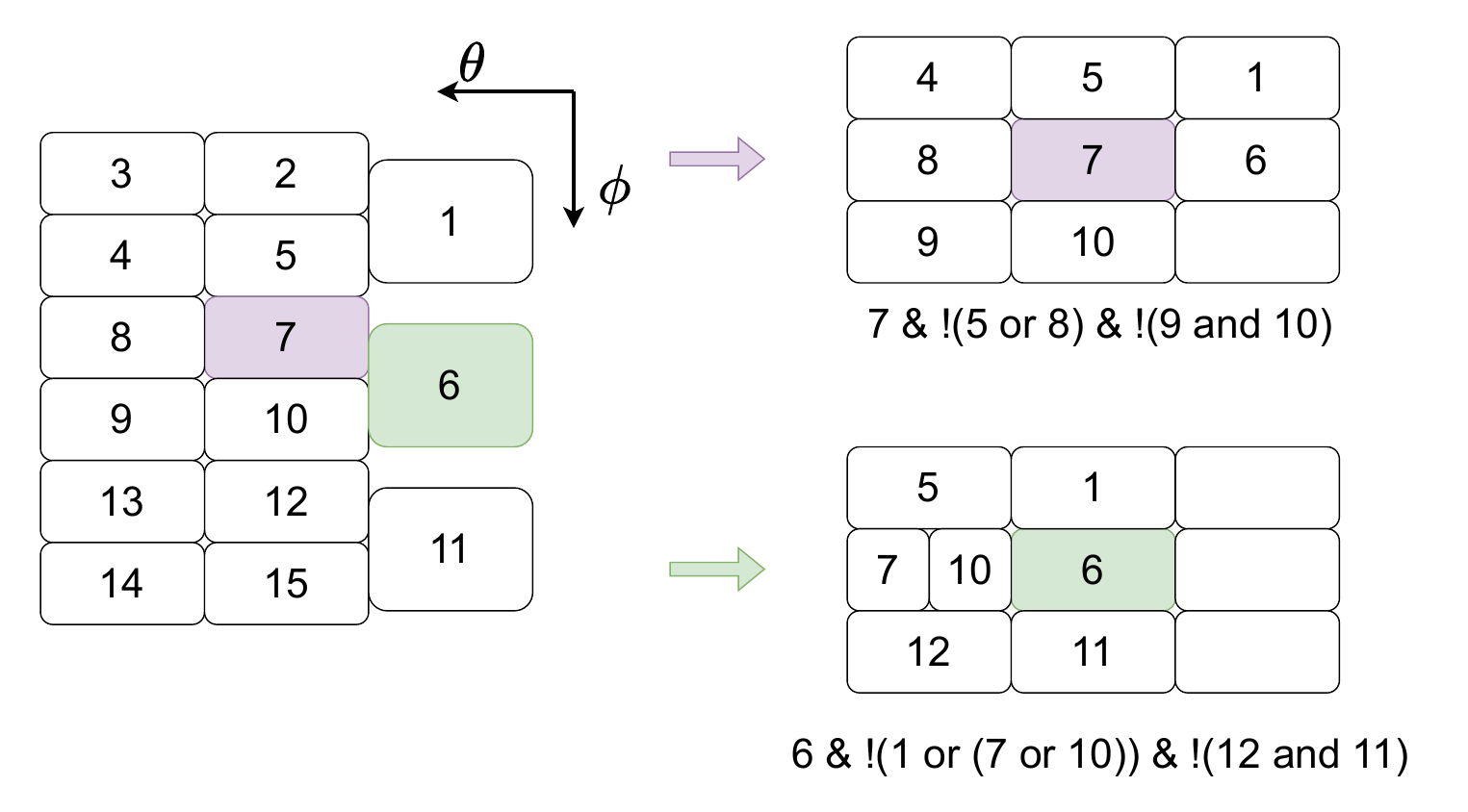}
    \caption{Illustration of the ICN logic in the forward endcap. 
    On the left the TCs in the forward endcap with their corresponding TC ID are shown in the $\theta$-$\phi$ plane. 
    On the right the ICN decision window and the corresponding decision logic is shown for the example TCs.
    }
    \label{fig:icn_logic_fwd_endcap}
\end{figure}

The electromagnetic calorimeter \trg is one of the three main subdetector \trg systems at Belle~II and provides real time reconstruction of calorimetric energy deposits from neutral and charged particles.
Its primary role is the reconstruction of photons and, in combination with the track \trg~\cite{Lai:2025gac}, the provision of complementary information for event reconstruction, particularly for low track multiplicity events where the track \trg efficiency is reduced.
In addition, the ECL \trg performs standalone identification of \eebkg{} events for fast instantaneous and integrated luminosity measurements~\cite{Kovalenko:2025tzc} and rejects a large fraction of these events to limit the overall \trg rate.
The existing ECL \trg logic, known as ICN~(Isolated Cluster Number) logic, is implemented on the \icnetm module and is based on the design from the Belle experiment~\cite{cheon_electromagnetic_2002}. 
Its primary function is to detect isolated energy depositions by identifying connected regions of trigger cells~(TCs), which are then reconstructed as \icnetmclusters. 
The \icnetm receives TC data from the Trigger Merger Modules (TMMs), which includes hit flags, energy, and timing. 
It processes the ECL in overlapping $3{\times}3$ TC windows in the $\phi$-$\theta$ plane with a step size of one TC, shown in \ref{fig:icn_logic}. 
Each window is evaluated using a deterministic decision logic that considers five specific TCs: the center TC (TC0), top center (TC1), middle left (TC2), and the two left TCs in the bottom row (TC3 and TC4).
In case the window spans a wider area than TCs available, for example in the backward endcap or on the outermost column in the barrel, the missing TCs are defined as not hit. 
In the forward endcap, where the crystal arrangement is geometrically irregular, the algorithm is adapted as shown in \cref{fig:icn_logic_fwd_endcap}.
For TCs in the innermost ring of the forward endcap, the two adjacent TCs in the $\theta$ direction are combined with an OR operation into TC2.
While the TCs, and all subsequent \trg logic, are read out and processed in 125\,ns windows, called \datawindows, each TC is held persistent for two \datawindows. 
This allows the ICN hit determination and clustering logic to operate over two adjacent \datawindows, corresponding to a 250~ns timing window, called \triggerwindow. 
 
The procedure to retrieve \icnetmcluster information is shown schematically in Fig.~\ref{fig:icn_logic}. 
The three detector regions are handled separately, which forbids an \icnetmcluster spanning the gap between the barrel and the forward or backward endcap. 
The following steps are performed in this order:
\begin{enumerate}[label=(\alph*)]
    \item Move $3\times3$ window to next TC. 
    \item A $3\times3$ window is flagged as an \icnhit if the following three conditions are satisfied:
    \begin{enumerate}[i)]        
        \item TC0 is hit.
        \item Neither TC1 nor TC2 is hit.
        \item TC3 and TC4 are not both hit.
    \end{enumerate}
    An \icnhit corresponds to TC0 if it fulfills the above conditions. 
    There are no requirements forbidding the existence of one or several \icnhits in the same $3\times3$ window, if this TC also fulfills the ICN conditions.
    \item Of all \icnhits, select up to six, based on TC-ID: first from the barrel (TC-IDs 81–512), then the forward endcap (TC-IDs 1–80), and finally the backward endcap (TC-IDs 513–576), following the order of increasing beam-background. No sorting by energy is applied.
    These selected \icnhits collectively form the \icncluster passed to the subsequent clustering stage.
    \item For each of the selected \icncluster (up to six), check if TC0 is the highest-energy TC within its evaluation window. The cluster energy is computed as the sum of all hit TCs in the window. 
    The cluster position is taken as that of the highest-energy TC, defined as the center of its front face oriented toward the interaction point.  
    \item If the highest-energy TC is not at the center, the window is shifted to center on it. The cluster energy is then recalculated, and the position is updated to that of the new center TC. 
    If a higher-energy TC appears within the shifted window, no further shifting is performed.
\end{enumerate}

\begin{figure*}[ht!]
    \centering

     \begin{subfigure}{0.25\textwidth}
        \centering
        \includegraphics[width=\linewidth]{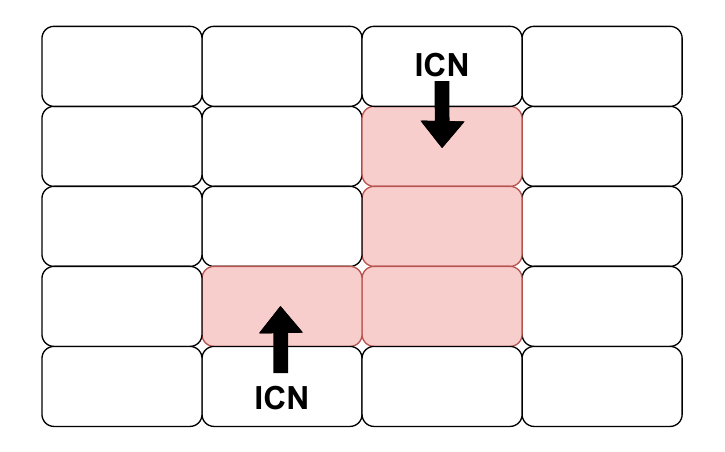}
        \caption{Window overlap with partial variation.}
        \label{fig:icn_logic_limitation_1}
    \end{subfigure}
    \hspace{0.9cm}
    \begin{subfigure}{0.25\textwidth}
        \centering
        \includegraphics[width=\linewidth]{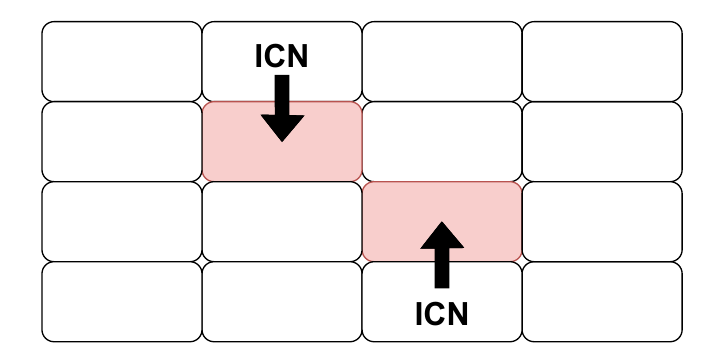}
        \caption{Exact duplication from full window match.}
        \label{fig:icn_logic_limitation_2}
    \end{subfigure}
    \caption{Illustration of duplicate hit generation in the ICN logic: (\subref{fig:icn_logic_limitation_1}) window overlap with partial variation and (\subref{fig:icn_logic_limitation_2}) exact duplication from full window match where two identical ICN hits are returned when the evaluation windows fully coincide. In case (\subref{fig:icn_logic_limitation_1}), the rotated variants only produce one hit. 
    }
    \label{fig:icn_logic_limitation}
\end{figure*}

The simplicity and speed of the ICN logic introduce a known limitation related to duplicate hit detection. 
Two cases may occur in which the logic incorrectly returns \icnhits with identical or nearly identical information, as illustrated in Fig.~\ref{fig:icn_logic_limitation}. 
In the first case, overlapping $3{\times}3$ windows lead to near identical \icnhits with slight variations in the reconstructed energy or position. 
In the second, both \icnhits are exactly identical. 
These duplicates typically result from two nearby particles hitting the ECL. 
While two \icnetmcluster are expected in such cases, the reconstructed parameters are incorrectly identical due to the \icnetmcluster position being taken from the highest-energy TC within the evaluation window. 
In the most frequent overlap case, the two-TC diagonal pattern, the recorded energy is twice the actual deposit, occurring at least once in about 31\% of events in simulated $B\bar B$ events with \textit{simulated beam backgrounds}~(see Sec.~\cref{sec:gnnetm_software_training}), though the rate is significantly lower in low-multiplicity final states.
However, with increasing background occupancy, similar patterns can be mimicked by a single particle, making such cases more likely. 
Since these duplicates are not removed before the final clustering stage and are treated as independent, they impact cluster number counting \trg lines.

In addition to energy and timing information, the \trg must also determine the collision time of the event to ensure correct bunch crossing identification. 
The input to the ICN algorithm is processed in one \triggerwindow, using two adjacent \datawindows as input. 
The collision time is determined by the highest-energy TC within the \triggerwindow. 
If two TCs have the same energy and belong to different \datawindows, the earlier TC is chosen. 
If they are in the same \datawindow, a deterministic ordering based on TC number is applied. 
The priority order is: 81–512 (barrel), 76–80 (forward edge), 1–75 (forward endcap), 573–576 (backward edge), and 513–572 (backward endcap).

To avoid boundary effects and possible events ambiguities from two adjacent \triggerwindows, no two consecutive \triggerwindows may result in trigger signals sent to the GDL. 
To enforce this, three adjacent \datawindows $W_{1, 2,3}$ are evaluated. 
Each \datawindow has an associated energy, $E_{1,2,3}$, defined as the sum of all TC energies within the \datawindow, and a timing $T_{1,2,3}$, defined as the timing of the highest-energy TC in that \datawindow.
From these, two possible \triggerwindows are formed: \triggerwindow $TW_A$ (\datawindows $W_1$ and $W_2$) and \triggerwindow $TW_B$ (\datawindows $W_2$ and $W_3$), with \datawindow $W_2$ shared between both. 
The event timing for \triggerwindow $TW_A$ is first determined using the procedure described previously.
If the event timing is given by $T_1$, \triggerwindow $TW_A$ is selected, and no trigger signal is allowed for \triggerwindow $TW_B$ to prevent consecutive \trgs. 
If the timing is $T_2$, the total energies of \triggerwindows $TW_A$ and $TW_B$ are compared. 
If $E_\mathrm{A} = E_1 + E_2 \geq E_\mathrm{B} = E_2 + E_3$, \triggerwindow $TW_A$ is selected. 
Otherwise, if $E_\mathrm{A} < E_\mathrm{B}$, \triggerwindow $TW_B$ is selected. 

In addition, a set of \trg bits is calculated directly on the ECL \trg hardware. 
These pure ECL \trg bits rely solely on information from the electromagnetic calorimeter and are computed on the \icnetm board. Once determined, they are sent to the GDL and processed by the same logic used for the final \trg decision.
To enable the correct selection of physics processes, the \icnetmcluster energies and positions must be converted from the laboratory frame to the center-of-mass frame. This transformation is performed using a lookup table that assigns each TC ID a polar angle, azimuthal angle, and energy conversion factor.
To suppress background-induced \trgs, many ECL \trg bits apply angular acceptance selections. 
TCs or \icnetmclusters located close to the beam pipe, which receive the highest background rates, are then excluded from the logic of these \trg bits.
The latency of the \icnetm logic, combining the ICN hit and clustering logic and the calculation of the \trg bits, is 554\,ns.

When a \trg signal is issued, the TCs of eight adjacent \datawindows are stored in raw data. 
In the majority of cases, the two \datawindows forming the \triggerwindow are the center two \datawindows, with three additional \datawindows stored before and after. 
The \triggerwindow can, however, be shifted in either direction. 
The information, which two \datawindows are part of the \triggerwindow, is also stored in data. 
The \icnetmclusters are only written out for the \triggerwindow.
The ECL crystal signals and the signals from all other subdetectors are read out within a 4$\,\mu$s window centered on the collision time for offline processing.
The timing of ECL crystals is then given relative to the collision time determined by the \trg.

\section{Design and training of the graph neural network level~1 calorimeter trigger}
\label{sec:gnnetm_software}

In the following section, we first describe the GNN training and evaluation data, architecture and the model inference.
We then describe the post-processing steps to choose the final cluster candidates and extract the cluster parameter information. 

\subsection{Training and evaluation datasets}
\label{sec:gnnetm_software_training}
Simulated \belletwo events are used for both the training and evaluation of the \trg algorithms. 
The interactions of final-state particles with the full detector geometry are modeled using \texttt{GEANT4}~\cite{Agostinelli:2002hh}. 
The resulting signals are processed together with a simulation of the detector response to generate digitized hits within the \belletwo Analysis Software Framework, \texttt{basf2}~\cite{basf21, basf22}.
Simulated beam background events~\cite{Liptak:2021tog, Natochii:2022vcs}, approximating 2021 collider conditions, are overlaid onto the signal particles, corresponding to an instantaneous luminosity of $\mathcal{L}_{\text{beam}} = 1.06 \times 10^{34}\,\text{cm}^{-2}\,\text{s}^{-1}$. 

Additional simulation of electronics noise models the behavior of the ECL readout chain. 
Beam backgrounds produce either cluster signatures or isolated crystal hits, typically from low-energy photons or neutrons. 
Electronics noise mostly alters the waveform baseline, degrading energy and timing resolution.

To train the network, the datasets are constructed to represent different cluster signatures over a wide energy range and over the full polar angle range. 
Two key challenges for improving the ECL \trg performance are:
\begin{enumerate}
    \item Increased beam backgrounds produce a higher number of reconstructed clusters, including many above 100\,MeV, that do not originate from the primary collision. The network must distinguish clusters from beam backgrounds from those produced by collision events. Since low-energy clusters are also characteristic of certain dark sector decays, a simple energy threshold is insufficient.
    \item The current \trg algorithm cannot separate clusters depositing energy in adjacent TCs, reducing efficiency for non-isolated photons, such as those originating from boosted neutral pions or light axion-like particles~\cite{Dolan:2017osp}.
\end{enumerate}
These samples are datasets that do not enforce conservation laws to avoid potential bias towards certain physics signatures.

For the first sample~\textit{(Poisson Uniform Photon Sample)}, between 1 and 6 photons are generated, with the number drawn from a uniform distribution. 
Photons are generated with a starting position at the nominal beam interaction point at time $t=0$. 
Each photon energy is sampled uniformly between 0.05 and 7\,GeV. 
The azimuthal angle $\phi$ is drawn uniformly between 0 and 360$^\circ$, covering the full $\phi$ range of the detector. 
The polar angle $\theta$ is sampled uniformly between 5$^\circ$ and 175$^\circ$, extending beyond the nominal ECL acceptance range to include signatures where particles are emitted close to the beam pipe that still produce energy depositions in the calorimeter.
Since the addition of beam background overlays in the simulation significantly increases the number of background-induced clusters, additional signal photons are generated to maintain a balanced ratio of signal and background clusters. 
Without this correction, we found a bias towards background-dominated event topologies and energy distributions during training.
Based on simulation studies, we determine the expected energy spectrum and the number of additional beam background-induced clusters as input to the generation of these additional signal photons. 
The energy probability density function is modeled as an exponential distribution, $f(E) = \exp(a - bE)$, where $E$ denotes the \offlinecluster energy. 
The \offlinecluster multiplicity per event is modeled as a Poisson distribution, $P(n) = \lambda^n / n! \exp(-\lambda)$, where $n$ denotes the number of \offlineclusters per event. 
The parameters $a$, $b$, and $\lambda$ are determined as $a = 5.0$, $b = 32.6$ and $\lambda = 3$. This prevents the network from learning a simple energy cut to distinguish between signal and background-induced clusters.

The second sample~\textit{(Non-Isolated Photon Sample)} is generated following the same procedure as the first sample: between 1 and 6 photons are generated at the nominal interaction point at time $t=0$, with energies sampled uniformly between 0.05 and 7\,GeV, azimuthal angles $\phi$ between 0 and 360$^\circ$, and polar angles $\theta$ between 5$^\circ$ and 175$^\circ$.
However, in addition to these photons, one extra photon pair is generated per event.
The energy of the pair is sampled uniformly between 0.05 and 7\,GeV, and both photons are assigned this same energy.
The opening angle between the two photons is sampled from a uniform distribution between 0.05 and 0.2 radians (2.86$^\circ$ to 11.45$^\circ$), enhancing the fraction of non-isolated cluster signatures. 
No additional signal photons based on beam background-induced cluster distributions are added for this sample.

For each sample, 40\,000 events were simulated for each photon multiplicity between 1 and 6, resulting in a total of 480\,000 events. 
Events without any TCs or without reconstructed offline ECL clusters are discarded, leaving 468\,000 events for training and evaluation. 
The \textit{Combined Photon Sample} consists of an equal number of events from each of the two samples. 
We use 90\% of our combined sample for training, and 10\% for validation of our models.

Each TC is assigned a training label using \offlineclusters as follows:
\begin{enumerate}
  \item For every TC, construct a pseudo-TC by summing the reconstructed energies of all ECL crystals mapped to that TC. 
  \item For each \offlinecluster $c$, compute the overlap energy
  \[
      E_{\text{overlap}}(c, \mathrm{TC}) = \sum_{i \in \mathrm{TC}} w_{i,c}\, E_i ,
  \]
  where $E_i$ is the reconstructed energy of crystal $i$ and $w_{i,c}$ is the fraction of that crystal's energy assigned to \offlinecluster $c$. 
  The TC is matched to the \offlinecluster $c$ with the largest overlap energy.
  \item Assign the TC to this \offlinecluster if $E_{\text{overlap}}(c, \mathrm{TC})$ exceeds the reconstructed background contribution in the pseudo-TC, otherwise no label is assigned. The background contribution is defined as
  \[
      E_{\text{bkg}}(\mathrm{TC}) = \sum_{i \in \mathrm{TC}} \bigl(1 - \sum_{c} w_{i,c}\bigr)\, E_i ,
  \]
  where the term in parentheses represents the fraction of crystal $i$ not assigned to any \offlinecluster.
  \item Regression targets for \gnncluster energy and position are taken from the offline \texttt{basf2} reconstruction. 
  \item A \offlinecluster is defined as \textit{signal} if the total simulated energy deposited by a single particle in all crystals of the \offlinecluster exceeds 20\% of the total \offlinecluster energy. Otherwise, the \offlinecluster is labeled as \textit{background}. This signal definition is simulation-dependent, whereas all previous steps rely solely on offline reconstruction.
\end{enumerate}

For the evaluation in \cref{sec:results_simulation}, three additional simplified simulated datasets are employed to reduce the effects of the dataset construction in the final evaluation metrics. 
For the optimization of the signal classifier threshold, a \textit{Uniform Photon Sample} without the additional simulated low-energy Poisson-distributed energies is used. The 1-6 simulated photons are simulated identically to the uniform photons in the \textit{Poisson Uniform Photon Sample} for the training dataset.
To evaluate the performance of the \gnnetm on single \offlineclusters in comparison to the \icnetm, a \textit{Single Photon Sample} dataset is simulated.
This sample contains one simulated photon with a generated energy sampled uniformly between 0.05 and 7\,GeV, a generated azimuthal angle $\phi$ between 0 and 360$^\circ$, and a polar angle $\theta$ between 5$^\circ$ and 175$^\circ$. 
For the performance evaluation of two close-by clusters, a \textit{(Overlap Diphoton Sample)} dataset is simulated. 
In this dataset, each event contains two photons, both having the same generated energy sampled uniformly between 0.05 and 5\,GeV. 
The opening angle between both photons is sampled from a uniform distribution between 0.05 and 0.2 radians, as in the~\textit{Non-Isolated Photon Sample}.
For the evaluation of the \gnnetm on events containing physics processes, we generate muon pair \mumubkg{} samples using the KKMC event generator \cite{jadach_precision_2000}. We additionally generate Bhabha events \eebkg{} using the BABAYAGA@NLO event generator \cite{balossini_matching_2006} to test the performance of \gnnetm on the most abundant process in the Belle~II detector. For the \eebkg{} events we require the polar angle $\theta$ of the generated particles to be within \qty{12.4}{\degree} and \qty{155.1}{\degree} to increase the probability of particles interacting with the detector. 
To evaluate the performance of the \trg algorithms on \mumubkg{} and \eebkg{} events in collision data, we use one data-taking period recorded on 27 December 2024 with the \gnnetm included in \belletwo data taking corresponding to an integrated luminosity of 0.059\,fb$^{-1}$, denoted \textit{Run A} in \cref{tab:run_descriptions}.
An evaluation under different background conditions is performed using six data taking periods, including \textit{Run A}, which is subdivided into \textit{Run A1} and \textit{Run A2} corresponding to two distinct background regimes observed during the run.
The corresponding run properties are summarized in \cref{tab:run_descriptions}.

Beam background conditions are quantified using the average number of out of time ECL crystals,
\begin{equation}
\begin{aligned}
\bar{N}_\mathrm{OOTC}
&= \left\langle N_\mathrm{OOTC} \right\rangle , \\
N_\mathrm{OOTC}
&= N\!\left(E > 7\,\mathrm{MeV},
\; \left| t - t_\mathrm{coll} \right| > 110\,\mathrm{ns}\right),
\end{aligned}
\label{eq:nootc}
\end{equation}
where $t_\mathrm{coll}$ denotes the offline determined collision time, $E$ and $t$ is the measured energy and time of a crystal, and the average is taken over all events in the corresponding run.
The value of $\bar{N}_\mathrm{OOTC}$ for each run is reported in \cref{tab:run_descriptions}.

\begin{table*}[ht!]
\caption{Duration, maximum instantaneous luminosity, integrated luminosity, and the average number of out-of-time ECL crystals of the six runs used for the evaluation on collision data. The \gnnetm was included in the \belletwo data taking for \textit{Runs A-D}.}  
\centering
\renewcommand{\arraystretch}{1.1}
\begin{tabular}{ccccc}
\hline
Run & Length & Max. inst. Luminosity (cm$^{-2}$s$^{-1}$) & Int. Luminosity (nb$^{-1}$) & $\bar{N}_\mathrm{OOTC}$ \\ \hline 
\textit{A} & 31\,min 52\,s & 4.33$\,\times\,10^{34}$ & 5.93$\,\times\, 10^4$ & - \\ 
\qquad\textit{A1} & \qquad12\,min 02\,s & - & - & 321.3 \\ 
\qquad\textit{A2} & \qquad19\,min 50\,s & - & - & 427.3 \\ 
\textit{B} & 10\,min 02\,s & 3.35$\,\times\,10^{34}$ & 1.09$\,\times\, 10^4$ & 259.3\\ 
\textit{C} & 6\,min 47\,s & 3.24$\,\times\,10^{34}$ & 7.02$\,\times\, 10^3$ & 227.5\\ 
\textit{D} & 10\,min 15\,s & 3.85$\,\times\,10^{34}$ & 2.04$\,\times\, 10^4$ & 400.5 \\ 
\textit{E} & 44\,min 51\,s & 2.52$\,\times\,10^{34}$ & 6.40$\,\times\,10^{4}$ & 174.3 \\ 
\textit{F} & 132\,min 16\,s & 1.90$\,\times\,10^{34}$ & 8.24$\,\times\,10^{4}$ & 130.9 \\ 
\hline
\end{tabular}
\label{tab:run_descriptions}
\end{table*}

\subsection{Performance metrics}
\label{sec:performance_metrics}
We define a \triggercluster as matched to an \offlinecluster if both the Euclidean 3D distance between the \triggercluster and the \offlinecluster is less than 40\,cm, and the energy ratio $R = E_{\text{trg}}/E_{\text{offline}}$ satisfies $0.01 \leq R \leq 2.0$, where $E_{\text{trg}}$ and $E_{\text{offline}}$ are the reconstructed energies of the \triggerclusters and \offlineclusters, respectively. 
The asymmetric bounds reflect the different failure modes of the trigger algorithms: the lower bound $R \geq 0.01$ rejects low-energy noise \triggerclusters, while the upper bound $R \leq 2.0$ is chosen to penalize the merging of two close-by clusters into a single \triggercluster.

If a \triggercluster matches multiple \offlineclusters, the \offlinecluster with the smallest distance is selected. 
If multiple \triggerclusters match the same \offlinecluster, the \triggercluster with the energy ratio closest to unity is selected.
We define the \textit{efficiency} $\varepsilon$ and \textit{purity} $\mathfrak{p}$ relative to \offlineclusters that are used as targets for training. 
\offlineclusters with very low energy below 80\,MeV or outside the \triggerwindow are excluded. 

We define the \textit{\trg cluster efficiency} as the ratio of the number of matched \triggerclusters to the number of all \offlineclusters
\begin{equation}
    \trgeff = \frac{N(\text{matched})}{N(\text{offline})}.
\end{equation}

We define the \textit{\trg cluster purity} as the ratio of matched \triggerclusters to the number of all \triggerclusters
\begin{equation}
    \trgpur = \frac{N(\text{matched})}{N(\text{trg})}.
\end{equation}

We define the reconstruction errors of the uncorrected energy by comparing the energy reconstructed by the \trg with the \offlinecluster energy
\begin{equation}
    \eresuncorr = \frac{E_{\text{trg}} - E_{\text{offline}}}{E_{\text{offline}}}.
\end{equation}

\begin{figure}[ht!]
    \centering
    \begin{subfigure}[b]{\halfwidth\textwidth}
         \centering
         \includegraphics[width=\textwidth]{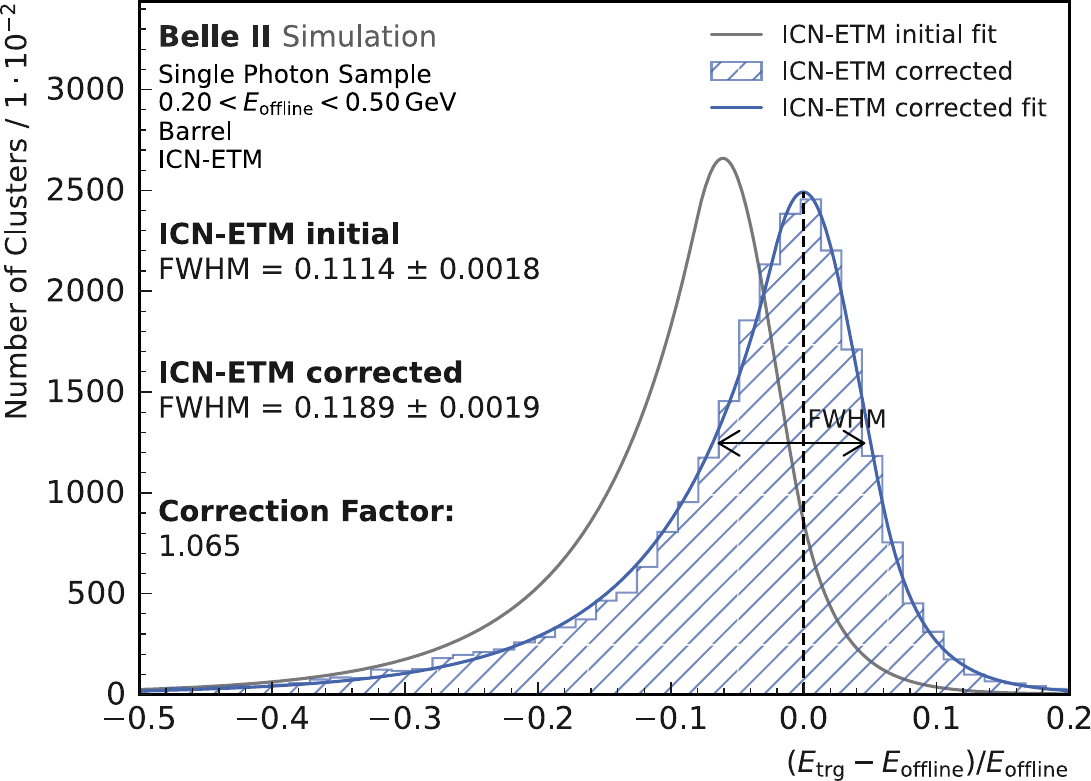}
         \caption{\icnetm}
         \label{fig:metrics_explanation:icnetm}
     \end{subfigure}
     \begin{subfigure}[b]{\halfwidth\textwidth}
         \centering
         \includegraphics[width=\textwidth]{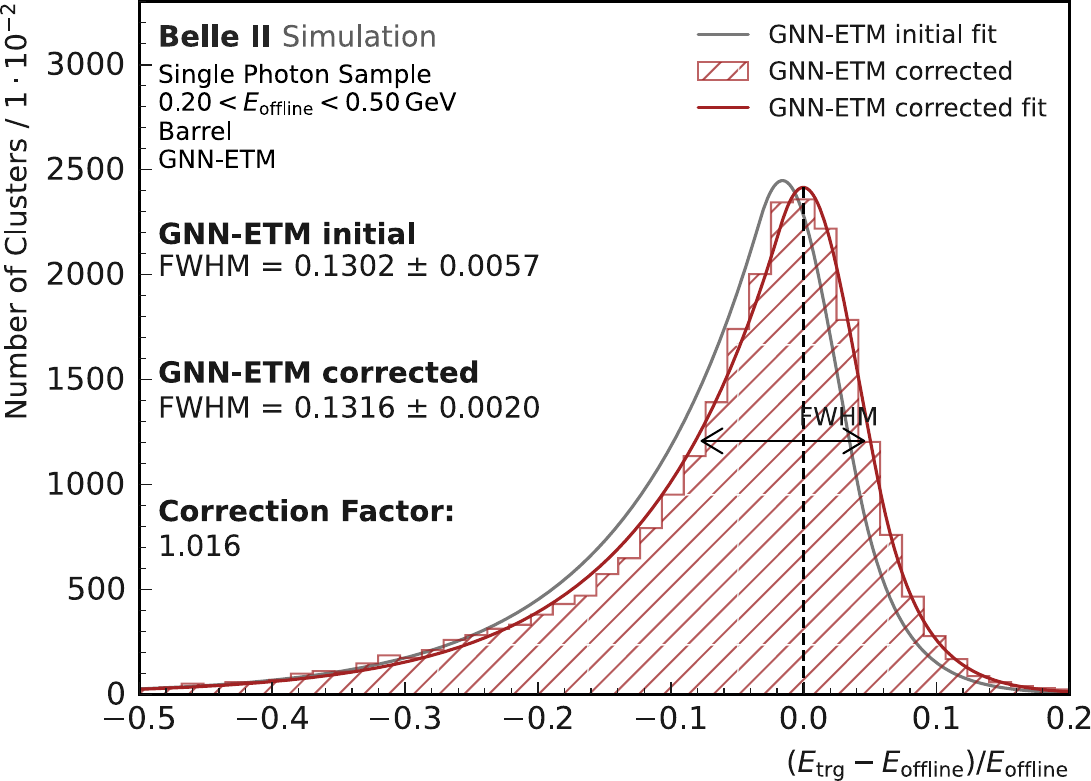}
         \caption{\gnnetm}
         \label{fig:metrics_explanation:gnnetm}
     \end{subfigure}
\caption{Example distribution of the relative reconstruction error of the \trg energy for \icnetm (\subref{fig:metrics_explanation:icnetm}) and for \gnnetm (\subref{fig:metrics_explanation:gnnetm}), and illustration of the bias correction and the FWHM.}
    \label{fig:metrics_explanation}
\end{figure}

This $\eresuncorr$ distribution is expected to be centered at zero for an unbiased reconstruction. 
However, both the \icnetm and the \gnnetm \trg algorithms are subject to potential biases from energy leakage and the presence of beam backgrounds, which can affect the reconstructed energy. 
The \offlinecluster energy is already bias-corrected to better than 0.5\%. 
The \trg reconstruction is therefore bias-corrected in analogy, using the following procedure. 
Evaluating the reconstruction algorithm on a large number of simulated photons with fixed energies yields peaking distributions in the uncorrected energy resolution \eresuncorr. 
A binned fit using a double-sided Crystal Ball density function is performed~\cite{Gaiser:1982yw,Skwarnicki:1986xj}. 
The mean $\mu$ of the fit is used to define a correction factor $f_{\text{corr}}(E_{\text{trg}}) = 1 - \mu$. 
All reconstructed energy values are shifted by applying this multiplicative factor\footnote{Additive and multiplicative correction functions $f_{\text{corr}}(E_{\text{true}})$ can be defined to give identical results if the true energy $E_{\text{true}}$ were known, and in principle both can be formulated as energy-dependent. 
In practice, however, only the reconstructed \trg energy $E_{\text{trg}}$ is available. 
Applying a correction as a function of $E_{\text{trg}}$ rather than $E_{\text{true}}$ necessarily introduces a slight broadening of the distribution, since upward and downward fluctuations are corrected with systematically different factors. 
This is true for both additive corrections $E_{\text{trg}}+\Delta(E_{\text{trg}})$ and multiplicative corrections $f_{\text{corr}}(E_{\text{trg}})\,E_{\text{trg}}$. 
The key advantage of the multiplicative form is that it reflects the physics of fractional energy leakage, which is nearly constant with energy, whereas absolute leakage grows roughly linearly with energy. 
As a result, $f_{\text{corr}}(E_{\text{trg}})$ can be interpolated more reliably.
For these reasons, we adopt the multiplicative form of $f_{\text{corr}}(E_{\text{trg}})$.} to correct for the difference between the fitted peak position and zero (see \cref{fig:metrics_explanation}).

The corrected relative energy residual \eres is calculated as
\begin{equation}
    \eres = \frac{E_{\text{trg}} \cdot f_{\text{corr}}(E_{\text{trg}}) - E_{\text{offline}}}{E_{\text{offline}}}.
\end{equation}

We then determine the \textit{corrected energy resolution} as the full width at half maximum (FWHM) of the final shifted distributions \eres.
For a normal distribution, the FWHM corresponds to 2.355 times the standard deviation, and we report the value as FWHM / 2.355. 
The uncertainty on the FWHM is calculated by propagating the parameter uncertainties using the full covariance matrix of the fit.
If necessary, the aforementioned energy bias correction factors $f_{\text{corr}}(E_{\text{trg}})$ must be implemented via lookup tables on the FPGAs to ensure real-time application within the \trg system.

For the positions, we use the unnormalised residuals in $x$, $y$, and $z$ as reconstruction errors, defined as the difference between the reconstructed \triggercluster position and the \offlinecluster position. 
No bias correction is applied. 
The \textit{position resolutions} are defined via the 90\% coverage
\begin{equation}
    r = P_{90\%}\left(\left|u - P_{50\%}(u)\right|\right),
\end{equation}
where $u \in \{x, y, z\}$ denotes any of the three coordinates, $P_q$ is the $q$-th percentile, and $P_{50\%}$ is the median. 
For a normal distribution, $r$ is equal to 1.65 times the standard deviation.

Each \gnncluster provides a classifier output $p_{\text{signal}}$ between 0 (background) and 1 (signal). 
No such classifier exists for the \icnetm. 
We define the \textit{signal retention rate} $R_S$ as
\begin{equation}
    R_S = \frac{N(\text{matched}, \text{signal}, p_{\text{signal}} > t_\mathrm{sig})}{N(\text{matched}, \text{signal})},
\end{equation}
and the \textit{background rejection rate} $R_B$ as
\begin{equation}
    R_B = \frac{N(\text{matched}, \text{background}, p_{\text{signal}} \leq t_\mathrm{sig})}{N(\text{matched}, \text{background})}.
\end{equation}
The resulting \gnnetm configuration with an applied signal classifier selection, tuned using a specific energy range and event topology to achieve a signal efficiency of about $R_S = 0.975$, is referred to as the \cgnnetm and is described in detail in \cref{sec:classifier}.
Separate cut values $t_\mathrm{sig}$ on the classifier output are chosen for the barrel, forward endcap, and backward endcap regions to achieve a given signal efficiency.

\subsection{Graph neural network architecture}
\label{sec:gnnetm_architecture}
\begin{figure}[ht!]
    \centering
    \includegraphics[width=0.6\textwidth]{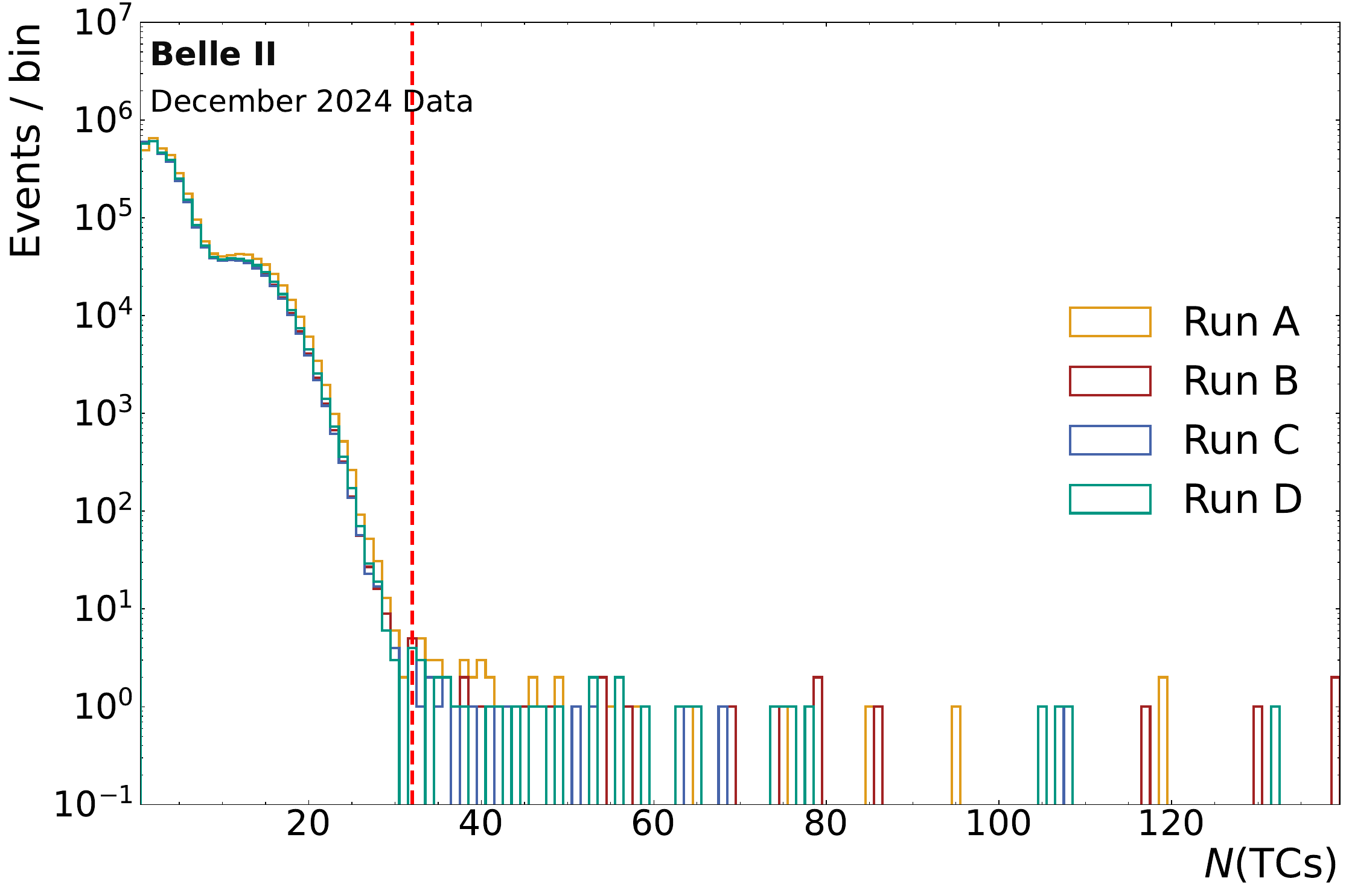}
\caption{Distribution of the number of active TCs per event in collision data recorded during December 2024 for \textit{Runs A-D}.
The red dotted line denotes the number of maximum input TCs for the \gnnetm. 
For each run, less than 0.002\% of events have $N(\text{TCs}) > 32.$}
    \label{fig:sparsity}
\end{figure}

Because of the rather small number of active TCs per event, the variable number of inputs, the lack of a natural ordering, and the non-uniform spatial layout in the endcaps, a GNN architecture is used. 
We refer to this model as \textit{CaloClusterNet}. 
Each node in the graph corresponds to a TC. 
The network is optimized for events with up to 32 input TCs, based on simulation studies and validated with collision data experiencing high beam background levels (see Fig.~\ref{fig:sparsity}). 
Inputs with fewer than 32 TCs are zero-padded, while those exceeding 32 are truncated without ordering, resulting in an arbitrary cut-off.

The model input features are the Cartesian coordinates $x$, $y$, and $z$ of each TC center, defined as the mean of the crystal centers it contains, along with the TC energy and its time relative to the highest-energy TC in the \triggerwindow.
We normalize the coordinate and time input values to a range between -1 and 1. 
The TC energy is divided by 8 to scale the majority of values to a range between 0 and 1. 
As outliers can reach energies of up to 20 GeV, the scaled TC energy can exceed 1 to allow for these inputs without clipping and to avoid squeezing low-energy values into a too small value range. 
Using Cartesian coordinates instead of polar coordinates removes the discontinuity at the $0^\circ/360^\circ$ boundary, which leads to increased training stability. 

CaloClusterNet predicts, for each \gnncluster, an energy scale factor, a position, and a background classifier. 
The scale factor is applied to the TC energy.
Because the number of clusters is not known a priori, the training objective employs an object condensation loss~\cite{Kieseler:2020wcq}. The network is implemented in \keras~\cite{keras:2015}.

\begin{figure*}[ht!]
    \centering
    \includegraphics[width=\textwidth]{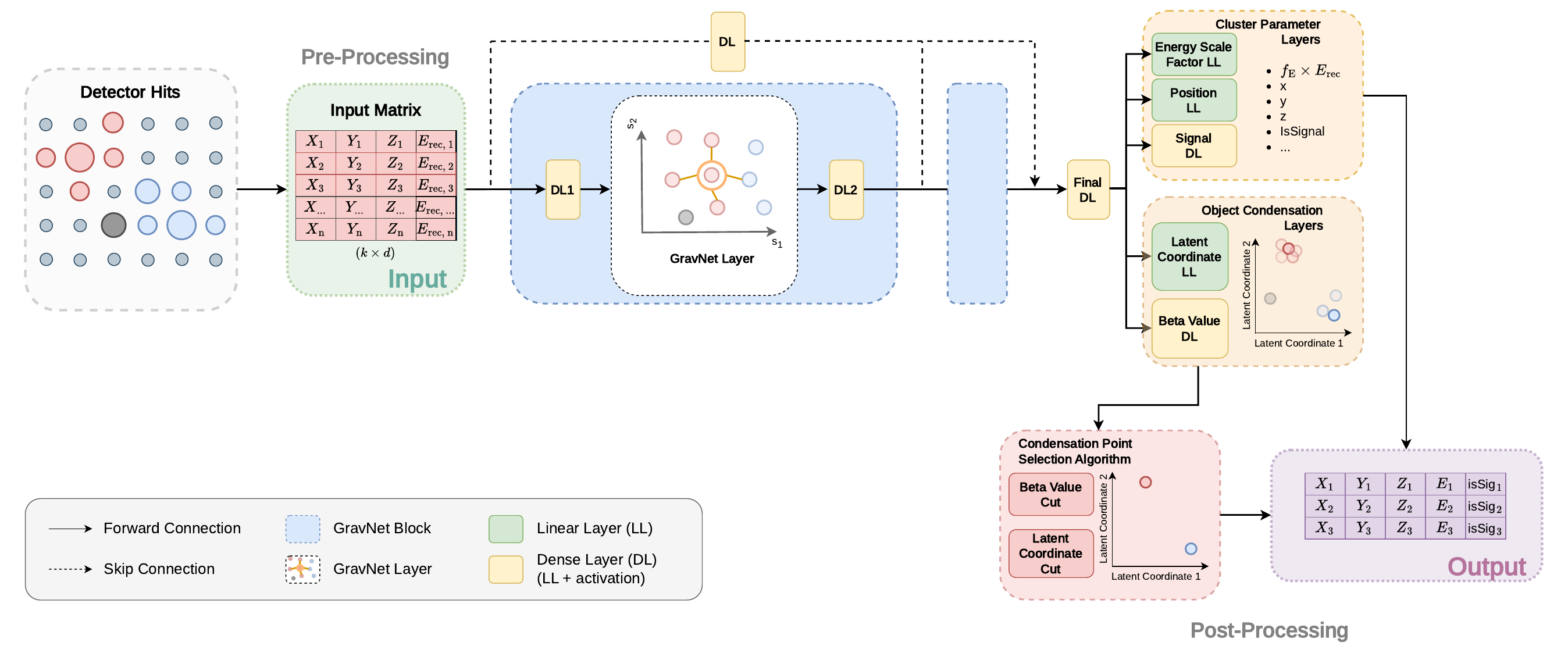}
    \caption{An illustration of the CaloClusterNet model architecture.}
    \label{fig:architecture}
\end{figure*}

The architecture of CaloClusterNet is shown in Fig.~\ref{fig:architecture} and consists of two GravNet blocks~\cite{Qasim:2019otl}. 
Each block contains two linear layers~(LL), one GravNetConv layer, and one dense layer~(DL). 
The output of each block is passed both to the next block and, via concatenation, to the final output layers.  

Within each GravNet block, a dense layer with rectified linear unit (ReLU) activation~\cite{Agarap:2018uiz} is applied first.
The subsequent GravNetConv layer maps the input node features into two learned spaces: a spatial representation $S$ and a feature space $F_{\text{LR}}$. Edges are then constructed by connecting each node to its $k$ nearest neighbors in $S$, determined from the Euclidean distance  
\begin{equation}
d = \sqrt{\sum_{i=1}^n (X_{i,j} - X_{i,k})^2},
\end{equation}  
where $X_j$ and $X_k$ denote the spatial coordinates of nodes $j$ and $k$. 
Message passing is performed by aggregating the features of connected nodes, with each feature being weighted by exp(${-f_\mathrm{exp} d}$) with $f_\mathrm{exp}$ being a tunable parameter to increase or decrease separation power. 
Afterwards, the aggregated features are concatenated with the original node features. 
Finally, the GravNetConv outputs undergo an additional dense transformation, enabling feature representations that support diverse value quantization schemes for FPGA deployment.

Following the final GravNet block, the extracted features are passed through parallel linear layers responsible for predicting cluster properties: 
one linear layer each for the energy scale factor, the cluster position, the signal classifier score, the latent space coordinate vector~(CCoords), and the $\beta$ value. 
The outputs for the $\beta$ value and the signal classifier are passed through a sigmoid activation function to constrain them between 0 and 1. 
The $\beta$ value and the CCoords are prediction values necessary for the object condensation loss explained below.
The target truth information for these predictions is taken from the \offlinecluster matched to this node.

The feature loss terms for the energy and position $L_{E,i}$ and $L_{\text{Pos},i}$ for each node $i$ are calculated using the absolute difference between truth and predicted value. For the signal classifier loss $L_{\text{signal},i}$, a binary cross-entropy loss is used.
The feature loss terms for energy, position, and signal classification are weighted to emphasize accurate predictions at condensation points. 
The weight factor is defined as
\begin{equation}
    \xi_i = (1 - n_i)\,\mathrm{artanh}(\beta_i) + q_{\text{min}},
\end{equation}
where $n_i = 1$ for TCs without an assigned training label and $n_i = 0$ for signal TCs, with the sum taken over all $N$ TCs in the event.\footnote{While the original paper uses $\mathrm{artanh}^2(\beta)$, $\mathrm{artanh}(\beta)$ displays the same concave behaviour and monotonous increase and is used for simplicity.} 
This construction prioritizes nodes with large $\beta$ values, ensuring that the representation points of each object carry the correct information.  
The total feature loss $L_F$ is given by
\begin{equation}
    L_F = \frac{1}{\sum_{i=1}^N \xi_i} \sum_{i=1}^N \left( L_{E,i} + L_{\text{Pos},i} + L_{\text{signal},i} \right) \xi_i.
\end{equation}
To avoid the trivial solution $\beta_i = 0$ for all $i$, a minimum charge $q_{\text{min}}=0.1$ is imposed.  
The total loss is then defined as the unweighted sum of the object condensation loss terms (attraction, repulsion, and $\beta$ components)~\cite{Kieseler:2020wcq} and the feature loss $L_F$.  

\begin{figure*}[ht]
     \centering
     \begin{subfigure}[t]{0.45\textwidth}
         \centering
         \includegraphics[width=\textwidth]{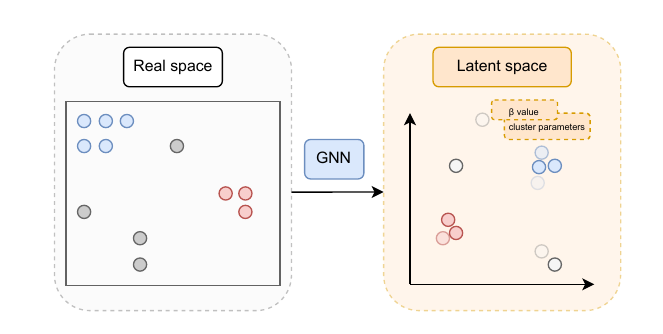}
         \caption{Translation from real space to latent cluster space.}
         \label{fig:gnn_1}
     \end{subfigure}
     \hfill
     \begin{subfigure}[t]{0.45\textwidth}
         \centering
         \includegraphics[width=\textwidth]{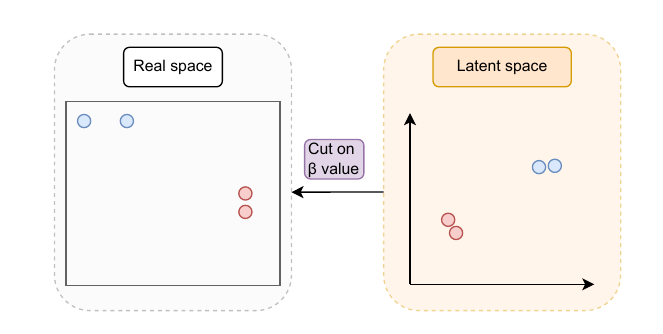}
         \caption{Condensation point candidate selection based on $\beta$ threshold.}
         \label{fig:gnn_2}
     \end{subfigure}\\

     \begin{subfigure}[t]{0.45\textwidth}
         \centering
         \includegraphics[width=\textwidth]{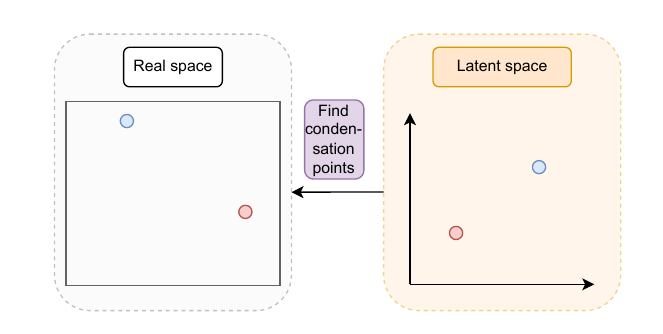}
         \caption{Condensation point selection based on isolation criteria.}
         \label{fig:gnn_3}
     \end{subfigure}
     \hfill
     \begin{subfigure}[t]{0.45\textwidth}
         \centering
         \includegraphics[width=\textwidth]{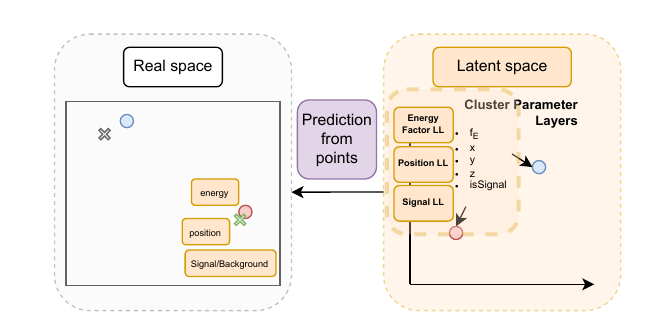}
         \caption{Parameter extraction for condensation points.}
         \label{fig:gnn_4}
     \end{subfigure}\\

    \caption{Cluster finding using object condensation: (\subref{fig:gnn_1}) Latent space, (\subref{fig:gnn_2}) condensation point candidate selection based on $\beta$ threshold, (\subref{fig:gnn_3}) condensation point selection based on isolation, and (\subref{fig:gnn_4}) parameter extraction.}
    \label{fig:gnn_educational}
\end{figure*}

The post-processing procedure, referred to as the condensation point selection algorithm, extracts \gnncluster information from the inference output of the trained model. 
It is illustrated in Fig.~\ref{fig:gnn_educational} and follows four steps:
\begin{itemize}
    \item[(a)] Each event is processed by the model, assigning each node a predicted position in the latent space, a $\beta$ value, and predictions for the \gnncluster energy, position, and background classifier.
    
    \item[(b)] Condensation point candidates are selected by applying a threshold $t_\beta$ on the $\beta$ values.
    
    \item[(c)] Isolated condensation points are identified among the candidates: the candidate with the highest $\beta$ value is selected first, and all other candidates within a distance $r < t_d$ in latent space are removed. This process is iterated until only isolated condensation points remain, each separated by a distance greater than $t_d$.
    
    \item[(d)] The predicted properties (energy scale factor, position, and signal probability) of each condensation point are used to define the corresponding \gnncluster.
\end{itemize}
Only the condensation points with their inferred parameters are used in the final \trg decision. 
It is not necessary to assign all TCs belonging to a cluster, as the relevant information is fully contained in the selected condensation points.

\subsection{Model compression}
\begin{figure*}[ht]
    \centering
    \includegraphics[width=\textwidth]{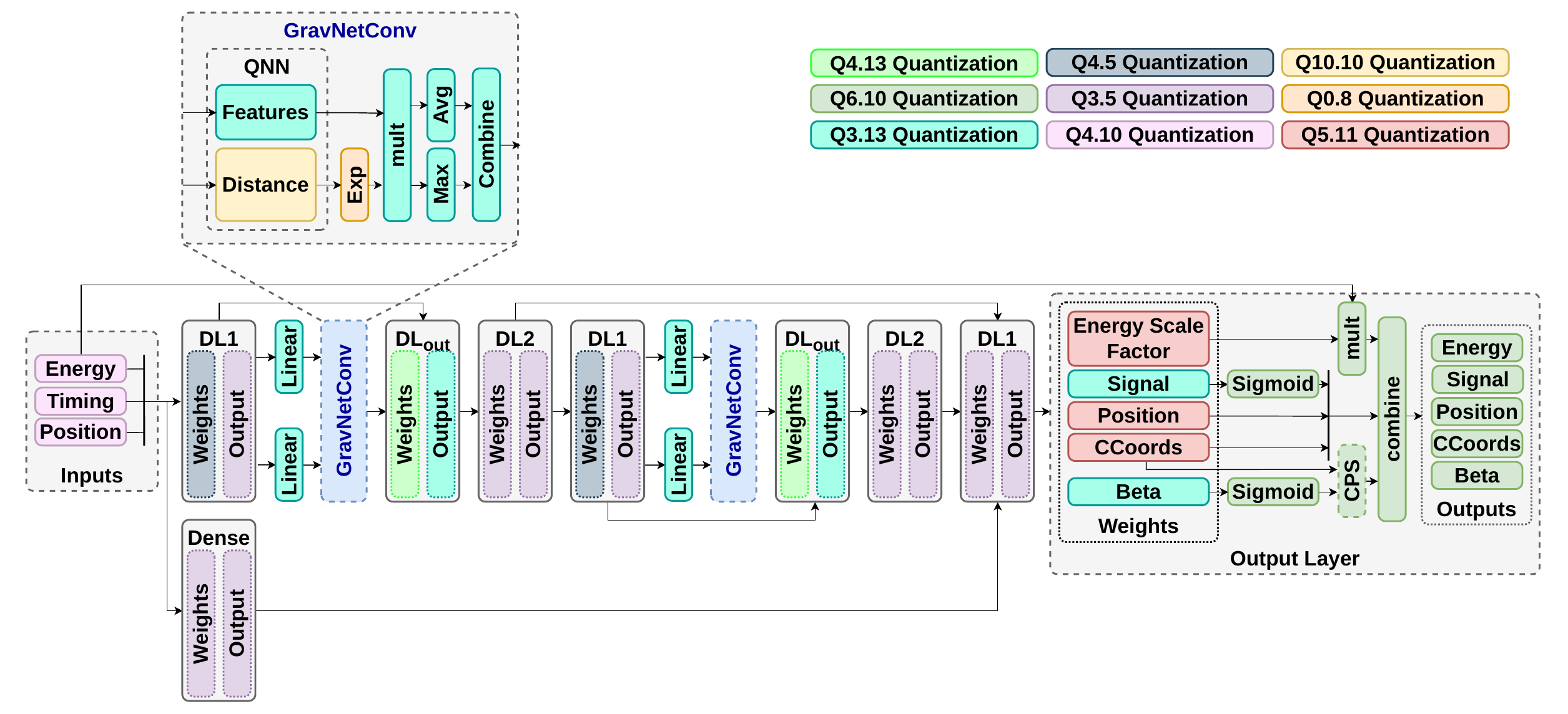}
    \caption{Quantized \gnnetm model showing layer-by-layer quantization details in \qkeras format.}
    \label{fig:quantized_model}
\end{figure*}

The CaloClusterNet model must be significantly compressed to allow implementation and online inference on an FPGA, given the limited resources and strict latency and throughput constraints. 
For full deployment in the \belletwo \trg system, the implementation targets the Universal Trigger Board (UT4) equipped with an AMD~Ultrascale~XCVU190 FPGA. 
The available resources on this device set an upper bound on the GNN architecture depth, limiting it to two GravNet blocks.  
As floating-point multiplications are prohibitively resource-intensive on FPGAs, all weights, biases, inputs, and outputs are quantized to fixed-point representations with limited precision and range. 
To minimize performance degradation, mixed-precision quantization-aware training is applied using \qkeras~\cite{coelho_automatic_2021}, with different quantization values assigned to each layer. 
Figure~\ref{fig:quantized_model} shows the final model configuration, including the quantization applied to each layer’s weights and biases as well as to the network inputs and outputs.  
The fixed-point representation is written in Q-format, denoted as Q3.5, meaning that 3 of the 8~bits (including the sign bit) are used for the integer part and 5 for the fractional part. 
The quantization ranges are designed bottom-up: for each layer, weights, biases, and outputs are initialized with a total width of 8 bits and a range of $[-4,4]$. 
Where required, the bit widths and parameter ranges are increased to ensure numerical stability and maintain model accuracy.
To ensure sufficient precision in the GravNetConv layers, the bit width is increased to 16~bit for all operations within the GravNet block, and the cluster parameter and object condensation layers. 
The distance calculation output in the GravNet layer is extended in range to prevent overflows. 
The exponential function output is limited to a maximum of 1, since the minimum distance before exponential weighting is 0.  

For the skip connections (direct links that bypass intermediate layers), the outputs of both GravNet blocks are appended to the original inputs, with the resulting vector serving as input to the final dense layer before the output layers.
This introduces two different quantization ranges: Q4.10 for the inputs and Q3.5 for the GravNet block outputs. 
To match these, an additional scaling dense layer reduces the input values to the Q3.5 range.  

All dense layers use ReLU activations, which require no additional quantization. 
Their output quantization matches that of the corresponding dense layer. 
The only other activation is the sigmoid function $\sigma(x) = (1 + e^{-x})^{-1}$, applied in the signal and $\beta$ output layers. 
For FPGA compatibility this is replaced by the linear sigmoid approximation provided by \qkeras, defined as $\sigma_{\text{QKERAS}}(x) = 0.1875x + 0.5$.  

The input and output values are quantized to optimize data handling. 
Since the full data readout in \texttt{basf2} is implemented in C++, the bit width is chosen in multiples of 8 to allow byte-wise access. 
Both input and output values therefore use a 16-bit (2-byte) representation. 
After normalization, only the TC energy can exceed 1, with outliers reaching up to 4. 
To fully cover this range, the input quantization is set to Q4.12, applied uniformly to all input values for simplicity. 
For the outputs, a Q6.10 format is used, ensuring a sufficiently large dynamic range for all variables.  

To further reduce the number of multiplications during inference, low-magnitude pruning~\cite{zhu2017prunepruneexploringefficacy} is applied with a pruning rate of 40\,\%. 
Attempts with higher pruning rates did not yield the intended sparsity. 
In addition, the distance calculations in the graph construction and condensation point selection steps use the L1 norm instead of the L2 norm~\cite{Qasim:2019otl}, further reducing the number of multiplications required.  

The model is trained with quantization-aware training and applies low-magnitude pruning after an initial warm-up phase of 10 epochs. 
All quantizations are applied during training except for those of the exponential and sigmoid functions. 
Quantizing the sigmoid function during training results in unstable behavior with large loss fluctuations, while quantizing the exponential function prevents convergence altogether. 
These two quantizations are therefore applied only after training, without any observed degradation in performance.  
The combined compression measures lead to a reduction in efficiency \trgeff, with the highest loss of up to 10 percentage points in the backward endcap, a purity loss of 2 percentage points, and slightly worse energy and position resolutions. 
The results of the quantized model are reported in \cref{sec:results}.  

\subsection{Hyperparameter optimization}

Hyperparameter optimization for a quantized model targeting FPGA deployment is constrained by hardware resources. 
Increasing the size of the GNN requires careful trade-offs with numerical precision, as using reduced bit widths for activations and weights permits wider layers while staying within FPGA resource limits.
The quantization of all parameters is fixed to the values defined in the previous section, and the number of GravNet blocks is limited to two. 
The optimization is performed with \texttt{Weights and Biases}~\cite{biewald_experiment_2020}, minimizing the full loss on the validation sample. 
The final hyperparameters and optimization results are summarized in \cref{tab:model_parameters_final_values}. 
Correlation factors are also reported, indicating how each hyperparameter relates to the final validation loss. 
Negative correlation factors imply that larger values lead to smaller validation losses, and thus better performance. 
Several hyperparameters reach the upper limit of the tested range, suggesting that larger models could further improve performance. 
Such configurations, however, are not feasible due to hardware constraints. 
The hyperparameter ranges are defined to realize the largest model that fits on the FPGA, based on system resource utilization.  

Improved hyperparameter optimization would jointly tune quantization values and pruning percentages alongside the model hyperparameters.
This requires an additional metric for FPGA resource consumption to prevent selecting models that exceed implementation limits. 
Further optimization in this direction is left for future work.

For the training, the learning rate is reduced by a factor of 2, if the total validation loss is not improved over 10 epochs. 
The training is stopped when the learning rate is reduced to 10$^{-7}$ and no improvement is seen over a further 50 epochs. 

In the post-training optimization, the tuning flexibility of the $\beta$ cut is reduced by the use of the linear sigmoid activation function. 
While training is performed with the standard sigmoid, which allows smooth slope adjustments near 0 and 1, the linear sigmoid pushes outputs directly to the extremes, thereby limiting fine-tuning. 
The optimal efficiency, with negligible purity loss, is achieved at $\beta = 0.04$. 
Lower values provide no further improvement, while higher values reduce efficiency.  
Adjustments to the latent space distance cut have negligible impact on overall performance due to the sparsity of the TCs, and it is fixed at 0.3.  

\begin{table*}[h]
\caption{Summary of the model parameters with their descriptions, hyperparameter search ranges, correlation coefficients with the overall validation loss~\cite{biewald_experiment_2020}, and the optimal values after optimization. 
$DL1$, $DL2$, and $DL_\mathrm{out}$ are defined in \cref{fig:quantized_model}, the other parameters in \cref{sec:gnnetm_architecture}.
A total of 400 training runs were performed on the \textit{Combined Photon Sample} (see \cref{sec:gnnetm_software_training}), each with a different set of hyperparameter values. 
For each run, 90\% of the data was used for training and 10\% for validation, with a reduced number of epochs. 
The best configuration was selected based on the minimum validation loss.}  
\centering
\renewcommand{\arraystretch}{1.1}
\begin{tabular}{lccc}
\hline
Hyperparameter & Optimization Range & Correlation & Result \\
\hline
Width of the dense layers $DL1$ & 4 - 16 & -0.277 & 16 \\
Width of the GravNet block output layer $DL_\mathrm{out}$ & 4 - 32 & -0.194 & 32 \\
Scaling factor for the exponential weighting $f_\mathrm{exp}$ & 1 - 10 & 0.098 & 10 \\
Size of representation space $S$ & 2 - 6 & -0.01 & 6 \\
Size of the feature space $F_\mathrm{LR}$ & 2 - 8 & -0.017 & 8 \\
Width of the dense layers $DL2$ & 4 - 16 & -0.061 & 16 \\
Number of dimensions for the latent space $N_\mathrm{LS}$ & 2 - 4 & 0.054 & 3 \\
Number of nearest neighbours in GravNet $k$ & 2 - 8 & 0.03 & 8 \\
\hline
\end{tabular}
\label{tab:model_parameters_final_values}
\end{table*}

\section{Hardware implementation and performance of the graph neural network level~1 calorimeter trigger}
\label{sec:gnnetm_hardware}

In the following, we describe the system architecture, outline our deployment approach, validate performance in terms of latency and resource utilization, and finally compare \qkeras, C simulation, and hardware implementation.

\subsection{System architecture}
In order to realize the online inference of our real-time algorithm, we develop a FPGA-based system architecture. 
Our architecture is based on the 4th generation of the Universal Trigger Board~(UT4).
The board features low-latency AXI-Stream interfaces~\cite{ARM:IHI0051B} to up- and downstream FPGA boards in the \trg chain as well as a slow control interface \texttt{Versa Module Eurocard} (VME)~\cite{IEEE:1014-1987}.
Additionally, it integrates the Belle2Link physical layer protocol~\cite{sun:2012,yamada:2015}, which allows us to read out data from the system synchronous to other components of the \trg system. 
We develop four components:
\begin{itemize}
    \item the Preprocessing Stage (see \cref{sec:gnnetm_hardware:preprocessing}),  
    \item the GNN Dataflow Accelerator (see \cref{sec:gnnetm_hardware:dataflow_accelerator}),  
    \item the Postprocessing Stage (see \cref{sec:gnnetm_hardware:postprocessing}), and  
    \item the Belle2Link Subsystem (see \cref{sec:gnnetm_hardware:belle2link}).  
\end{itemize}  

The functionality of these components is explained in detail below.

\begin{figure}
    \centering
    \includegraphics[width=0.7\linewidth]{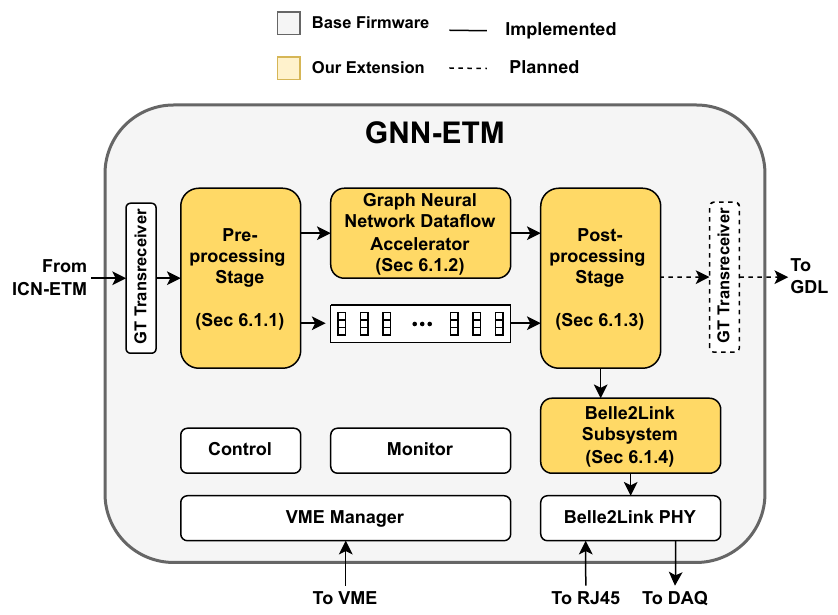}
    \caption{Overview of our \gnnetm system architecture. 
    Existing components are shown in gray, while components introduced in this work are shown in yellow. We indicate the planned extension of the interface to the Global Decision Logic. 
    The connection exists but is currently unused.}
    \label{fig:gnnetm_hardware:architecture}
\end{figure}

\subsubsection{Preprocessing stage}
\label{sec:gnnetm_hardware:preprocessing}
\begin{figure}
    \centering
    \includegraphics[width=0.7\linewidth]{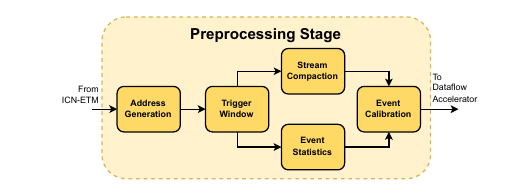}
    \caption{System overview of the preprocessing stage.}
    \label{fig:gnnetm_hardware:preprocessing}
\end{figure}

The preprocessing stage of the \gnnetm consists of five central modules shown in \cref{fig:gnnetm_hardware:preprocessing}.
We implement it as a parametrizable hardware generator, and describe the modules in detail below:

\begin{enumerate}
    \item \textbf{Address Generation}: Calculates the address of each incoming TC. This step is necessary for the subsequent sparsity compression.  
    \item \textbf{Trigger Window}: Retains all TCs from the previous \SI{125}{\ns} timeframe, thereby extending the time considered by the \trg algorithm to \SI{250}{\ns}. 
    The Flash ADC Analog Module (see \cref{fig:icn_chain}) guarantees that TCs are only sent once in a given \SI{250}{\ns} timeframe eliminating the need for handling duplications.
    \item \textbf{Stream Compaction}: Compresses the sparse input matrix containing $N_{TCs}=576$ rows into a smaller matrix with a maximum of $N^{max}_{TCs} = 32$ rows, while also storing the identifier of each TC. 
    To minimize latency overhead, we developed a hierarchical stream compaction approach that processes multiple input streams in parallel.  
    \item \textbf{Event Statistics}: Identifies the TC with the highest energy in each \triggerwindow, as well as its timing. This information is subsequently used by the Event Calibration module.  
    \item \textbf{Event Calibration}: Performs a fast online calibration of the timing and a LUT-based transformation of input features. First, the calibrated TC timing in each \triggerwindow is computed relative to the timing of the most energetic TC in that \triggerwindow. Second, the TC location ($x,y,z$) is retrieved from memory to form the input vector for the subsequent CaloClusterNet inference step.  
\end{enumerate}

\subsubsection{Graph neural network dataflow accelerator}
\label{sec:gnnetm_hardware:dataflow_accelerator}

The Graph Neural Network Dataflow accelerator is composed of processing elements~(PEs), connected by first-in-first-out~(FIFO) buffers. 
During the deployment, network layers are mapped to one or more PEs in the accelerator.
Each PE is pipelined with an initiation interval $I_{\text{init}}$ and a parallelism factor $P_\mathrm{par}$. 
For example, a PE with \hbox{$I_{\text{init}} = 16, P_{par} = 2$} accepts a new event every 16 clock cycles and is instantiated twice in parallel. 
This configuration can process up to $N = 32$~TCs per event, satisfying the required throughput. 
PEs execute computations in sequence, which simplifies latency and throughput analysis. 
We further require all PEs to behave as single-rate actors, consistent with FINN~\cite{blott:2018} and hls4ml~\cite{fastml_hls4ml}.

\begin{figure*}
    \centering
    \includegraphics[width=0.95\linewidth]{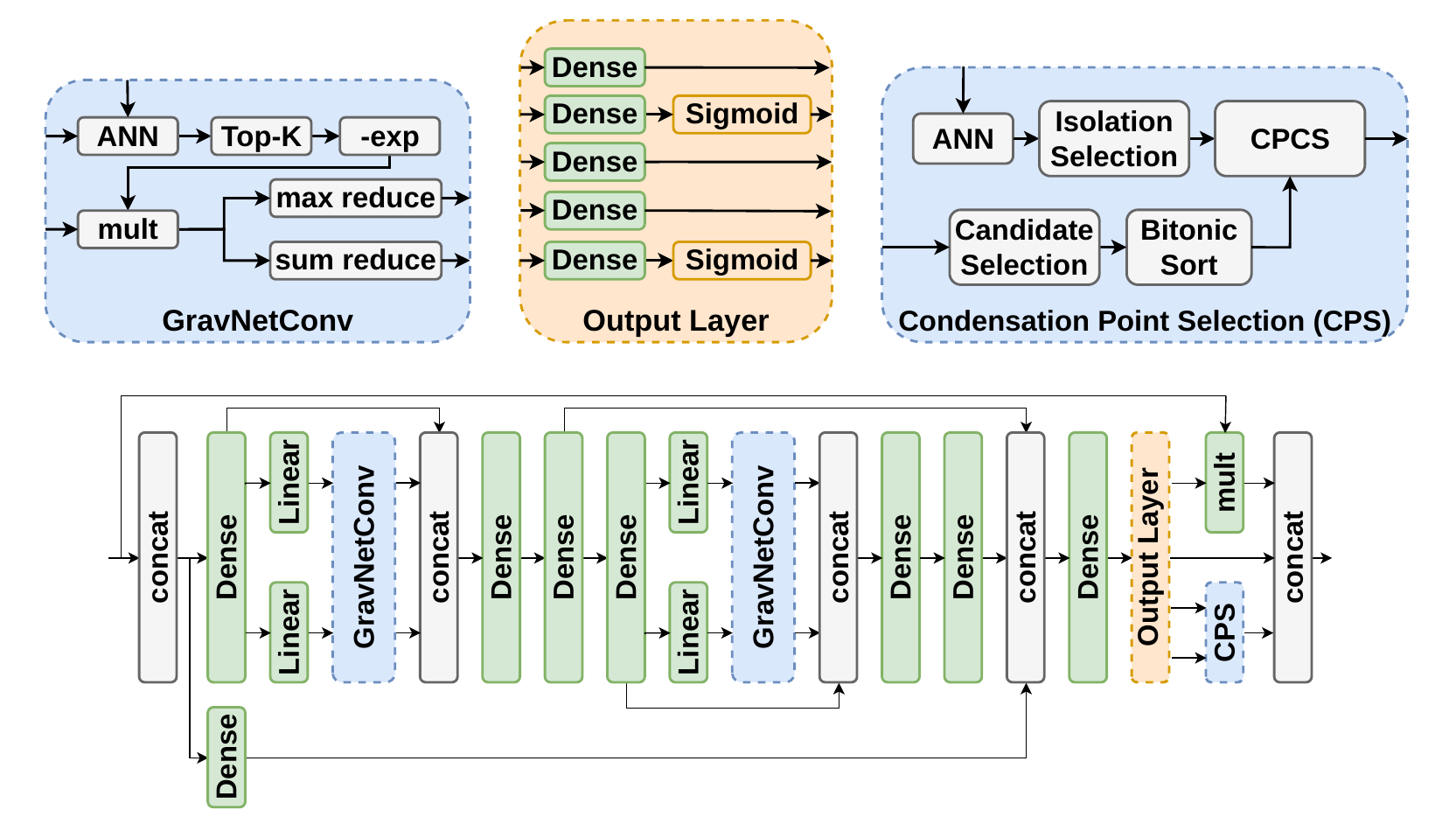}
    \caption{Mapping of CaloClusterNet onto the dataflow architecture. 
Solid boxes represent hardware modules, while dashed boxes group multiple PEs and are included for visualization only.}
    \label{fig:gnnetm_hardware:dataflow}
\end{figure*}

An overview of the complete accelerator is shown in \cref{fig:gnnetm_hardware:dataflow}. 
Computations are generally performed from left to right. 
External data interfaces are implemented as AXI-Streams in which all features are concatenated. 
Blocks with solid lines represent deployed PEs on the FPGA, whereas blocks with dashed lines indicate hierarchical containers comprising multiple PEs. 
Each PE is implemented in HLS~C as an architecture template that represents a dedicated operator in the neural network. 
For clarity, PEs are named identically to the operator they implement.  

The following provides an overview of the PEs used in the \gnnetm:  
For dense layers, matrix–vector multiplication, activation function, and rescaling are fused into a single pipelined hardware module.  
Linear layers correspond to dense layers without an activation function.  
Output layers differ in their activation: they either contain no activation function (Energy, Position, CCoords)
or a sigmoid activation function (Signal, $\beta$).  
Sigmoid activation functions are implemented as separate PEs, using the same linear approximation as hls4ml~\cite{fastml_hls4ml}.  
The \texttt{mult} PE multiplies the energy factor output with the input energy, ensuring that the network returns a calibrated cluster energy in \gev.  

Two operators of higher algorithmic complexity, GravNetConv and the Condensation Point Selection~(CPS) algorithm, are split into multiple PEs and are described in more detail below.

\textbf{GravNetConv}  
The GravNetConv operator is implemented as a PE, shown in the upper left of \cref{fig:gnnetm_hardware:dataflow}. 
This module combines the graph building and message passing steps of the GravNet layer:  

\begin{itemize}
    \item \textbf{Graph building:} Implemented using an all-nearest-neighbor (\texttt{ANN}) algorithm followed by a hierarchical Top-K sort (\texttt{Top-K}).
    \item \textbf{Message passing:} Sorted features are retrieved from ping–pong buffers and multiplied with exponentially weighted distances (\texttt{exp}, \texttt{mult}).  
    \item \textbf{Neighborhood aggregation:} Performed using \texttt{max reduce} and \texttt{sum reduce} operators.  
\end{itemize}

This dynamic graph construction contrasts with previous FPGA implementations that  relied on static graphs~\cite{Neu:2023sfh}. 
By constructing edges in a learned latent space rather than from fixed detector geometry, dynamic graph construction enables the use of event-dependent features such as TC energy and timing, at the cost of increased computational complexity introduced by the All-Nearest-Neighbour algorithm. 
A direct comparison with a static graph baseline is beyond the scope of this work and left for future study.

\textbf{Condensation Point Clustering}
We implement multiple PEs for the CPS operator as depicted in the upper right part of \cref{fig:gnnetm_hardware:dataflow}. 
This processing element implements a clustering algorithm operating in feature space. 
Cluster seed isolation is first determined using an \texttt{ANN} step, followed by an \texttt{Isolation Selection}. 
Candidate cluster points are then ranked by a priority value  in the \texttt{Candidate Selection} and sorted in the \texttt{Bitonic Sort} after which the final Condensation Point Candidate Selection~(\texttt{CPCS}) is applied. 

In the following, we describe our implementation of the CPCS algorithm in detail, as it is the core compute step in the CPS.
The corresponding pseudo-code is given in \cref{alg:gnnetm_hardware:CPCS}. 
It calculates the subset of clusters from the respective condensation point candidates.
It is expected that condensation point candidates are stored strictly in order, thus each candidate is uniquely identified by its memory address (\textit{id}). 
The PE requires the following two inputs for every query candidate $q_i$:
(1) A bitmask $\{0,1\}^{N}$ containing the isolation criteria between $q_i$ and all other candidates.  
(2) A boolean flag indicating if the $\beta$ criteria is fulfilled for the query candidate $q_i$.
These values are stored in the arrays \textit{isolations} and \textit{candidates}. 
In addition, an ordered list of query \textit{ids} is required. This way, we ensure that query candidates with higher $\beta$ value are prioritized.
The output is given as a bitmask $\{0,1\}^{N}$, indicating that the respective query candidate in the memory location has been selected as a cluster condensation point.

We highlight the simplicity of our implementation, as the computationally intensive steps have been offloaded to previous pipeline stages inside the CPS PE. 
As a result, our realization of the condensation point clustering is able to select up to $P_\mathrm{par}$ clusters per clock cycle, resulting in an initiation interval of $I_{\text{init}} = \lceil\frac{N}{P_{par}}\rceil$. 
Notably, there are no limitations in the number of clusters to be selected, as long as $P_\mathrm{par}$ is chosen appropriately. 
Our solution is statically pipelined and thus complies with our hard real-time requirements.

\begin{algorithm}
    \caption{\\Condensation Point Candidate Selection (CPCS)}

    \label{alg:gnnetm_hardware:CPCS}

    \algblock{ParFor}{EndParFor}
    \algnewcommand\algorithmicparfor{\textbf{parallel\ for}}
    \algnewcommand\algorithmicpardo{\textbf{do}}
    \algnewcommand\algorithmicendparfor{\textbf{end\ for}}
    \algrenewtext{ParFor}[1]{\algorithmicparfor\ #1\ \algorithmicpardo}
    \algrenewtext{EndParFor}{\algorithmicendparfor}
    
    \begin{algorithmic}[1]
        \Procedure{CPCS}{\textit{isolations},\textit{candidates},\textit{ids}}                    
            \State \textit{cps} $\gets$ $\{0\}^{N}$
            \State \textit{flags} $\gets$ \textit{candidates}
            \For{$i \gets 0,\,I_{init}-1$}
                \ParFor{$p \gets 0,\, P-1$}
                    \State \textit{id} $\gets$ \textit{ids.pop()}
                    \State \textit{cps[id]} $\gets$ \textit{flags[id]}
                    \State \textit{flags} $\gets$ \textit{flags} \& \textit{isolations[id]}
                \EndParFor
            \EndFor
            \State \Return \textit{cps}
        \EndProcedure
    \end{algorithmic}
\end{algorithm}
 
\subsubsection{Postprocessing stage}
\label{sec:gnnetm_hardware:postprocessing}

This module is responsible for merging multiple data streams. 
During this process, the features of all TCs and clusters are encoded using their corresponding unique identifiers. 
This encoding ensures that a complete coordinate-offset format is available for subsequent processing stages, meaning that each cluster is unambiguously associated with the unique identifier of its corresponding TC.

\subsubsection{Belle2Link subsystem}
\label{sec:gnnetm_hardware:belle2link}

\begin{figure}
    \centering
    \includegraphics[width=0.7\linewidth]{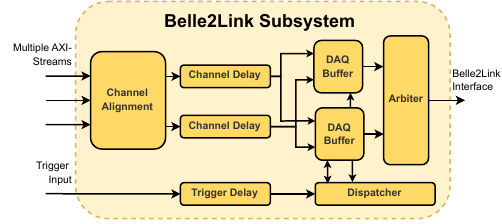}
    \caption{Overview of the Belle2Link Subsystem.} 
    \label{fig:gnnetm_hardware:belle2link}
\end{figure}

To interface our firmware with the existing \trg system at \belletwo, we developed a reusable module. 
The dataflow is illustrated in \cref{fig:gnnetm_hardware:belle2link}, proceeding from left to right. 
The module accepts AXI-Stream interfaces as inputs and generates a Belle2Link interface~\cite{sun:2012,yamada:2015} as output. 
It is derived from the original \icnetm design~\cite{cheon_electromagnetic_2002}, but has been ported to Chisel~\cite{bachrach:2012} in a more generalized form.  

Our module performs data-level synchronization on multiple AXI-Streams and builds dynamically sized Belle2Link packets based on the design configuration. 
It is composed of six submodules:  

\begin{itemize}
    \item \textbf{Channel Alignment}: Buffers incoming streams to synchronize data. 
    The module performs self-synchronization based on the first received valid signal. 
    In \gnnetm, it automatically synchronizes identifiers, TCs, and clusters.  
    \item \textbf{Channel Delays}: Provides a ring buffer that allows the user to introduce a programmable delay at run time. 
    This module is essential to synchronize \icnetm and \gnnetm.  
    \item \textbf{Trigger Delay}: Enables a programmable delay for the trigger signal.  
    \item \textbf{DAQ Buffers}: Store event data as part of the acquisition logic.  
    \item \textbf{Dispatcher}: Controls the state machine of all \texttt{DAQ Buffers}, allowing multiple events to be recorded even if the transmission of a previous packet via Belle2Link has not yet finished.  
    \item \textbf{Arbiter}: A round-robin arbiter managing access to the Belle2Link bus. 
    The current version supports up to 12 concurrent transmissions.  
\end{itemize}

The Belle2Link Subsystem is ready to be used in other FPGA-based \trg systems at \belletwo to record data. 

\subsection{Deployment}
\label{sec:gnnetm_hardware:deployment}

\begin{figure}
    \centering
    \includegraphics[width=0.7\linewidth]{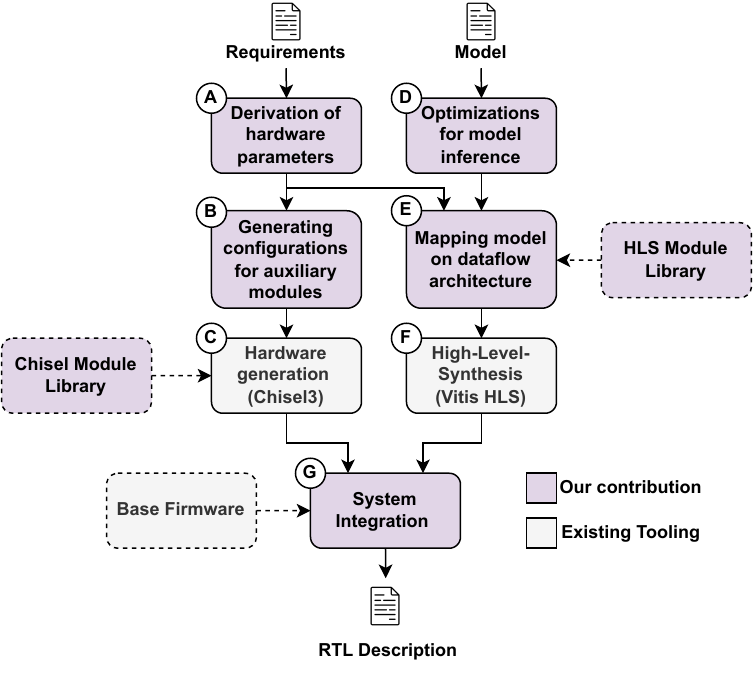}
    \caption{Schematic overview of the deployment process, starting from system requirements and model definition and ending with the register-transfer description of the \gnnetm.}
    \label{fig:gnnetm_hardware:deployment}
\end{figure}

Deploying GNNs on FPGAs is particularly challenging in low-latency, high-throughput applications such as those in the \trg and DAQ systems of high-energy particle detectors. 
In this section, we describe the integration of a dynamic GNN model into an end-to-end processing pipeline designed to meet strict system requirements on latency and throughput. 
An overview of the deployment approach is provided in \cref{fig:gnnetm_hardware:deployment}.

Our methodology requires the following inputs:  
\begin{itemize}
    \item a model description of the target GNN,  
    \item its quantized weight files, and  
    \item a specification of system requirements such as throughput, latency, and hardware constraints.  
\end{itemize}

Based on these system requirements, we derive implementation-specific parameters, such as the required level of parallelism $P_\mathrm{par}$ and the initiation intervals $I_\mathrm{init}$ for each hardware module on the target FPGA~\Circled{A}.
From these system requirements, we also determine configuration parameters for auxiliary modules~\Circled{B}. 
These configurations serve as input to both the hardware generation~\Circled{C} and the dataflow mapping~\Circled{E}, ensuring compatibility across otherwise independent design steps.

For hardware generation~\Circled{C}, we utilize Chisel~\cite{bachrach:2012} as hardware construction language. 
As described above, we developed custom hardware generators for key components, including the preprocessing and postprocessing stages, as well as the Belle2Link subsystem. 
These generators are configured using a YAML file produced in step~\Circled{B}, allowing flexible specification of design-time parameters. 
The resulting Verilog code, produced by the Chisel toolchain, is then packaged as a FuseSoC~\cite{fusesoc:2015,fusesoc:2019} IP core and prepared for system-level integration~\Circled{G}.

The deployment methodology of the GNN model begins with the optimization of its inference compute graph~\Circled{D}. 
This step involves three key transformations:
\begin{itemize}
    \item an 8-bit lookup table is generated at design time for the function $e^{-|x|}$ used in the GravNetConv layer, tailored to the fixed-point datatypes. The table covers the interval $x \in [0,0.5]$, while values with $x > 0.5$ are mapped to zero;
    \item the sigmoid activation function is replaced with a hard sigmoid, as implemented in \texttt{hls4ml}~\cite{fastml_hls4ml}, to reduce resource usage and latency;
    \item the required bit widths of internal registers, such as accumulators, are determined to prevent overflow and truncation of results.
\end{itemize}

Next, we deploy the optimized inference dataflow graph by mapping its operators to corresponding HLS templates from our custom architecture library~\Circled{E}. 
This mapping step, currently performed manually, produces HLS C code. 
After mapping, FIFO buffers are sized to ensure stall-free execution across the inference pipeline. 
In step~\Circled{F}, the high-level model is then converted into a synthesizable RTL description using AMD Vitis~HLS~2024.1~\cite{vitis:2024:1}.  

Once RTL descriptions for all individual modules are generated, the system is integrated into the base firmware~\Circled{G}.  

\subsection{Performance analysis}

We implement the \gnnetm architecture as an RTL design targeting the UT4 system with an AMD~Ultrascale~XCVU190~FPGA, using AMD~Vivado~2024.1~\cite{vivado:2024:1}. 
Cycle-accurate RTL simulations are performed with ModelSim 2023.4~\cite{modelsim:2023:4}. 
To meet the \belletwo \trg requirement of 7.945 million \datawindows per second, the pipelined dataflow architecture is divided into modules $m$, each required to satisfy the throughput constraint  

\begin{equation}
    \label{eq:gnnetm_hardware:throughput}
    R_{\text{thr}} \overset{!}{\leq} \Bigl\lfloor\frac{f_{m}}{I_{\text{init}}}\Bigr\rfloor,
\end{equation}

where $R_{\text{thr}}$ is the target throughput, $f_{m}$ the operation frequency, and $I_{\text{init}}$ the initiation interval.  

The \gnnetm operates with a \SI{127.216}{\mega\hertz} system clock, except for the preprocessing stage, which runs at \SI{254.232}{\mega\hertz}. 
These frequencies are derived from the SuperKEKB RF reference clock by dividing by four and two, respectively.
This results in an initiation interval $I_{\text{init}} = 16$. 
Given the number of non-zero TCs per event $ N > I_{\text{init}}$, we set the parallelism factor to $P_{\text{par}} = \lceil\frac{N}{I_{\text{init}}}\rceil = 2$ to satisfy system throughput.

We report the end-to-end latency of the \gnnetm architecture in \cref{fig:gnnetm_hardware:latency}, measured from the AXI-Stream input received from \icnetm to the AXI-Stream output of the postprocessing stage.
This path, highlighted in \cref{fig:gnnetm_hardware:architecture}, represents the critical real-time segment within the \trg system.
Latency values for each stage are obtained via cycle-accurate RTL simulation and are validated with timing results on the real hardware platform. 
The total end-to-end inference latency is 3.168\,$\mu$s, dominated by the graph neural network dataflow accelerator with 2.052\,$\mu$s, followed by the preprocessing stage with 0.621\,$\mu$s. The postprocessing stage has a latency of 0.495\,$\mu$s. 
This latency exceeds the maximum allowed latency of 1.221\,$\mu$s for an active \trg decision by approximately a factor of three.
To meet the requirement in future deployments, we plan to double the processing frequency, remove one GravNet layer, and reconfigure the module chain to eliminate the additional latency introduced by routing data through the \icnetm. 
Nevertheless, the current setup already enables evaluation of our algorithm on the actual hardware platform within the experiment.  

\begin{figure}
    \centering
    \small
    \begin{tikzpicture}
        \begin{axis}[
            xbar stacked,
            xlabel={Latency (\si{ns})},
            xmin=0, xmax=3500,
            ytick={1},
            yticklabels={\empty},
            enlarge x limits=0.02,
            enlarge y limits=0.2,
            width=0.75\linewidth, 
            height=3cm,
            bar width=8pt,
            ymajorgrids = true,
            minor x tick num=5,
            legend style={
                at={(1.02,1)},
                anchor=north west,
                draw=none,
                legend cell align=left
            },
            legend image code/.code={%
                \draw[#1] (0cm,-0.1cm) rectangle (0.2cm,0.1cm);
            }
        ]
            \addplot+[style={kit-orange,fill=kit-orange,mark=none}] coordinates {(621,1)};
            \addplot+[style={kit-cyan,fill=kit-cyan,mark=none}] coordinates {(2052,1)};
            \addplot+[style={kit-purple,fill=kit-purple,mark=none}] coordinates {(495,1)};

            \addlegendentry{Preprocessing Stage}
            \addlegendentry{CaloClusterNet}
            \addlegendentry{Condensation Point Selection}

        \end{axis}
    \end{tikzpicture}   
    \caption{End-to-end latency for the complete inference chain on the UT4 with an AMD~Ultrascale~XCVU190 FPGA.}
    \label{fig:gnnetm_hardware:latency}
\end{figure}

The overall system resource utilization on the UT4 platform is presented in \cref{fig:gnnetm_hardware:utilization}, including flip-flop registers~(FFs), lookup tables~(LUTs), digital signal processors~(DSPs), and block RAMs~(BRAMs). 
The design uses approximately \SI{51}{\percent} of available LUTs and utilizes all DSP resources. 
BRAM usage remains below \SI{9}{\percent}, primarily due to the Belle2Link subsystem.  

Most DSPs are used by the trainable dense layers of the neural network, particularly those computed at \SI{16}{bit} precision, which benefit most from a mapping on DSPs. 
In contrast, \SI{8}{bit} dense layers are largely mapped to distributed logic. 
The GravNetConv operator accounts for a substantial share of FF and LUT usage due to its $\mathcal{O}(N^2)$ complexity in computing pairwise distances. 
By comparison, the Condensation Point Selection stage requires relatively few resources. 
Although it also computes pairwise distances, it operates in a lower-dimensional space (3 vs.\ 6 dimensions) compared to the GravNetConv layer.  

Overall, the design occupies \SI{82.34}{\percent} of the FPGA’s configurable logic blocks~(CLBs). 
The disproportionately high CLB usage relative to LUTs and FFs indicates routing congestion. 
We tested alternative network configurations with LUT utilization of up to \SI{60}{\percent}, but timing closure becomes increasingly difficult as overall resource usage grows.
To close the timing for this specific version, we used the \textit{Congestion\_SpreadLogic\_high} implementation strategy in Vivado.

\begin{figure}
    \centering
    \small
    \begin{tikzpicture}
        \begin{axis}[
            width  = 0.75\linewidth, 
            height = 5cm,
            major x tick style = transparent,
            ybar stacked,
            bar width=16pt,
            ymajorgrids = true,
            symbolic x coords={FF,LUT,DSP,BRAM},
            xtick = data,
            scaled y ticks = false,
            enlarge x limits=0.2,
            enlarge y limits=0.1,
            ymin=0,
            ymax=100,
            ytick={0, 20, 40, 60, 80, 100},
            yticklabels={0~\%,20~\%,40~\%,60~\%,80~\%,100\%},
            minor y tick num=5,
            legend style={
                at={(1.02,1)},
                anchor=north west,
                draw=none,
                legend cell align=left
            },
            legend image code/.code={%
                \draw[#1] (0cm,-0.1cm) rectangle (0.2cm,0.1cm);
            }
        ]
            \addplot+[style={kit-gray,fill=kit-gray,mark=none}] coordinates {(BRAM, 1.56) (DSP, 0.00) (FF, 2.15) (LUT, 5.25)};
            \addplot+[style={kit-red,fill=kit-red,mark=none}] coordinates {(BRAM, 5.73) (DSP, 0.00) (FF, 0.40) (LUT, 1.45)};
            \addplot+[style={kit-orange,fill=kit-orange,mark=none}] coordinates {(BRAM, 0.95) (DSP, 0.00) (FF, 1.50) (LUT, 2.69)};
            \addplot+[style={kit-cyan,fill=kit-cyan,mark=none}] coordinates {(BRAM, 0.00) (DSP, 87.56) (FF, 11.79) (LUT, 19.71)};
            \addplot+[style={kit-cyan!50,fill=kit-cyan!50,mark=none}] coordinates {(BRAM, 0.00) (DSP, 12.44) (FF, 9.47) (LUT, 17.59)};
            \addplot+[style={kit-purple,fill=kit-purple,mark=none}] coordinates {(BRAM, 0.00) (DSP, 0.00) (FF, 2.91) (LUT, 3.83)};

            \addlegendentry{Base Firmware}
            \addlegendentry{B2Link Subsystem}
            \addlegendentry{Preprocessing Stage}
            \addlegendentry{Dense}
            \addlegendentry{GravNetConv}
            \addlegendentry{Condensation Point Selection}

        \end{axis}
    \end{tikzpicture}
    \caption{Utilization of system resources on the UT4 with an AMD~Ultrascale~XCVU190 FPGA.}
    \label{fig:gnnetm_hardware:utilization}
\end{figure}

\subsection{Validation}
\label{sec:hardware_validation}
\begin{figure*}
    \centering
    \includegraphics[width=0.7\linewidth]{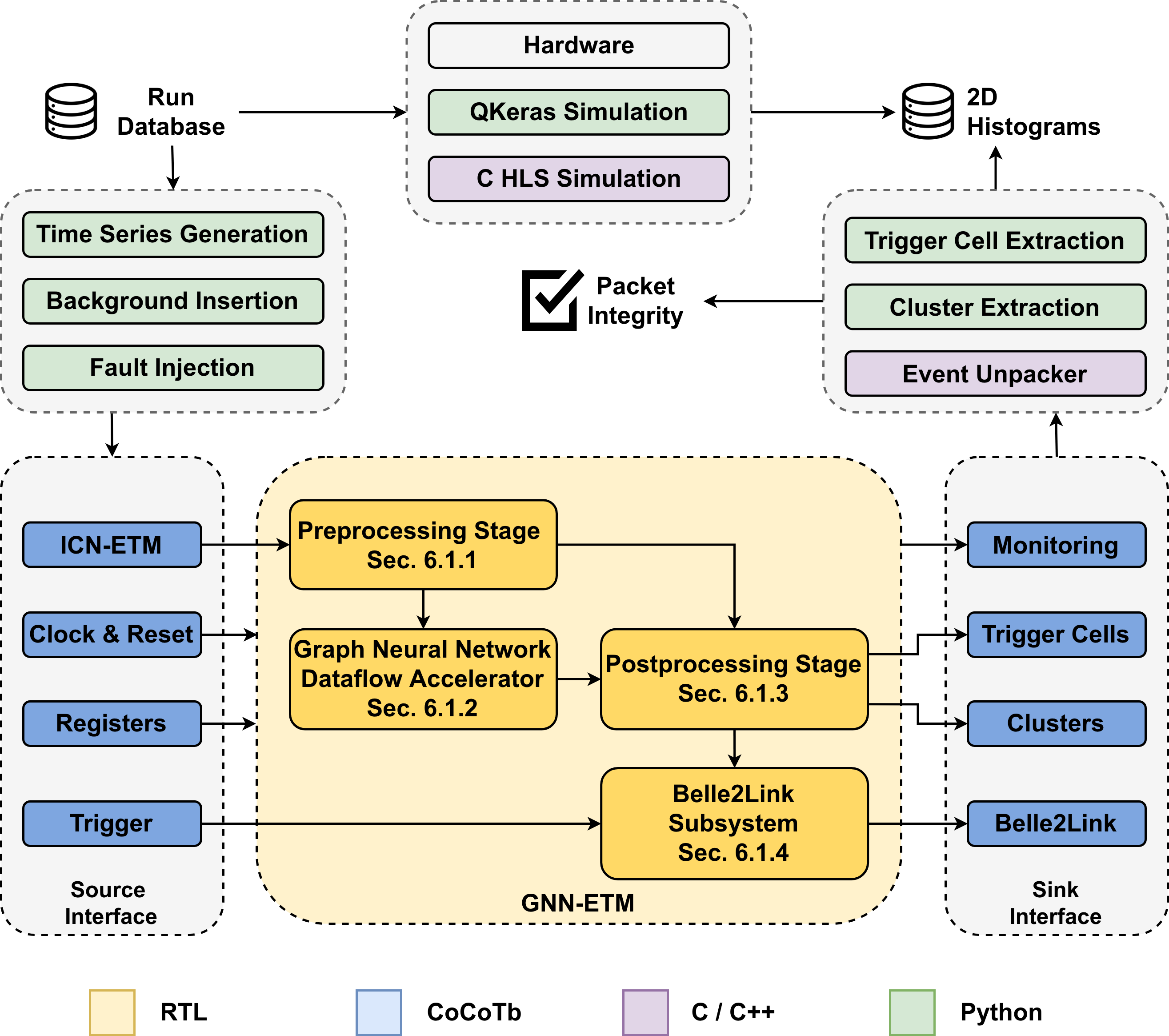}
    \caption{Full simulation framework for the \gnnetm validation. The workflow for comparing the simulations on all three levels of abstraction and the hardware output itself is shown, with their respective software used for implementation.}
    \label{fig:gnnetm_hardware:validation}
\end{figure*}

We verify the correct functionality of \gnnetm through comprehensive simulations across multiple abstraction levels. 
An overview of the simulation framework is provided in \cref{fig:gnnetm_hardware:validation}. 
Based on either technical input samples, simulated data or previously recorded events, we conduct simulations on three levels of abstraction:

\begin{enumerate}
    \item \textbf{Quantized model simulation:} Reference results are computed using the quantized model implemented in \texttt{QKeras}. This simulation is also used during quantization-aware training, as described in \cref{sec:gnnetm_software_training}.  
    \item \textbf{C-level transaction simulation:} This simulation, implemented in Vitis~2024.1, evaluates neural network inference and the condensation point selection algorithm. It excludes the preprocessing and postprocessing stages as well as the Belle2Link subsystem.  
    \item \textbf{Cycle-accurate RTL simulation:} Conducted for the \gnnetm, excluding only hard IP cores such as GTY transceivers and clocking resources. Input events from the run database are converted into time-series datastreams compatible with the AXI-Stream interface. The simulation allows injection of random TCs as simulated background or erroneous AXI-Stream packets.  
    It is executed in ModelSim~2023.4~\cite{modelsim:2023:4} with CoCoTb~\cite{cocotb:2025} and the CoCoTb-AXI~\cite{cocotb-axi:2023} extension. The Belle2Link subsystem driver replicates the behavior of the Belle2Link protocol. Packets received by the driver are unpacked with a C implementation of the experiment’s software unpacker, enabling complete end-to-end validation of the processing chain.  
\end{enumerate}

\begin{figure*}[ht!]
     \centering
     \begin{subfigure}[b]{\thirdwidth\textwidth}
         \centering
         \includegraphics[width=\textwidth]{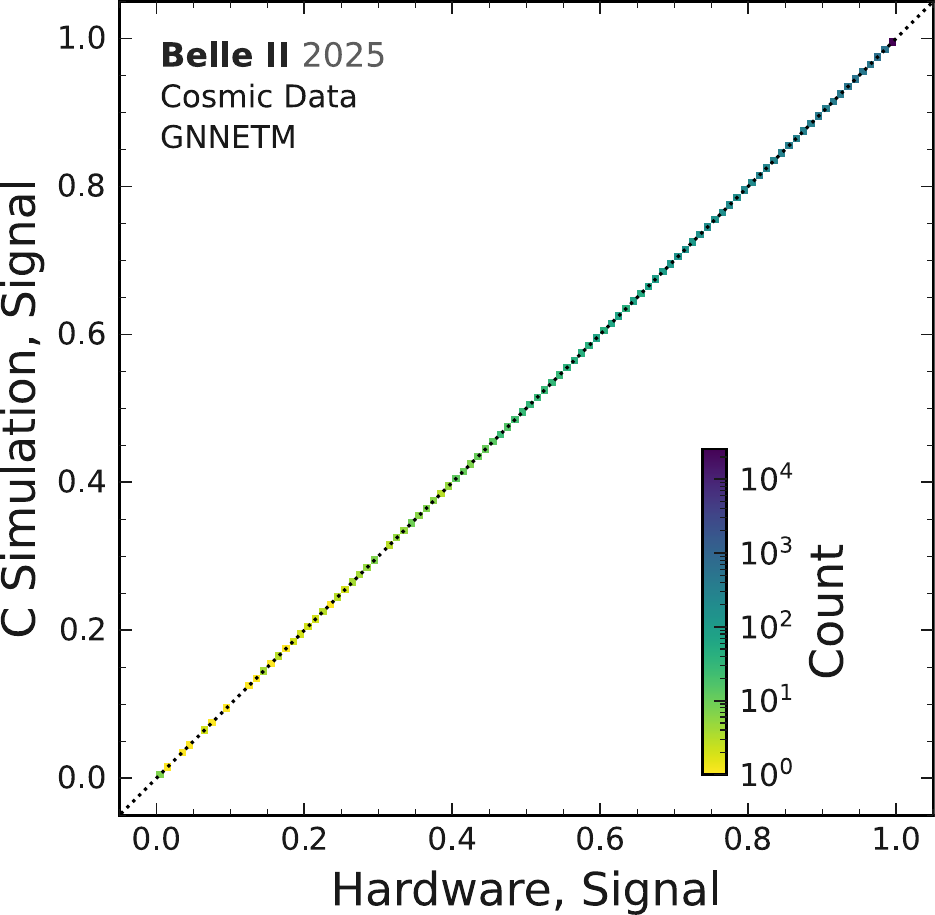}
         \caption{C-Simulation vs. Hardware}
         \label{fig:gnnetm_hardware:validation:signal:csim_hw}
     \end{subfigure}\hfill
        \begin{subfigure}[b]{\thirdwidth\textwidth}
         \centering
         \includegraphics[width=\textwidth]{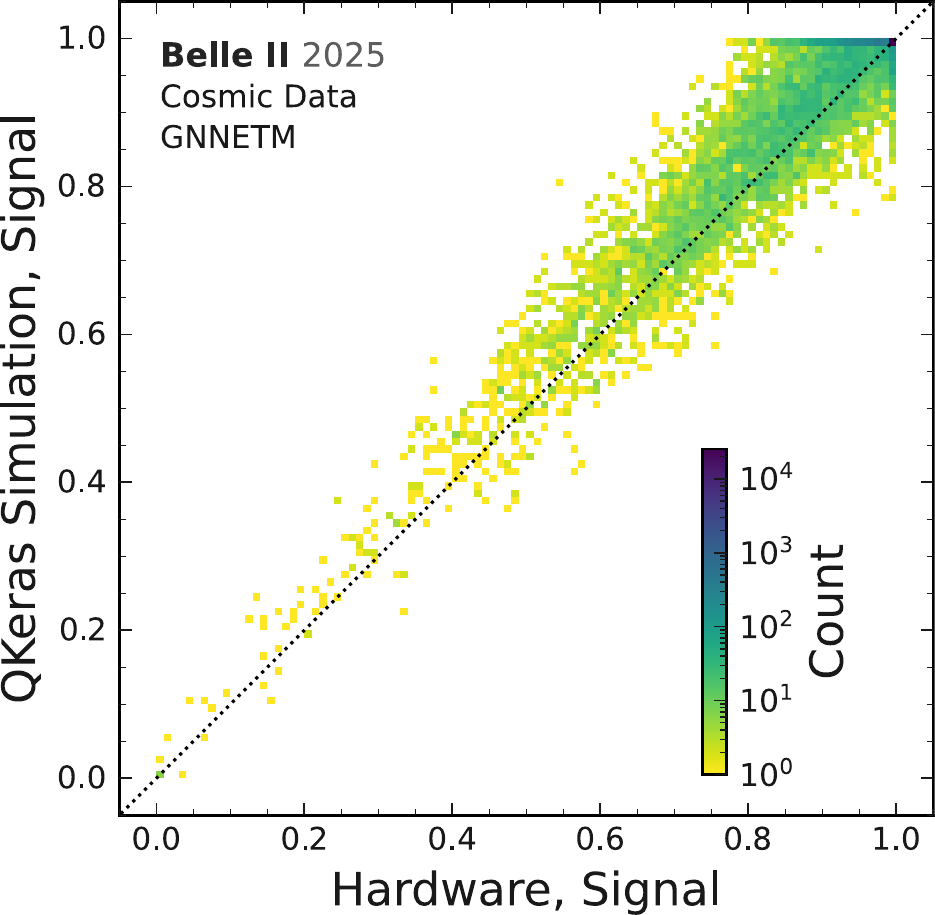}
         \caption{QKeras vs. Hardware, before}
        \label{fig:gnnetm_hardware:validation:signal:qkeras_hw:a}
     \end{subfigure}\hfill
        \begin{subfigure}[b]{\thirdwidth\textwidth}
         \centering
         \includegraphics[width=\textwidth]{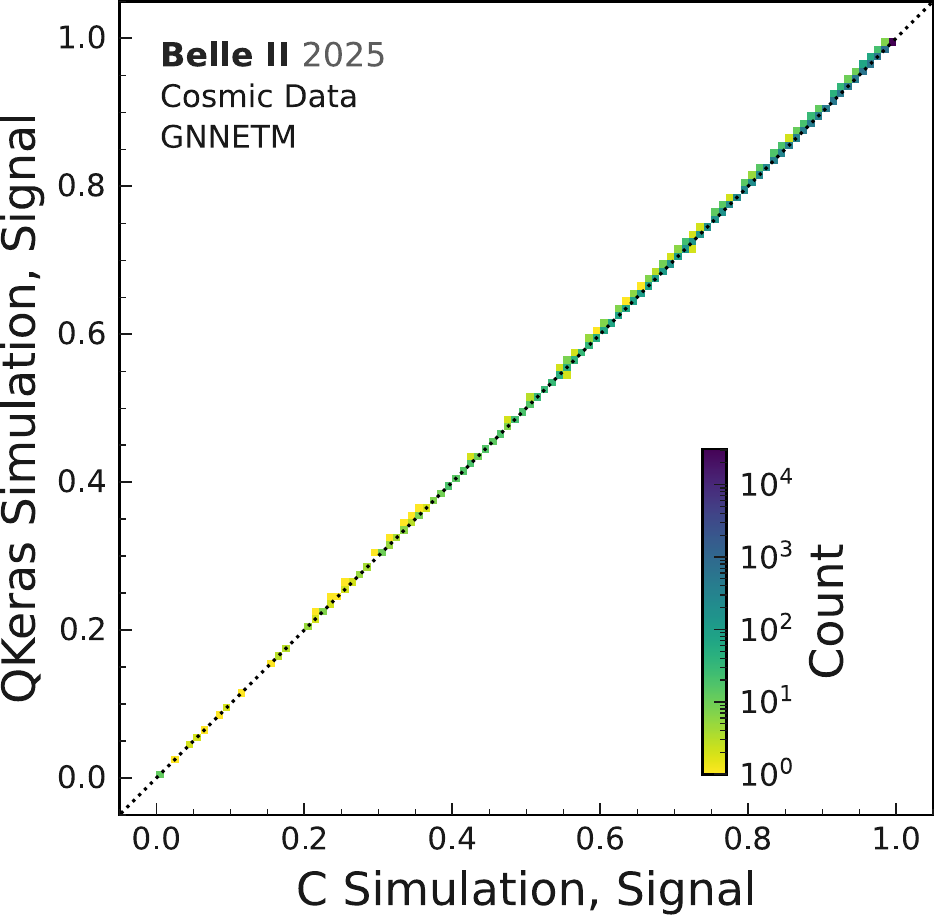}
         \caption{QKeras vs. Hardware, after}
    \label{fig:gnnetm_hardware:validation:signal:qkeras_hw:b}
     \end{subfigure}
\caption{2D histograms of the predicted signal classifier values from C-simulation vs. hardware (\subref{fig:gnnetm_hardware:validation:signal:csim_hw}), \qkeras vs. hardware with the original agreement
(\subref{fig:gnnetm_hardware:validation:signal:qkeras_hw:a}) and \qkeras vs. C-simulation after improvements to both C-simulation and \qkeras (\subref{fig:gnnetm_hardware:validation:signal:qkeras_hw:b}). The agreement between C-simulation and hardware is bitwise correct.}
\label{fig:validation:signal_all}
\end{figure*}

To validate our C-based HLS simulation, we compare it with cosmic-ray data recorded in the absence of colliding beams. 
Such cosmic runs are regularly collected to monitor and calibrate the system. 
In this study, we use \num{10000} events recorded on the hardware, referred to as \textit{Cosmic Data}. 
Since multiple \triggerwindows are available per event and only those containing at least one TC are retained, the evaluation is based on \num{18325} \triggerwindows in total. 
The comparison, shown in \cref{fig:gnnetm_hardware:validation:signal:csim_hw}, demonstrates that the two distributions are in excellent agreement. 
We therefore conclude that the C-simulation model provides a reliable basis for algorithmic performance analysis.

Validation of the \qkeras simulation against the hardware is performed via the C-based simulation, which serves as an exact reference. 
\qkeras is essential for fast iteration, as it enables rapid testing of design changes and retraining of models without repeating the time-consuming hardware implementation. 
Because data types, rounding, and quantization must be configured manually, the initial agreement with the hardware was poor, as shown in \cref{fig:gnnetm_hardware:validation:signal:qkeras_hw:a}, limiting the reliability of performance studies.
To improve this agreement, we adapted both the hardware implementation and the \qkeras model. 
Input feature scaling was originally fused into the first dense layer (DL1, see \cref{fig:quantized_model}), which caused a mismatch.
Moving the scaling into the preprocessing stage (\cref{sec:gnnetm_hardware:preprocessing}) removed one quantization step and resolved this issue. 
We also corrected truncation of intermediate results in the processing elements by adjusting bit widths. 
On the \qkeras side, rounding in the inference step was set to \texttt{floor} to match hardware behavior, and a dedicated function was implemented to model the LUT-based exponential function (\cref{sec:gnnetm_hardware:deployment}).
The improved agreement between \qkeras and the C-simulation is shown in \cref{fig:gnnetm_hardware:validation:signal:qkeras_hw:b}. 
As the agreement between the C-simulation and the hardware is in excellent agreement, the improved agreement also propagates to the comparison between \qkeras and the hardware.
The remaining differences stem from our unstable sorting implementation: although the same input order yields deterministic results, the ordering cannot be propagated consistently across hardware, C-simulation, and \qkeras. 
In practice, this could be resolved by employing a stable sorting algorithm if exact agreement between all implementations is required.
While employing a stable sorting algorithm would enforce exact agreement across all implementations, we retain the current approach as the observed differences have negligible impact on the physics performance results.

The network’s large depth makes it sensitive to small perturbations, and even a single-bit change in 16-bit fixed-point inputs can substantially alter the final predictions.
This is particularly relevant for the regression targets, where small deviations may alter the condensation point selection or degrade the quality of the energy and position estimates. 
To ensure results that match hardware performance as closely as possible, all subsequent evaluations are therefore based on the C-simulation.

\FloatBarrier
\section{Physics performance}
\label{sec:results}

In this section, we present a comparison of the performance between the \gnnetm and the \icnetm \trg algorithms. 
We first discuss the performance metrics (see \cref{sec:performance_metrics}) for simulated samples in Sec.~\ref{sec:results_simulation}.
We then validate the system using real high rate collision data recorded under standard \belletwo operating conditions in Sec.~\ref{sec:results_data}.

\FloatBarrier
\subsection{Simulation studies}
\label{sec:results_simulation}
We evaluate the clustering performance of the \gnnetm and the \icnetm algorithm using the \textit{Single Photon Sample} described in Sec.~\ref{sec:gnnetm_software_training} processed with the C-level transaction simulation described in \cref{sec:hardware_validation}. To allow for a fair comparison between both algorithms for the cluster finding efficiency, purity, and resolutions, only events containing exactly one \offlinecluster are selected. For this evaluation, we do not use any cut on the predicted signal classifier value to distinguish signal from background clusters.

\begin{figure*}[ht!]
     \centering
     \begin{subfigure}[b]{\thirdwidth\textwidth}
         \centering
         \includegraphics[width=\textwidth]{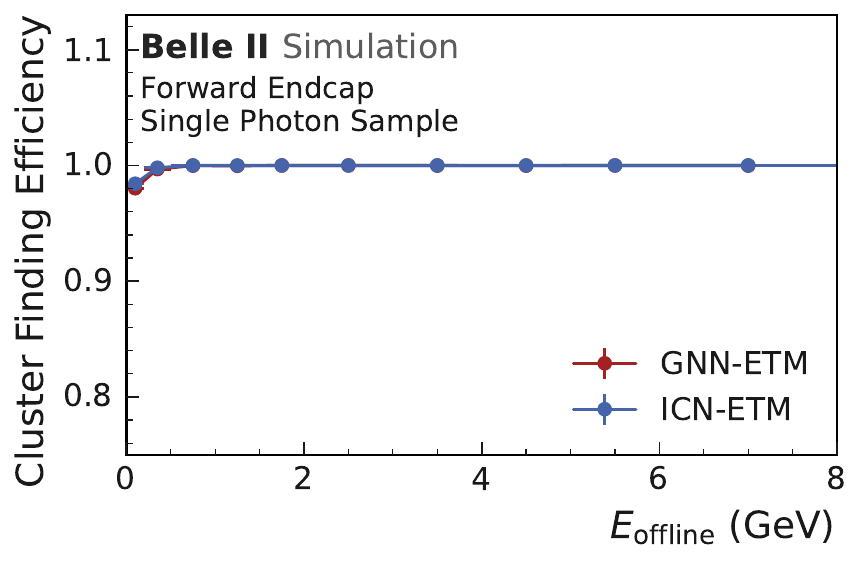}
         \caption{Forward endcap.}
         \label{fig:simulated_performance_eff:a}
     \end{subfigure}\hfill
        \begin{subfigure}[b]{\thirdwidth\textwidth}
         \centering
         \includegraphics[width=\textwidth]{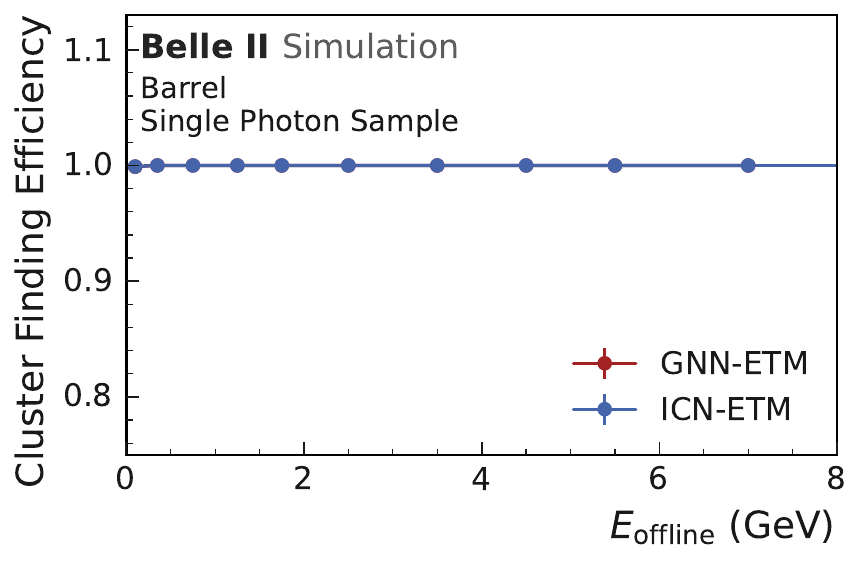}
         \caption{Barrel.}
         \label{fig:simulated_performance_eff:b}
     \end{subfigure}\hfill
        \begin{subfigure}[b]{\thirdwidth\textwidth}
         \centering
         \includegraphics[width=\textwidth]{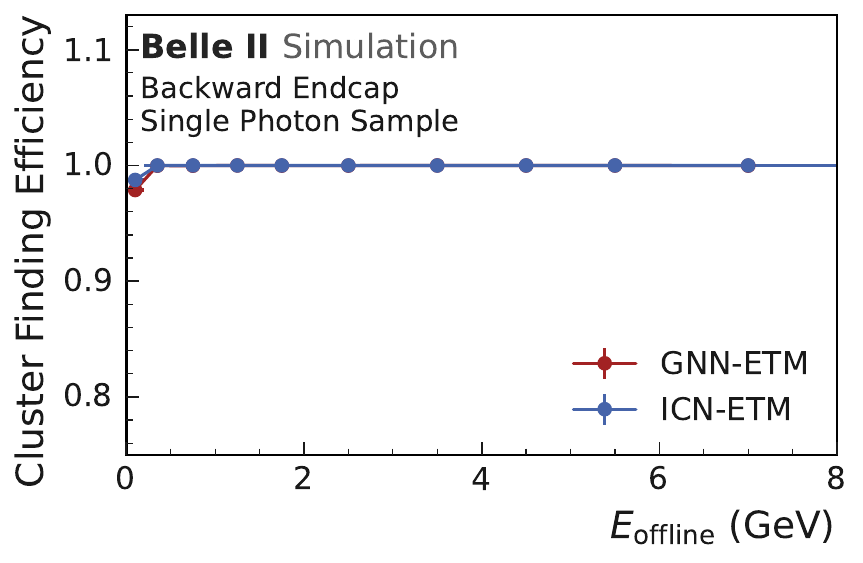}
         \caption{Backward endcap.}
         \label{fig:simulated_performance_eff:c}
     \end{subfigure}
\caption{Cluster finding efficiency as a function of \offlinecluster energy $E_{\text{offline}}$ for the \baseline~(blue) and the \gnnetm~(red) in the (\subref{fig:simulated_performance_eff:a}) forward endcap, (\subref{fig:simulated_performance_eff:b}) barrel, and (\subref{fig:simulated_performance_eff:c}) backward endcap, evaluated on the \textit{Single Photon Sample} with only one \offlinecluster per event. 
Markers are connected by solid lines to guide the eye. 
Vertical error bars indicate statistical uncertainties and are in most cases smaller than the marker size. 
Horizontal error bars show the bin width. 
The uncertainties of the two \trg algorithms are correlated since they are evaluated on the same simulated events. 
Results are shown without any signal classifier cut.}
\label{fig:simulated_performance_eff}
\end{figure*}

\begin{figure*}[ht!]
     \centering
     \begin{subfigure}[b]{\thirdwidth\textwidth}
         \centering
         \includegraphics[width=\textwidth]{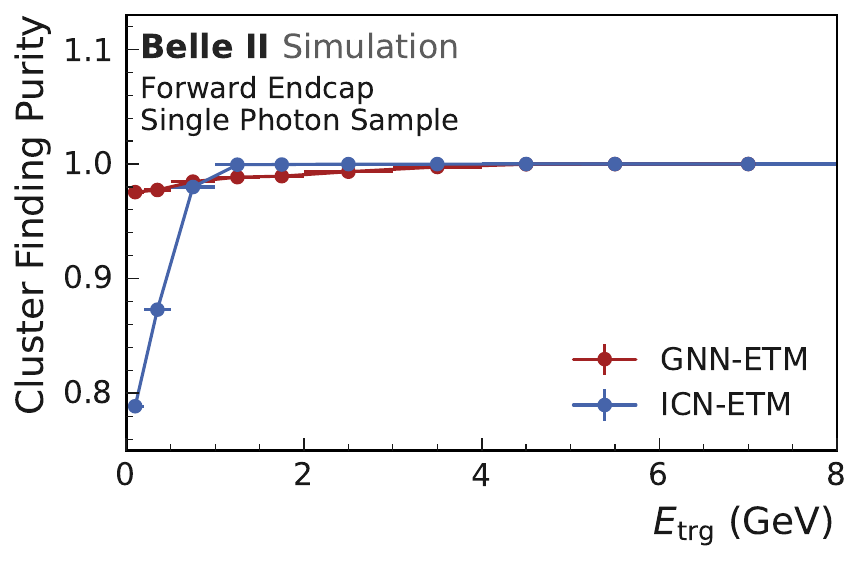}
         \caption{Forward endcap.}
         \label{fig:simulated_performance_pur:a}
     \end{subfigure}\hfill
        \begin{subfigure}[b]{\thirdwidth\textwidth}
         \centering
         \includegraphics[width=\textwidth]{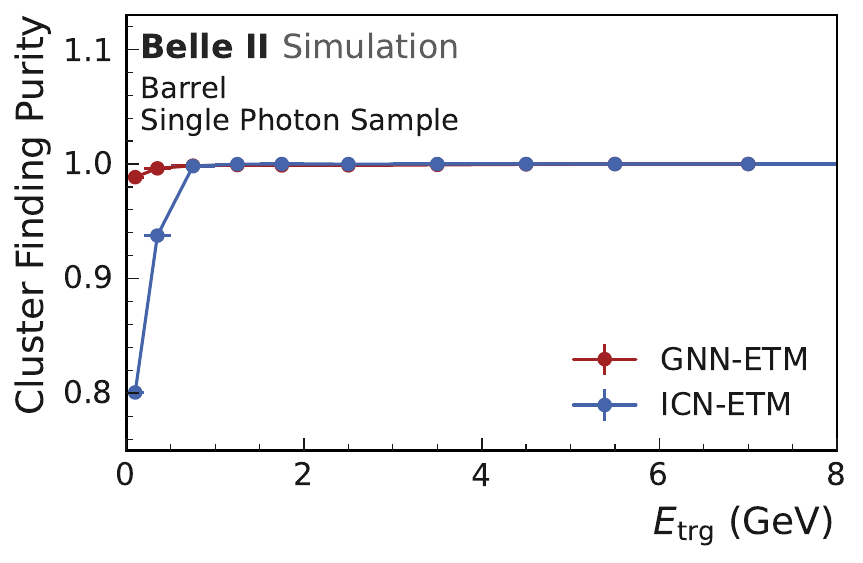}
         \caption{Barrel.}
         \label{fig:simulated_performance_pur:b}
     \end{subfigure}\hfill
        \begin{subfigure}[b]{\thirdwidth\textwidth}
         \centering
         \includegraphics[width=\textwidth]{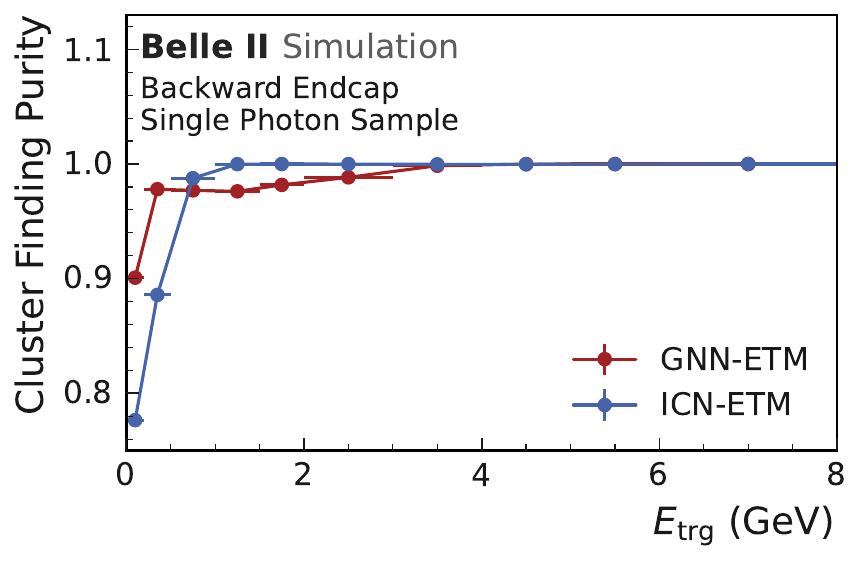}
         \caption{Backward endcap.}
         \label{fig:simulated_performance_pur:c}
     \end{subfigure}
\caption{Cluster finding purity as a function of \triggercluster energy $E_{\text{trg}}$ for the \baseline~(blue) and the \gnnetm~(red) in the (\subref{fig:simulated_performance_pur:a}) forward endcap, (\subref{fig:simulated_performance_pur:b}) barrel, and (\subref{fig:simulated_performance_pur:c}) backward endcap, evaluated on the \textit{Single Photon Sample} with only one \offlinecluster per event. 
Markers are connected by solid lines to guide the eye. 
Vertical error bars indicate statistical uncertainties and are in most cases smaller than the marker size. 
Horizontal error bars show the bin width. 
The uncertainties of the two \trg algorithms are correlated since they are evaluated on the same simulated events. 
Results are shown without any signal classifier cut.}
\label{fig:simulated_performance_pur}
\end{figure*}

The cluster finding efficiency as a function of the \offlinecluster energy $E_{\text{offline}}$ is shown in \cref{fig:simulated_performance_eff}, and the corresponding purity as a function of the \triggercluster energy $E_{\text{trg}}$ is shown in \cref{fig:simulated_performance_pur}.
For both the \gnnetm and \icnetm, the efficiency is 1 for all bins in the barrel and all except the lowest energy bin in the forward and backward endcap. 
The slight drop in efficiency for both algorithms for low-energetic \offlineclusters in the endcaps is due to beam background energy depositions leading to an overestimation of the \triggercluster energy in comparison to the \offlinecluster. 
If the beam background energy is sufficiently high, the \triggercluster fails to match the \offlinecluster.

For the purity, both algorithms reach a purity of 1 for high-energetic clusters for all three detector regions. 
However, the \icnetm shows systematically lower purity in the low-energy region due to its inability to cluster over the gaps between the detector regions (see \cref{sec:belle2_icn}). 
While the offline reconstruction allows for \offlineclusters with crystals in both the barrel and one of the endcap regions, and subsequently has transferred this ability to the \gnnetm by setting the correct target, the \icnetm always returns two \icnetmclusters in the case of TCs existing in two regions. 
As the reconstructed position of the \offlinecluster is nearly always closer to the higher energy deposition in the detector, the higher-energetic \icnetmcluster will be the primary match (see Sec.~\ref{sec:gnnetm_software_training}).
The loss in purity for lower energy bins in the \gnnetm likely originates from cases where the model returns two low-energy \gnnclusters instead of one high-energy \gnncluster. 
We suspect that this behavior is influenced by close-by clusters present in parts of the training sample, which arise from increased beam background in the endcaps and from photons converting in inactive material between the beam pipe and the ECL endcaps.

\FloatBarrier
\FloatBarrier
\subsubsection{Energy and angular resolutions}
\begin{figure*}[ht!]
     \centering
        \begin{subfigure}[b]{\thirdwidth\textwidth}
         \centering
         \includegraphics[width=\fullwidth\textwidth]{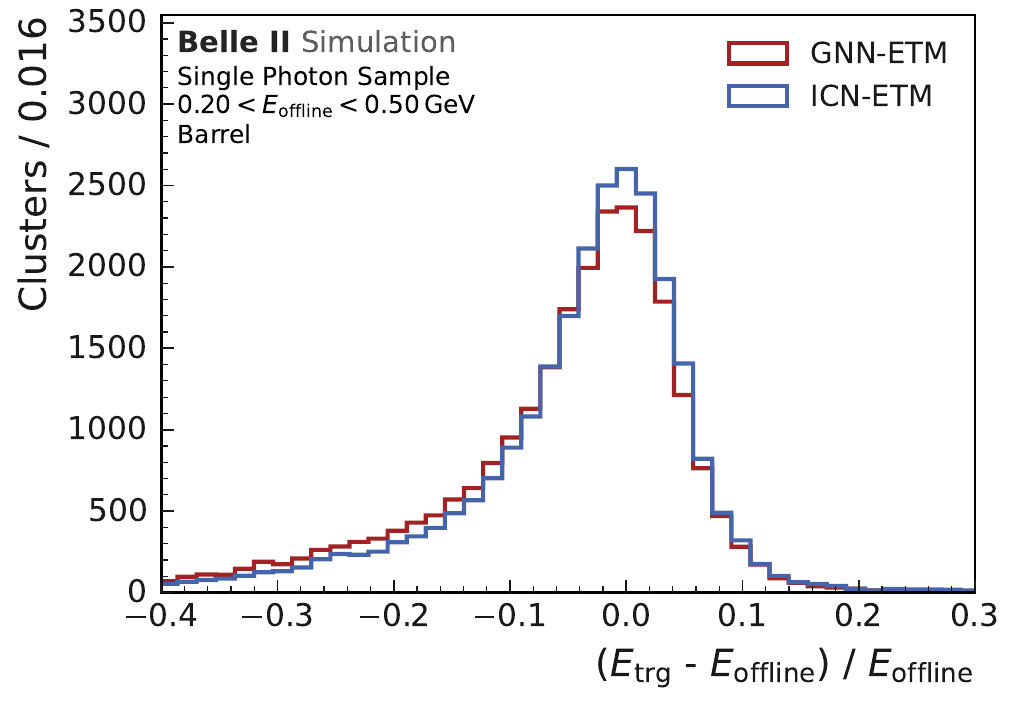}
         \caption{Energy resolution.}
         \label{fig:simulated_resolution_exemplary:energy}
     \end{subfigure}\hfill
     \begin{subfigure}[b]{\thirdwidth\textwidth}
         \centering
         \includegraphics[width=\fullwidth\textwidth]{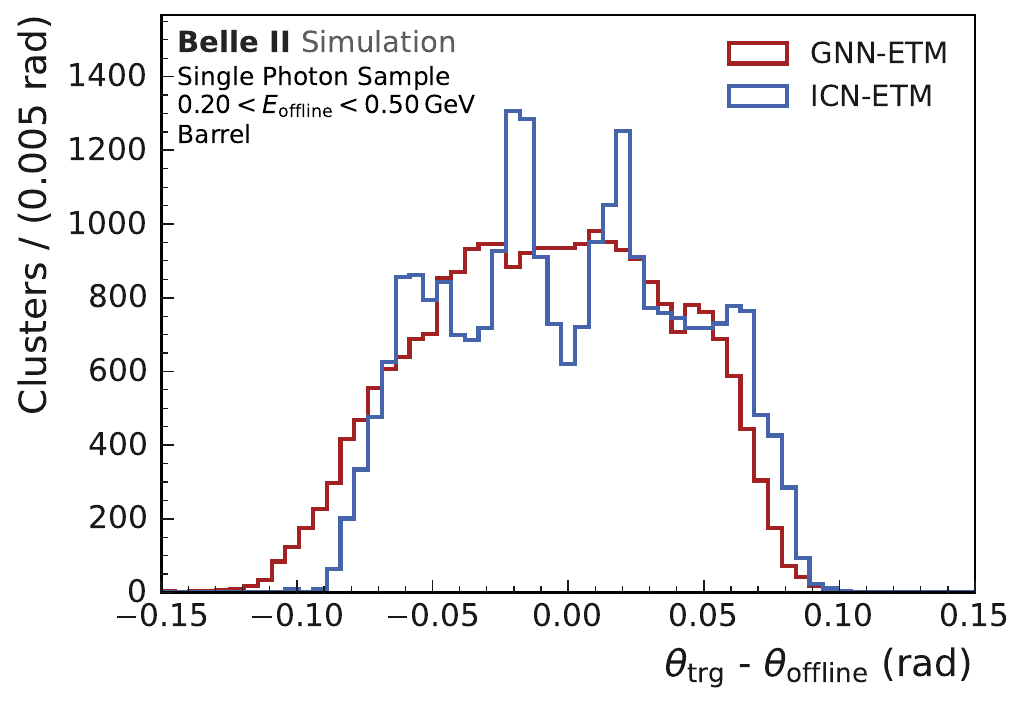}
         \caption{$\theta$ resolution.}
         \label{fig:simulated_resolution_exemplary:theta}
     \end{subfigure}\hfill
     \begin{subfigure}[b]{\thirdwidth\textwidth}
         \centering
         \includegraphics[width=\fullwidth\textwidth]{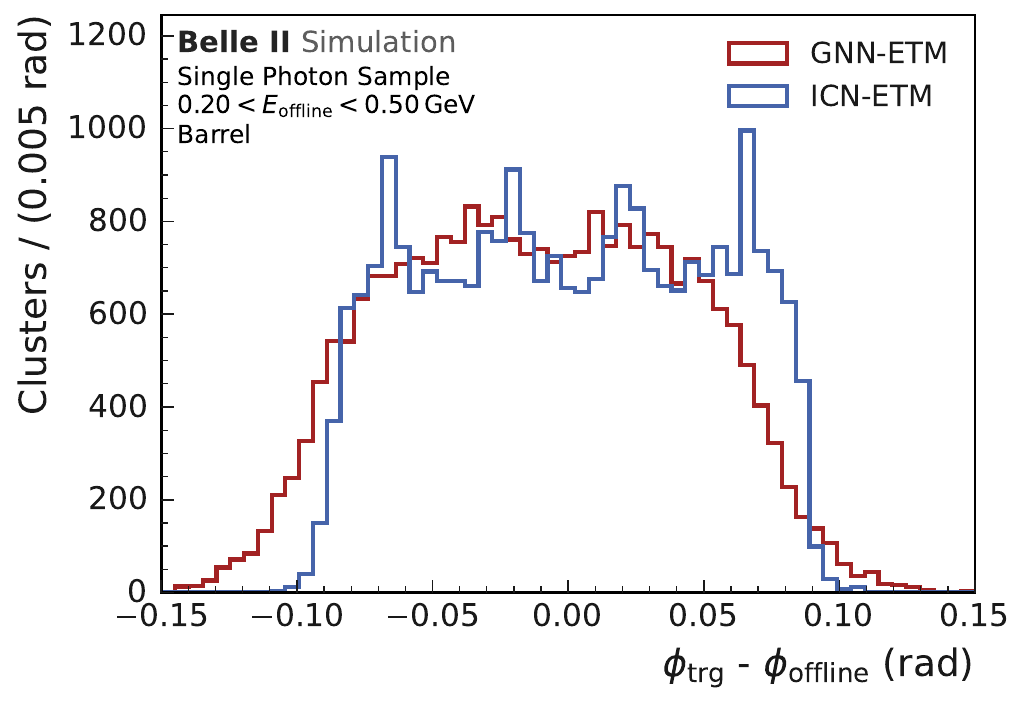}
         \caption{$\phi$ resolution.}
         \label{fig:simulated_resolution_exemplary:phi}
     \end{subfigure}
\caption{Comparison of (\subref{fig:simulated_resolution_exemplary:energy}) energy and (\subref{fig:simulated_resolution_exemplary:theta} and \subref{fig:simulated_resolution_exemplary:phi}) angular resolutions between the \gnnetm~(red) and the \baseline~(blue) for \offlineclusters in the energy range \hbox{200--500\,MeV} in the barrel, using the \textit{Single Photon Sample} with only one \offlinecluster per event. 
Only \triggerclusters reconstructed by both algorithms and matched to an \offlinecluster are considered.}\label{fig:simulated_resolution_exemplary}
\end{figure*}

We evaluate the energy and angular resolutions of the \gnnetm in comparison to the \icnetm algorithm. 
The resolution is computed only for \triggerclusters found by both algorithms that can be associated with an \offlinecluster, ensuring a fair comparison based on identical input. 
As an example, the energy resolution, and the angular resolutions in $\theta$ and $\phi$ are shown in \cref{fig:simulated_resolution_exemplary} for the \offlinecluster range 200--500\,MeV, using the \textit{Single Photon Sample} containing exactly one \offlinecluster per event.  
The $\theta$ and $\phi$ resolutions obtained with the \icnetm exhibit a discrete, spiky structure, which reflects the choice of TC centers as \icnetmcluster positions.
In contrast, the \gnnetm inference does not rely on discrete positions, resulting in smoother distributions but with longer tails.

\begin{figure*}[ht!]
    \centering
    \begin{subfigure}[b]{\thirdwidth\textwidth}
        \centering
        \includegraphics[width=\textwidth]{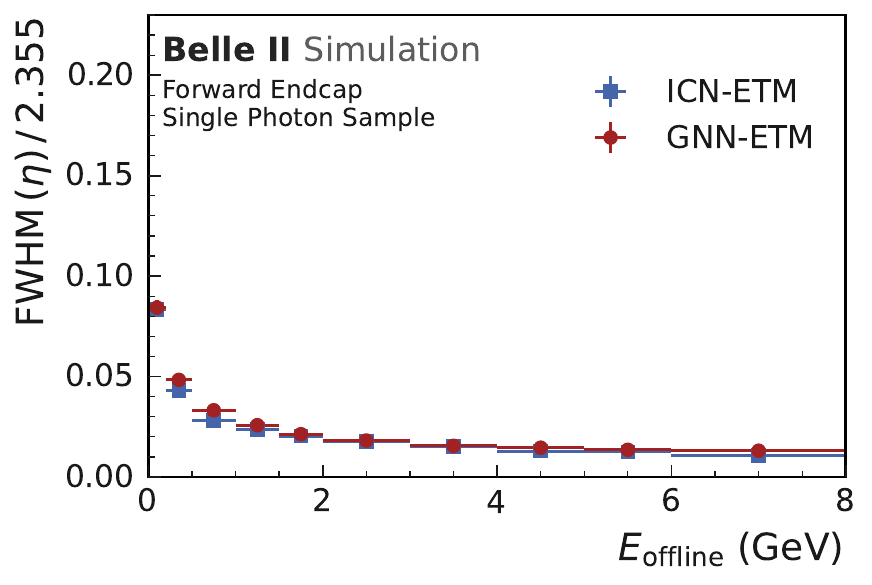}
        \caption{Forward endcap.}
        \label{fig:resolution_energy_fwd}
    \end{subfigure}\hfill
    \begin{subfigure}[b]{\thirdwidth\textwidth}
        \centering
        \includegraphics[width=\textwidth]{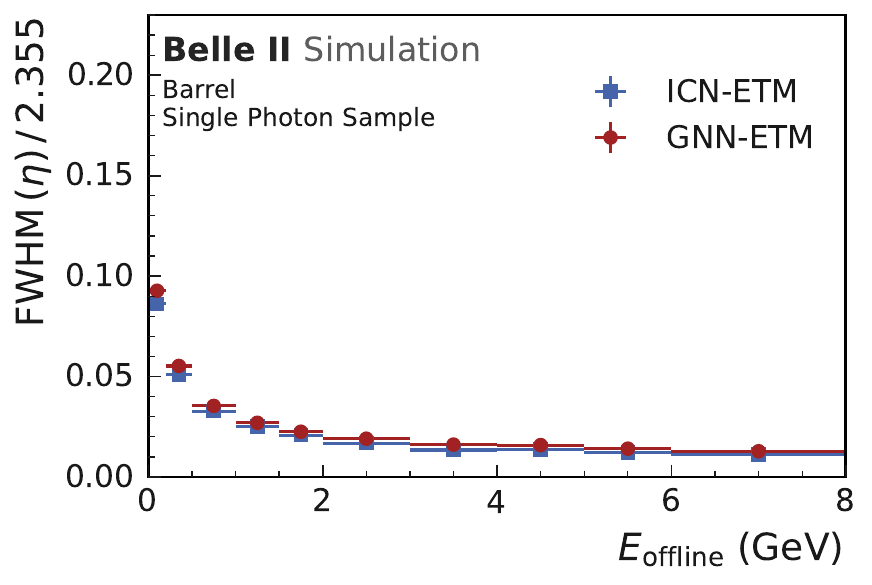}
        \caption{Barrel.}
        \label{fig:resolution_energy_barrel}
    \end{subfigure}\hfill
    \begin{subfigure}[b]{\thirdwidth\textwidth}
        \centering
        \includegraphics[width=\textwidth]{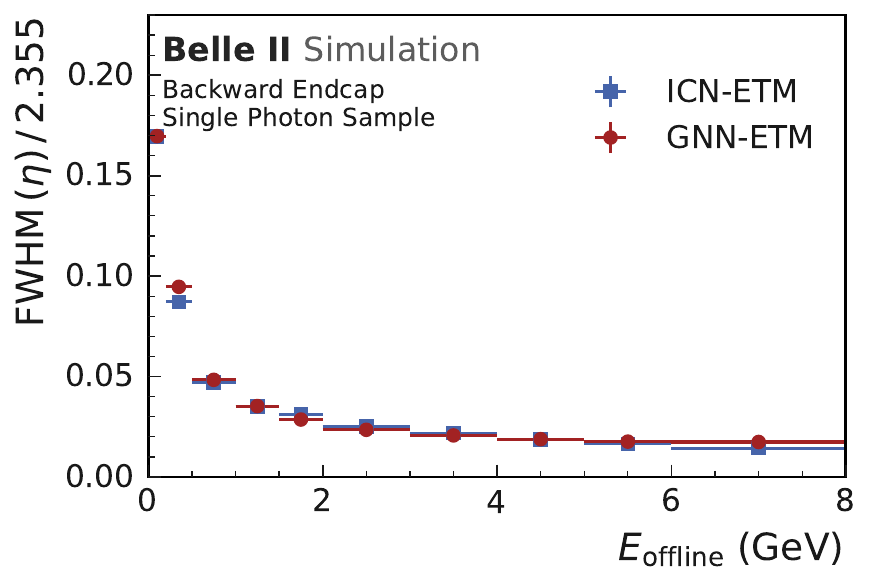}
        \caption{Backward endcap.}
        \label{fig:resolution_energy_bwd}
    \end{subfigure}

    \vspace{1em}
    \begin{subfigure}[b]{\thirdwidth\textwidth}
        \centering
        \includegraphics[width=\textwidth]{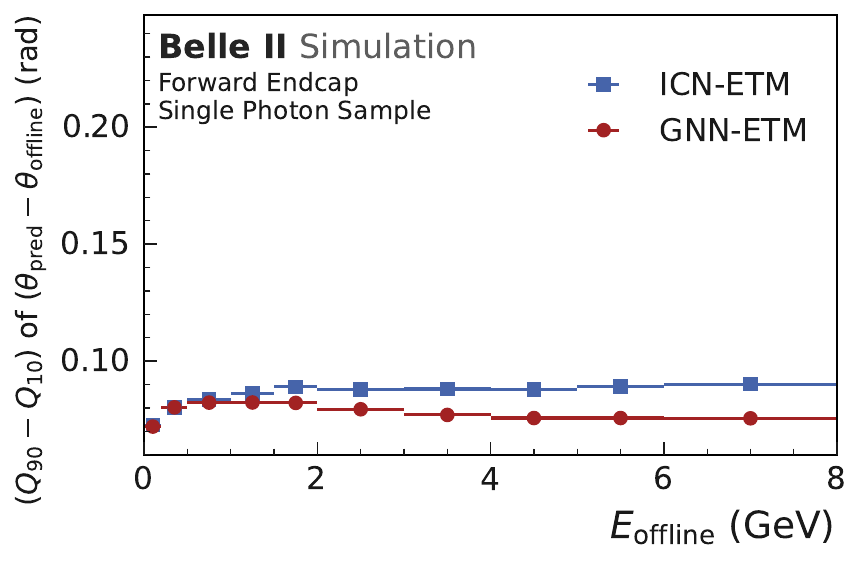}
        \caption{Forward endcap.}
        \label{fig:resolution_theta_fwd}
    \end{subfigure}\hfill
    \begin{subfigure}[b]{\thirdwidth\textwidth}
        \centering
        \includegraphics[width=\textwidth]{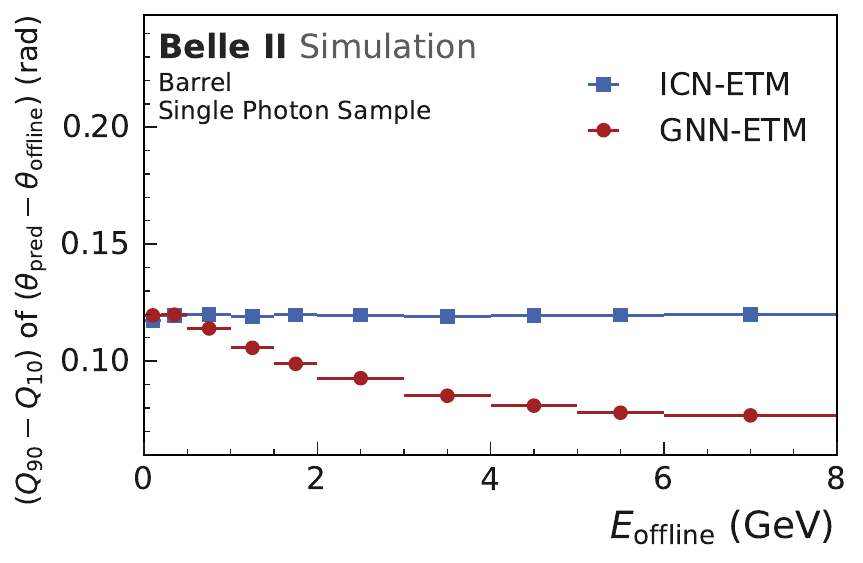}
        \caption{Barrel.}
        \label{fig:resolution_theta_barrel}
    \end{subfigure}\hfill
    \begin{subfigure}[b]{\thirdwidth\textwidth}
        \centering
        \includegraphics[width=\textwidth]{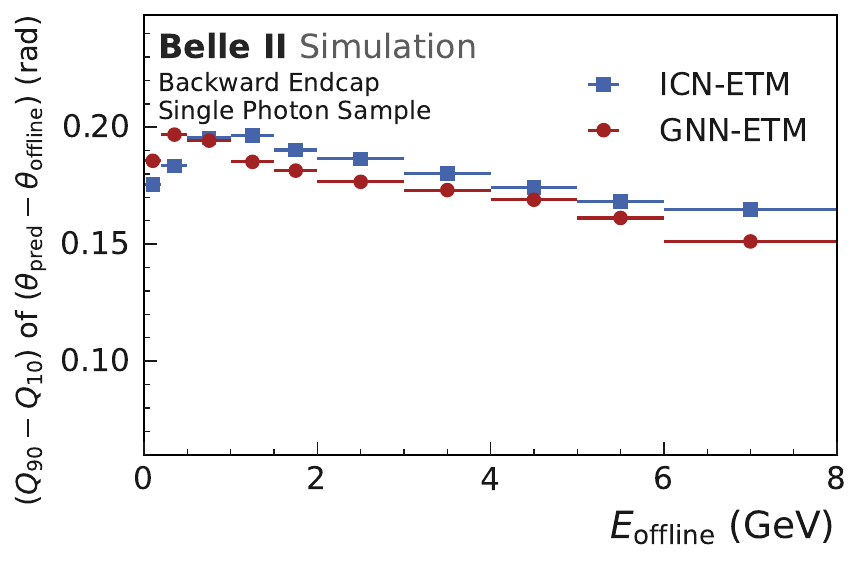}
        \caption{Backward endcap.}
        \label{fig:resolution_theta_bwd}
    \end{subfigure}

    \vspace{1em}
    \begin{subfigure}[b]{\thirdwidth\textwidth}
        \centering
        \includegraphics[width=\textwidth]{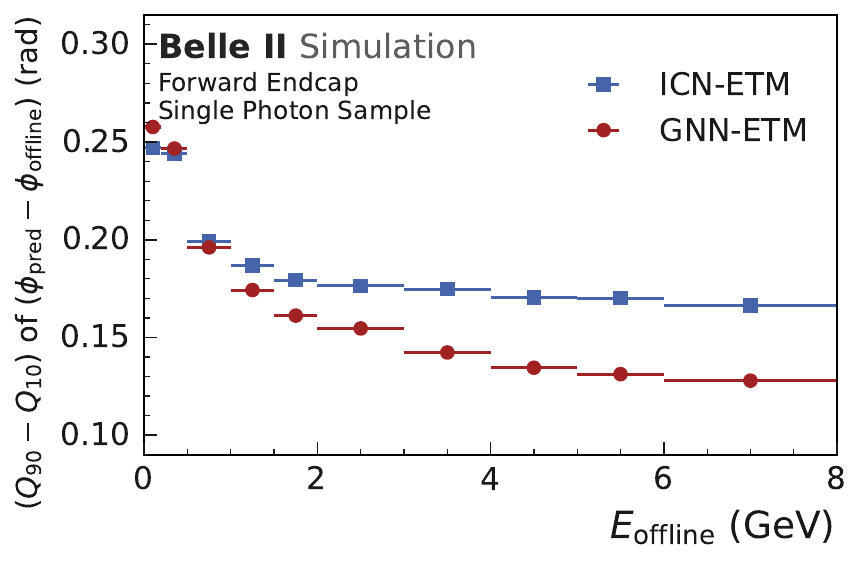}
        \caption{Forward endcap.}
        \label{fig:resolution_phi_fwd}
    \end{subfigure}\hfill
    \begin{subfigure}[b]{\thirdwidth\textwidth}
        \centering
        \includegraphics[width=\textwidth]{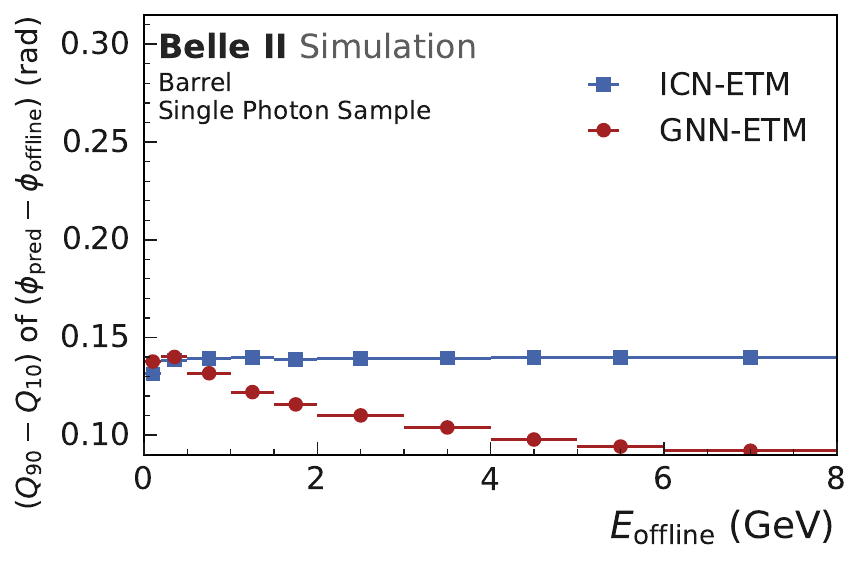}
        \caption{Barrel.}
        \label{fig:resolution_phi_barrel}
    \end{subfigure}\hfill
    \begin{subfigure}[b]{\thirdwidth\textwidth}
        \centering
        \includegraphics[width=\textwidth]{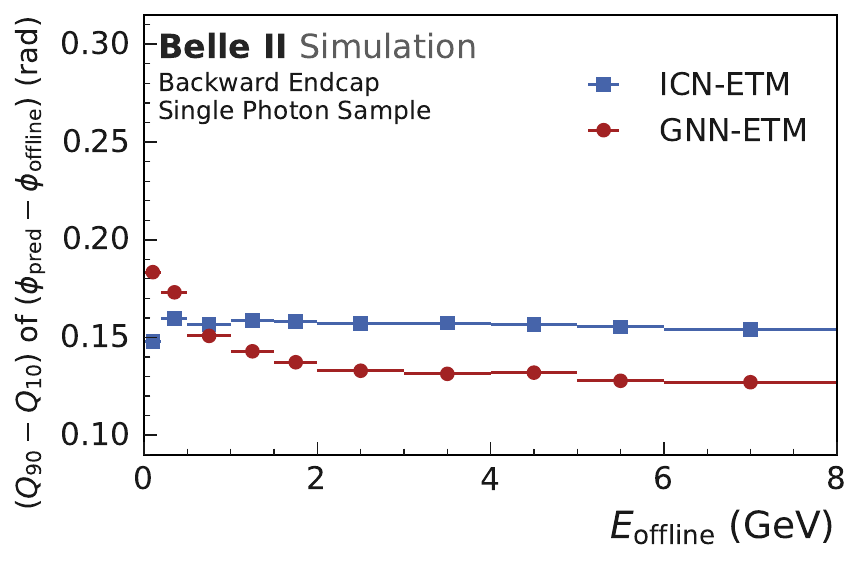}
        \caption{Backward endcap.}
        \label{fig:resolution_phi_bwd}
    \end{subfigure}

    \caption{Corrected relative energy (top row), polar angle $\theta$ (middle row), and azimuthal angle $\phi$ (bottom row) resolutions of the \gnnetm and the \baseline as a function of \offlinecluster energy $E^{\text{offline}}$, shown separately for the forward endcap (left column), barrel (middle column), and backward endcap (right column). 
    Only \triggerclusters found by both algorithms and matched to an \offlinecluster are included.
    Resolutions are evaluated using the \textit{Single Photon Sample} with only one \offlinecluster per event.
    Results are shown without any classifier cut.
    Vertical error bars indicate statistical uncertainties.
    The uncertainties of the two \trg algorithms are correlated since they use the same simulated events.}
    \label{fig:simulated_resolution_energy_theta_phi}
\end{figure*}

The energy resolution as a function of \offlinecluster energy is shown in \cref{fig:simulated_resolution_energy_theta_phi} for the forward endcap (\cref{fig:resolution_energy_fwd}), barrel (\cref{fig:resolution_energy_barrel}), and backward endcap (\cref{fig:resolution_energy_bwd}). 
As expected, the resolution for both algorithms improves approximately as $1/\sqrt{E_{\text{offline}}}$, with the \gnnetm performing slightly worse than the \icnetm across all energies.
However, the \gnnetm requires much smaller energy corrections $f_{\text{corr}}(E_{\text{trg}})$, while the \icnetm systematically underestimates the cluster energy by about 10\% before bias-correction.

The corresponding angular resolutions in $\theta$ and $\phi$ are also shown in \cref{fig:simulated_resolution_energy_theta_phi}.
Here, the \gnnetm is able to improve the angular resolutions for higher energies significantly, due to the added spatial information given by multiple TCs per \triggercluster.
For low-energetic clusters, where the energy deposition is contained within a singular TC, the \gnnetm performs slightly worse than the \icnetm due to the longer tails in the \gnnetm distribution.
The backward endcap shows the worst angular resolution in both $\theta$ (\cref{fig:resolution_theta_bwd}) and $\phi$ (\cref{fig:resolution_phi_bwd}), a result of the higher beam background level present in the backward endcap and the lower granularity of the TCs.
In $\theta$, the average improvement of the \gnnetm position resolution over the \icnetm position resolution for the barrel over all \offlinecluster energies is 18\,\%, while for $\phi$, the improvement is 17\,\%.

\FloatBarrier
\subsubsection{Signal classifier evaluation}
\label{sec:classifier}
The evaluation and threshold determination of the signal classifier are performed on the \textit{Uniform Photon Sample}, using a subset of events that contain exactly two signal and one background \offlinecluster (2S1B events). 
Only isolated \offlineclusters are considered, with a minimum distance to the closest other offline ECL cluster of 45\,cm.
This selection removes the efficiency losses due to overlapping \offlineclusters, which are studied separately in the next section.
The performance of the signal classifier depends mildly on the event topology, in particular on the number and energy of signal and background \offlineclusters.
The subset of 2S1B events is chosen to reduce the dependency on the timing information:
Timing is the most discriminating variable used by the signal classifier to distinguish between signal and background \triggerclusters.
Background \offlineclusters are generally lower energetic than signal \offlineclusters. 
In events with one signal and one background offline ECL cluster~(1S1B events), if the signal cluster is also low energetic, the energies of the signal and background clusters become comparable. 
In this case, the highest energetic TC used to compute the timing may originate from either cluster, leading to an ambiguous timing assignment and degraded classification performance of the \gnnetm signal classifier.
In events with additional \offlineclusters, which often have higher energies, the highest energetic TC is more reliably associated with these clusters.
This results in a more stable timing assignment and improved classifier performance. 
In such topologies, the classifier may also exploit weak correlations between \offlinecluster position and energy.

To prevent the network from learning a trivial energy-based separation between signal and background, we use the \textit{Poisson Uniform Photon Sample} during training.
In this sample, the energy distributions of signal and background \offlineclusters per event are made similar by including an increased number of low energy signal \offlineclusters.
Discrimination based on cluster shape information is not available to the network, as the low granularity of the TCs does not provide sufficient spatial resolution to resolve detailed energy deposition patterns.
The evaluation is performed on 2S1B events, as events with higher \offlinecluster multiplicities are typically triggered regardless of the detailed timing assignment.

\begin{figure}[t]
    \centering
    \includegraphics[width=0.52\linewidth]{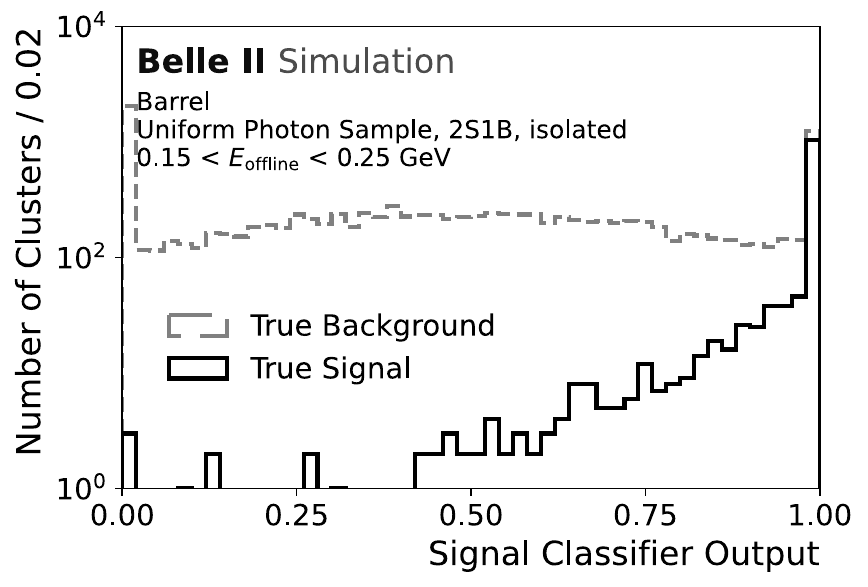}
\caption{Classifier output of the \gnnetm for signal~(black) and background~(gray) clusters in the energy range $0.15 < E_{\text{offline}}< 0.25$\,GeV, evaluated on the \textit{Uniform Photon Sample} with two true signal and one true background \offlinecluster per event (2S1B), and a minimum \offlinecluster distance of 45\,cm.}
    \label{fig:simulated_classifier_output}
\end{figure}

\begin{figure*}[t]
     \centering
     
     \begin{subfigure}[b]{\thirdwidth\textwidth}
         \centering
         \includegraphics[width=\textwidth]{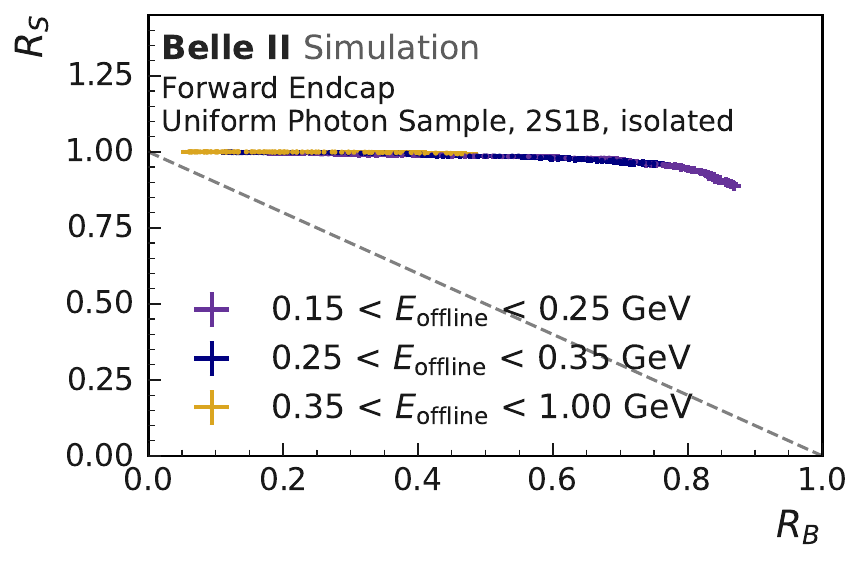}
         \caption{Forward endcap.}
         \label{fig:roc:a}
     \end{subfigure}\hfill
        \begin{subfigure}[b]{\thirdwidth\textwidth}
         \centering
         \includegraphics[width=\textwidth]{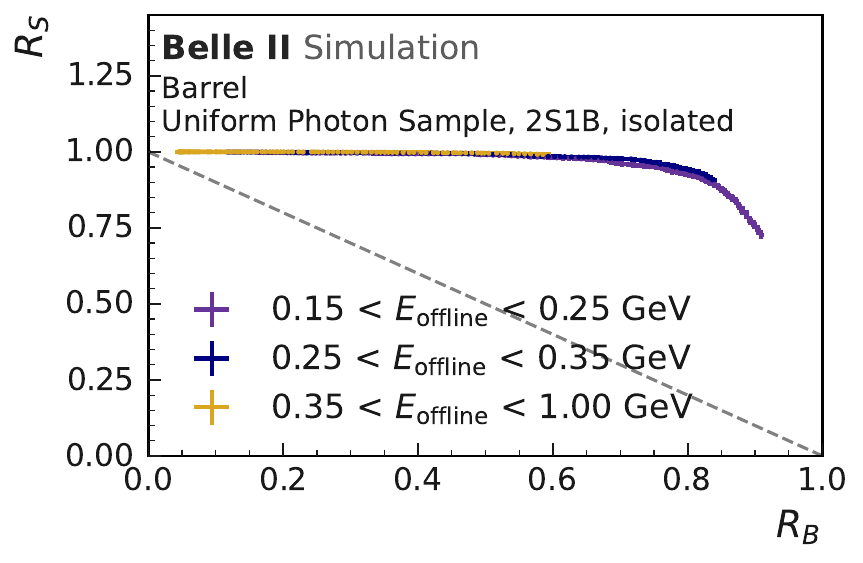}
         \caption{Barrel.}
         \label{fig:roc:b}
     \end{subfigure}\hfill
        \begin{subfigure}[b]{\thirdwidth\textwidth}
         \centering
         \includegraphics[width=\textwidth]{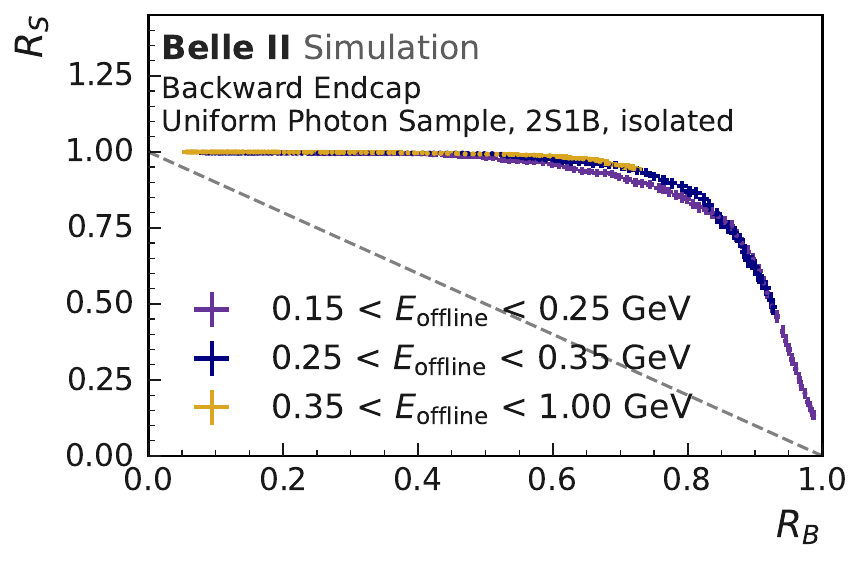}
         \caption{Backward endcap.}
         \label{fig:roc:c}
     \end{subfigure}
\caption{Signal retention $R_{S}$ versus background rejection $R_{B}$ of the \gnnetm~classifier, shown separately for the (\subref{fig:roc:a}) forward endcap, (\subref{fig:roc:b}) barrel, and (\subref{fig:roc:c}) backward endcap, evaluated on the \textit{Uniform Photon Sample} with two signal and one background \offlinecluster~(2S1B) per event, and a minimum \offlinecluster distance of 45\,cm. 
Each panel contains three \offlinecluster energy $E_{\text{offline}}$ ranges: \hbox{0.15--0.25\,GeV}~(violet), \hbox{0.25--0.35\,GeV}~(blue), and \hbox{0.35--1.0\,GeV}~(yellow). 
The dashed grey line indicates the ROC curve corresponding to a random classifier.
Vertical error bars indicate the statistical uncertainty on $R_{S}$, horizontal error bars indicate the statistical uncertainty on $R_{B}$.
For \offlinecluster energies between 0.35 and 1.0~GeV, the number of background clusters is too small to probe the highest background-rejection values, and the ROC curves in this region therefore terminate at lower values of $R_B$.
}
\label{fig:simulated_performance_roc}
\end{figure*}

The classifier output of the \gnnetm for signal and background \offlineclusters for the \textit{Uniform Photon Sample} on 2S1B events containing only isolated \offlineclusters is shown in \cref{fig:simulated_classifier_output}.
Figure~\ref{fig:simulated_performance_roc} shows the signal retention $R_S$ and background rejection $R_B$ for different classifier thresholds.

For \triggerclusters matched to \offlineclusters with energies between 0.25 and 0.35\,GeV in the barrel region, the \gnnetm can reject up to 72\% of all background while retaining 97.5\% of the signal.
For the lower-energy region between 0.15 and 0.25\,GeV, for the same signal efficiency of 97.5\%, the background rejection still reaches up to 66\% in the barrel region. 
In the backward endcap region, the \gnnetm can reject 52\% of background clusters for energies between 0.15 and 0.25\,GeV and 61\% for energies between 0.25 and 0.35\,GeV, while keeping a signal efficiency of 97.5\% for each energy region.
The classifier performance for low-energy \triggerclusters is of particular importance, as the majority of background clusters are found at low energies.
For \offlineclusters with energies between 0.35 and 1\,GeV, the signal efficiency is close to 100\% but the low number of background \offlineclusters within that region does not allow to validate the highest background rejections.

The cutoff of values at the right side of the ROC curves, especially visible in the forward endcap and the barrel region, is a result of the linear sigmoid approximation of the activation function of the signal output.
Values close to 0 (1) given by the standard sigmoid function are set to exactly 0 (1) by the linear approximation, pushing the distribution to the minimum (maximum) values and disallowing further differentiation between different clusters.

\begin{figure*}[ht]
     \centering
\begin{subfigure}[b]{\thirdwidth\textwidth}
         \centering
         \includegraphics[width=\textwidth]{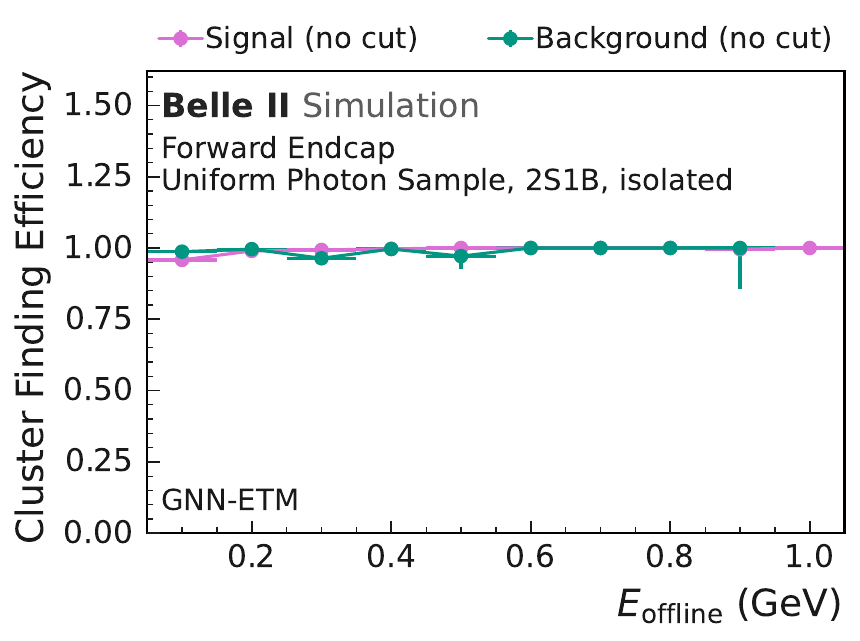}
         \caption{Forward endcap.}
         \label{fig:nocut_signal_background_eff:a}
     \end{subfigure}\hfill
        \begin{subfigure}[b]{\thirdwidth\textwidth}
         \centering
         \includegraphics[width=\textwidth]{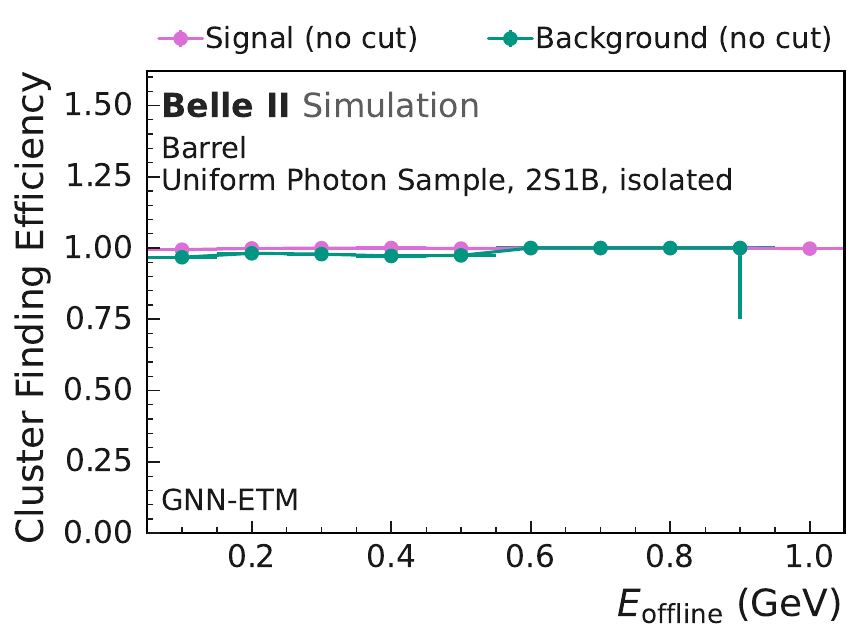}
         \caption{Barrel.}
         \label{fig:nocut_signal_background_eff:b}
     \end{subfigure}\hfill
        \begin{subfigure}[b]{\thirdwidth\textwidth}
         \centering
         \includegraphics[width=\textwidth]{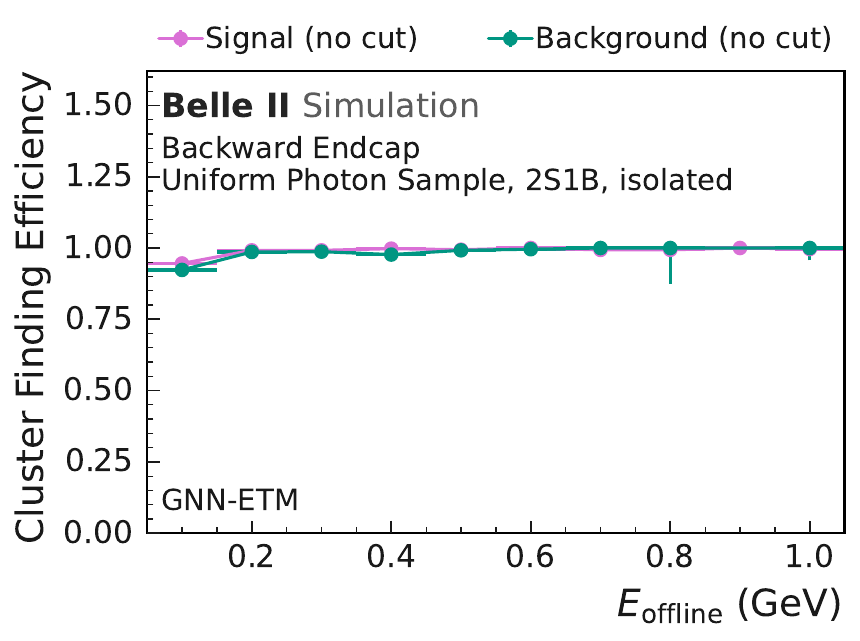}
         \caption{Backward endcap.}
         \label{fig:nocut_signal_background_eff:c}
     \end{subfigure}

     \vspace{1em}
     \begin{subfigure}[b]{\thirdwidth\textwidth}
         \centering
         \includegraphics[width=\textwidth]{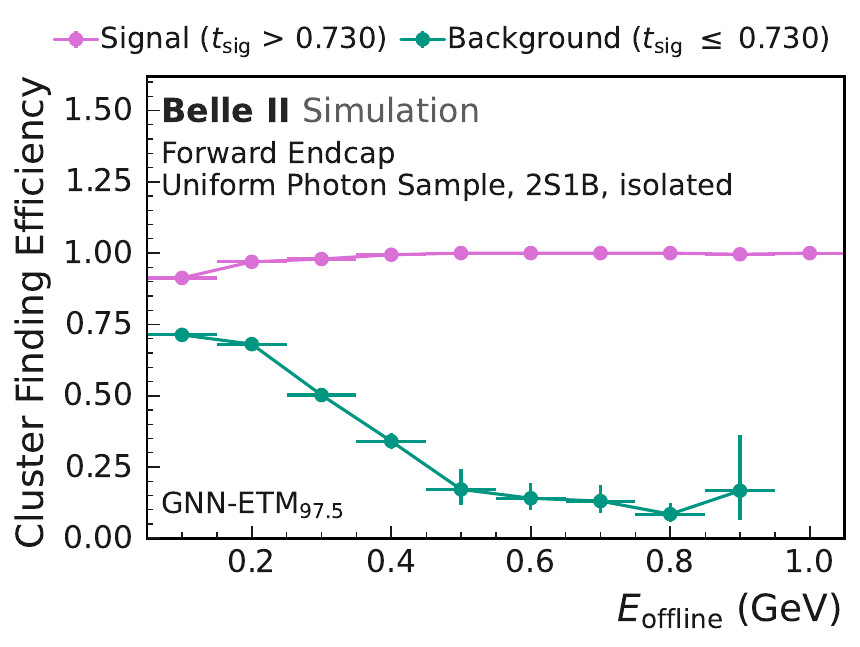}
         \caption{Forward endcap.}
         \label{fig:signal_background_eff:a}
     \end{subfigure}\hfill
        \begin{subfigure}[b]{\thirdwidth\textwidth}
         \centering
         \includegraphics[width=\textwidth]{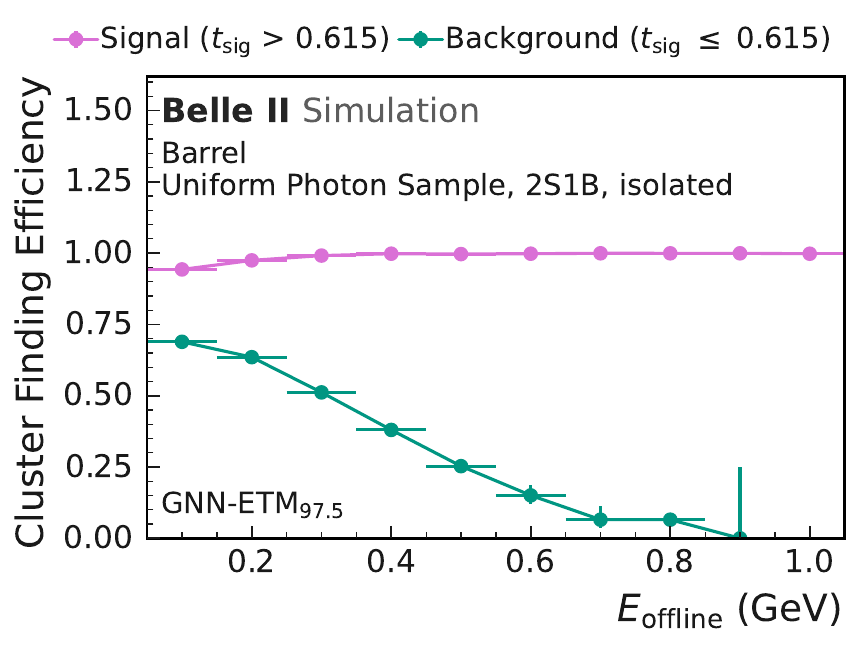}
         \caption{Barrel.}
         \label{fig:signal_background_eff:b}
     \end{subfigure}\hfill
        \begin{subfigure}[b]{\thirdwidth\textwidth}
         \centering
         \includegraphics[width=\textwidth]{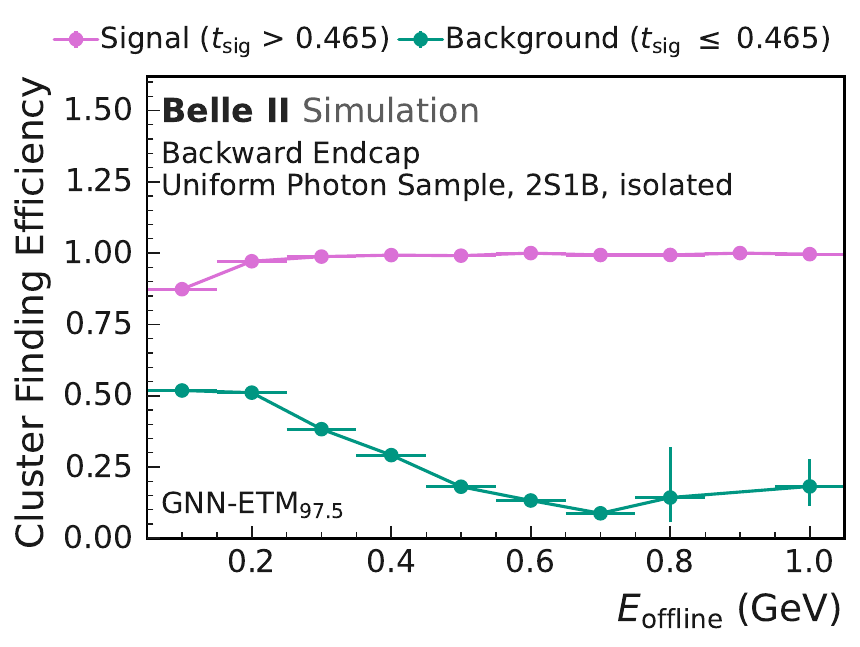}
         \caption{Backward endcap.}
         \label{fig:signal_background_eff:c}
     \end{subfigure}
\caption{Cluster finding efficiency as a function of offline cluster energy $E_\mathrm{offline}$ for true signal (violet) and true background (green) clusters, evaluated for the \gnnetm and the \cgnnetm using the \textit{Uniform Photon Sample}. 
Events contain two signal and one background offline cluster~(2S1B) and require a minimum cluster separation of 45\,cm.
The upper row shows the \gnnetm cluster finding efficiency without applying a signal classifier selection. 
The lower row shows the \cgnnetm efficiency when a signal (background) cluster is defined by a matched \gnnetm cluster with classifier output above (below) the threshold $t_\mathrm{sig}$, which is tuned to achieve a signal efficiency of 97.5\% in the energy range 0.15–0.25\,GeV~(see text for details). 
Vertical error bars indicate statistical uncertainties and are in most cases smaller than the marker size, while horizontal error bars show the bin width.}
\label{fig:simulated_signal_background_efficiency}
\end{figure*}

In \cref{fig:simulated_signal_background_efficiency}, the impact of the signal classifier on the cluster finding efficiency is evaluated using a fixed classifier threshold as implemented in hardware.
For each detector region, the signal classifier threshold $t_\mathrm{sig}$ is defined such that a signal efficiency of 97.5\% is achieved for \offlineclusters with 0.15 \textless{} $E_\mathrm{offline}$ \textless{} 0.25\,GeV.
The resulting \gnnetm configuration with this region-dependent classifier cut is referred to as the \cgnnetm.

A \trg cluster is classified as signal if the classifier output is greater than $t_\mathrm{sig}$. 
A \trg cluster is considered successfully found only if it is correctly matched to a true signal or background \offlinecluster.
Achieving 97.5\% overall signal efficiency for the energy range of 0.15 \textless{} $E_\mathrm{offline}$ \textless{} 0.25\,GeV results in a background rejection rate of 50\% to 70\%, depending on the detector region.
The background rejection is highest in the forward endcap and lowest in the backward endcap.
This behavior is consistent with increasing number of beam background energy depositions from forward to backward endcap, leading to a more difficult classification of true signal \offlineclusters with higher percentages of beam background energy depositions. 
Since $t_\mathrm{sig}$ is defined per region, this results in region dependent background rejection rates.
The background rejection rate decreases with increasing \offlinecluster energy.
This is partly due to the exponential energy distribution of background \offlineclusters, which strongly suppresses the number of high energy background examples available during training. 
As a consequence, high energetic \gnnclusters are more likely to be classified as signal.
In addition, the TC timing is defined relative to the highest energetic TC in the event.
High energetic background \triggerclusters are therefore more likely to define the event timing and obtain a timing of zero. 
Since the classifier relies strongly on timing information, such \triggerclusters are preferentially classified as signal, further reducing the background rejection at high energies.

\FloatBarrier
\subsubsection{Non-isolated photon signatures}
\begin{figure*}[ht!]
    \begin{subfigure}[b]{\thirdwidth\textwidth}
        \centering
        \includegraphics[width=\textwidth]{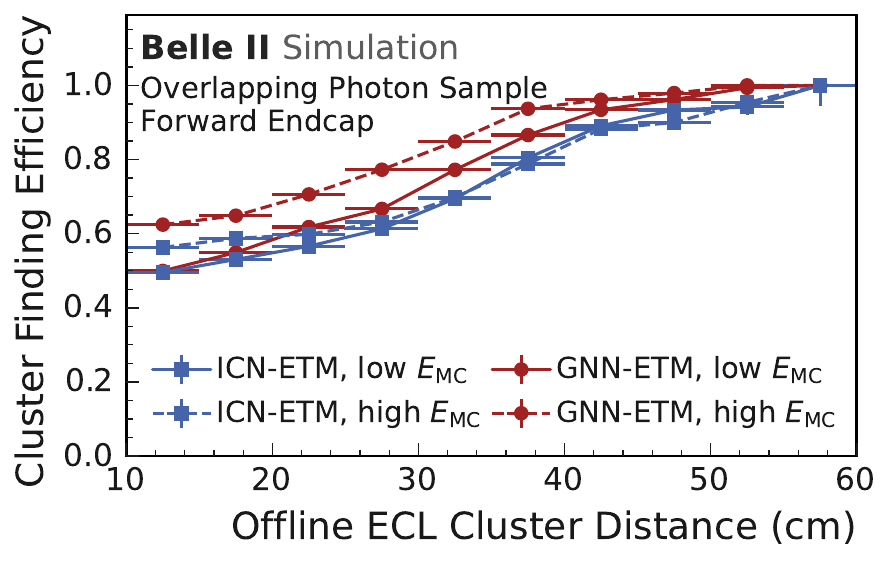}
        \caption{Forward endcap.}
        \label{fig:opening_angle_fwd}
    \end{subfigure}
    \begin{subfigure}[b]{\thirdwidth\textwidth}
        \centering
        \includegraphics[width=\textwidth]{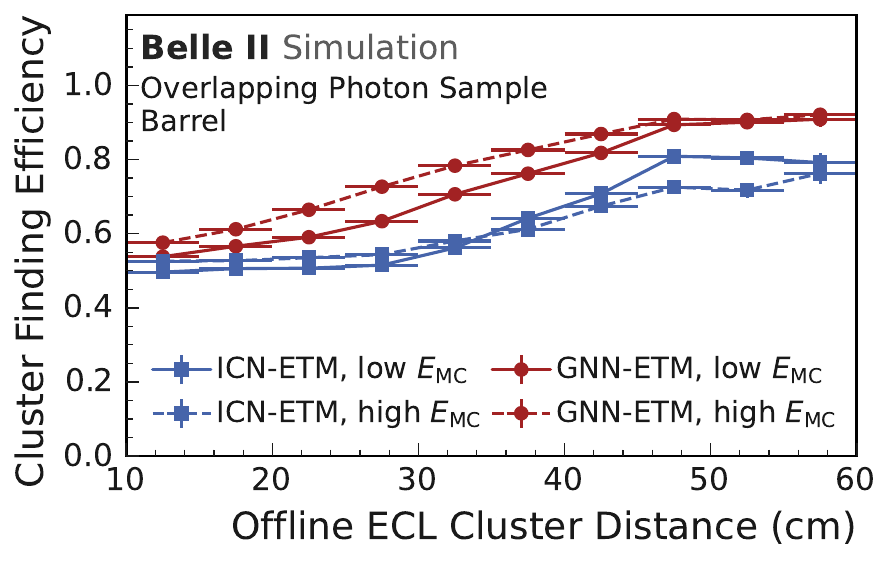}
        \caption{Barrel.}
        \label{fig:opening_angle_barrel}
    \end{subfigure}\hfill
    \begin{subfigure}[b]{\thirdwidth\textwidth}
        \centering
        \includegraphics[width=\textwidth]{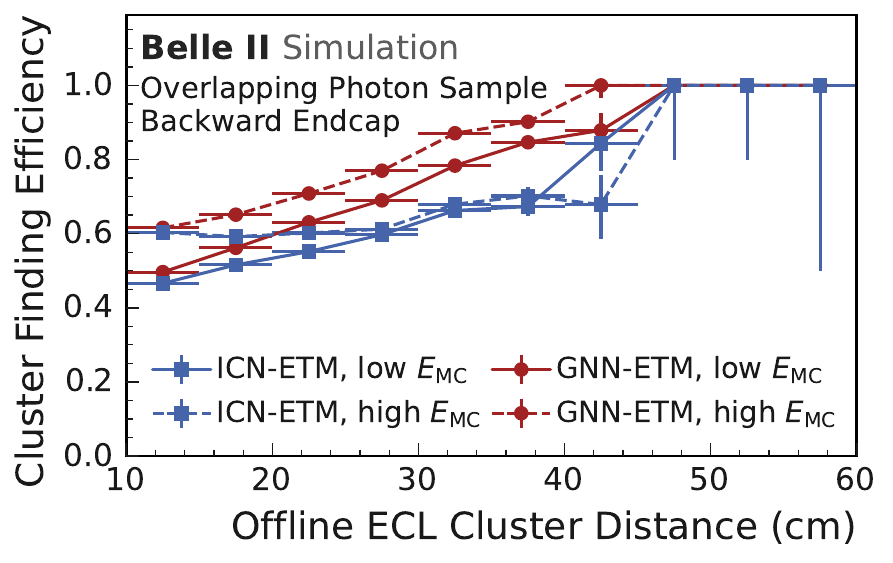}
        \caption{Backward endcap.}
        \label{fig:opening_angle_bwd}
    \end{subfigure}
    \caption{Cluster finding efficiency as a function of the Cartesian distance between the reconstructed position of the two offline ECL clusters, evaluated using the \emph{Overlap Diphoton Sample} shown separately for the (\subref{fig:opening_angle_fwd}) forward endcap, (\subref{fig:opening_angle_barrel}) barrel, and (\subref{fig:opening_angle_bwd}) backward endcap. 
    The cluster finding efficiency is evaluated for simulated photon energies $1 < E_\mathrm{MC}<2\,\text{GeV}$, denoted low $E_\mathrm{MC}$, and $4 < E_\mathrm{MC}<5\,\text{GeV}$, denoted high $E_\mathrm{MC}$.
    Vertical error bars that indicate statistical uncertainties are smaller than the marker size.
    Distances above about 50\,cm in the endcaps are only possible in the azimuthal angle $\phi$ direction due to the small polar angle $\theta$ coverage of the endcaps.
    The uncertainties of the two \trg algorithms are correlated since they use the same simulated events.}
    \label{fig:simulated_performance_opening_angle}
\end{figure*}

To evaluate the cluster finding efficiency for non-isolated photon signatures, we use the \emph{Overlap Diphoton Sample} described in Sec.~\ref{sec:gnnetm_software_training}. 
To evaluate the performance on exactly two offline reconstructed clusters, we select events where two offline reconstructed clusters are each assigned to at least one TC as a label to ensure a possibility of distinguishing between both offline reconstructed clusters on \trg level.
Additionally, we require that both clusters originate from different particles.
The resulting efficiency as a function of the Euclidean distance between the reconstructed positions of the two offline ECL clusters is shown in \cref{fig:simulated_performance_opening_angle}.
The cluster finding efficiency is shown for two different simulated photon energy ranges $E_\mathrm{MC}$, a low energy range 1 \textless{} $E_\mathrm{MC}$ \textless{} 2\,GeV, and a high energy range  4 \textless{} $E_\mathrm{MC}$ \textless{} 5\, GeV to illustrate the effect of the offline ECL cluster size on the cluster separation capabilities of \icnetm and \gnnetm.
The \gnnetm shows a higher cluster finding efficiency than the \icnetm for all three detector regions and all opening angles. 
The efficiency for the \gnnetm is higher for higher offline cluster energies, as higher energies result in more information available for the algorithm on \trg level, as more TCs have an energy deposit above 100\,MeV.
The \gnnetm can in principle use this additional cluster shape information and return two \trg clusters.
In comparison, the efficiency of the \icnetm decreases more strongly with increasing energy. 
The higher efficiency observed in the endcaps for large cluster separations originates from the geometry of the TCs. 
In the barrel, each TC covers four crystals in both directions, so two offline clusters separated by about 40 cm typically lie in adjacent TCs. 
In the endcaps, however, TCs are only two to three crystals wide in $\phi$, meaning that offline clusters separated by the same distance are usually divided by one or two empty TCs. 
As a result, offline clusters with a separation of around 40\,cm are more likely to be reconstructed as distinct \trg clusters, leading to a higher cluster-finding efficiency in the endcaps than in the barrel.
An increase in cluster-finding efficiency of up to 20~percentage points is observed for the \gnnetm, with the improvement growing at higher distances, highlighting the advantage of the algorithm over the \icnetm design.

\FloatBarrier
\subsection{Data studies}
\label{sec:results_data}

The clustering performance is evaluated using simulated signal samples of the \mumubkg{} and \eebkg{} processes, consisting of 1 million and 10 million events, respectively. 
For the evaluation on data, \textit{Run A} is used~(see \cref{tab:run_descriptions}).

\subsubsection{Performance on \texorpdfstring{\mumubkg{}}{e+e- -> mu+mu-} events}

\begin{figure*}[ht!]
     \centering
     \begin{subfigure}[b]{\thirdwidth\textwidth}
         \centering
         \includegraphics[width=\textwidth]{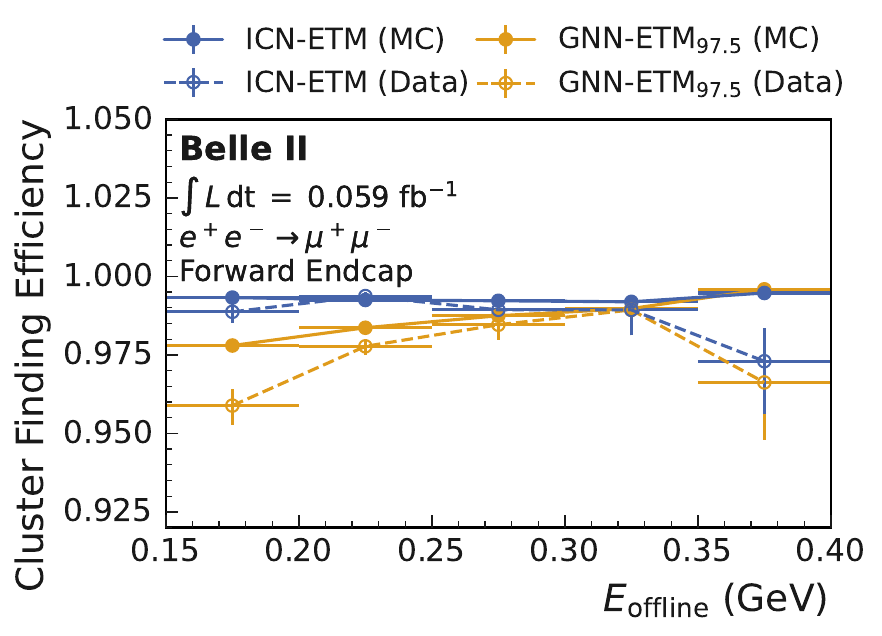}
         \caption{Forward endcap.}
         \label{fig:mumu_performance_eff:a}
     \end{subfigure}\hfill
        \begin{subfigure}[b]{\thirdwidth\textwidth}
         \centering
         \includegraphics[width=\textwidth]{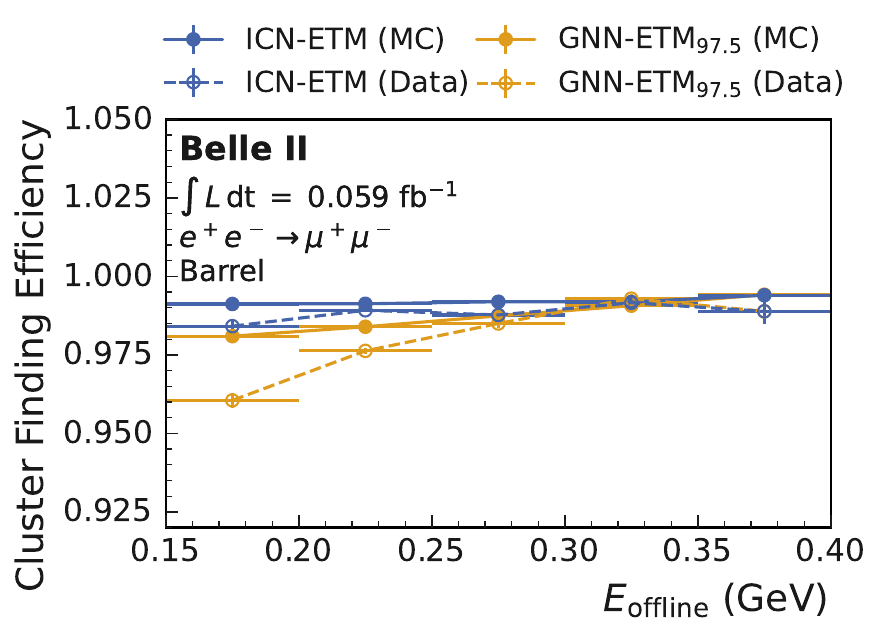}
         \caption{Barrel.}
         \label{fig:mumu_performance_eff:b}
     \end{subfigure}\hfill
        \begin{subfigure}[b]{\thirdwidth\textwidth}
         \centering
         \includegraphics[width=\textwidth]{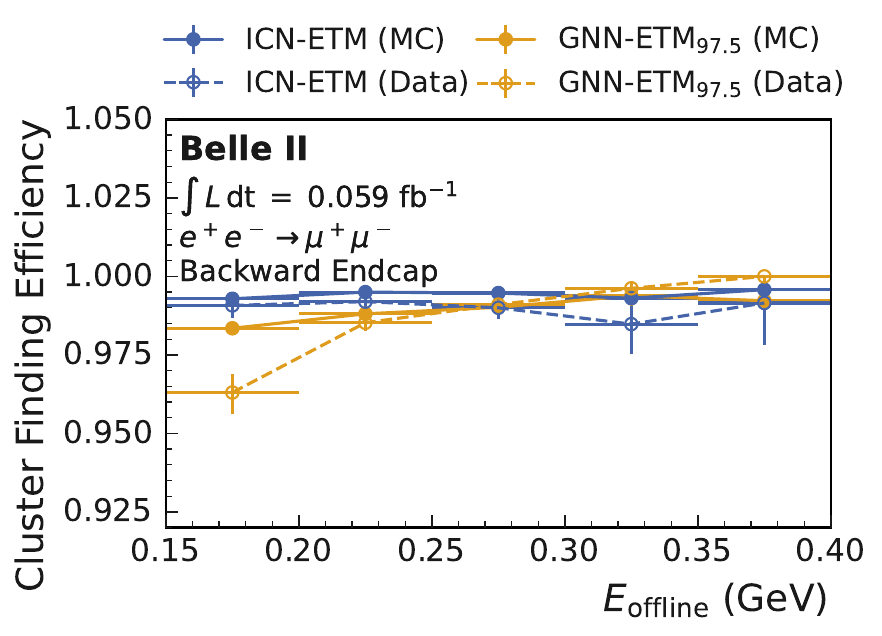}
         \caption{Backward endcap.}
         \label{fig:mumu_performance_eff:c}
     \end{subfigure}
\caption{Cluster finding efficiency for \mumubkg{} events as function of \offlinecluster energy $E_{\text{offline}}$ for the \baseline~(blue) and the \cgnnetm~(orange) in the (\subref{fig:mumu_performance_eff:a}) forward endcap, (\subref{fig:mumu_performance_eff:b}) barrel, and (\subref{fig:mumu_performance_eff:c}) backward endcap for simulated events (filled markers) and for \textit{Run A} (unfilled markers). 
Markers are connected by solid or dashed lines to guide the eye.
The vertical error bars that show the statistical uncertainty are usually smaller than the marker size.
The horizontal error bars indicate the bin width.
The uncertainties of the two \trg algorithms are correlated since they use the same events.
The larger efficiency loss observed in data for \cgnnetm compared to simulation is attributed to higher beam-induced backgrounds, see text for a detailed discussion.}
\label{fig:mumu_performance_eff}
\end{figure*}

We use \mumubkg{} events to study low-energy \triggerclusters produced by muons, which deposit on average around 200~MeV in the calorimeter.
The event selection is the same as for the Belle~II luminosity measurement~\cite{Belle-II:2024vuc} except that the polar angle acceptance selection on the particle candidates is removed to include \offlineclusters in the endcaps.
All \offlineclusters matched to muon tracks are selected, and the efficiency and resolution of both \cgnnetm and \icnetm on these \offlineclusters are reported.
For the \cgnnetm, the same signal classifier selection as shown in \cref{fig:simulated_signal_background_efficiency}, determined on the \textit{Uniform Photon Sample} for the three regions separately, is applied. 
In \cref{fig:mumu_performance_eff}, the cluster finding efficiency for the muon \offlineclusters is shown for all three detector regions and for both data and simulated events. 
Both algorithms have a cluster finding efficiency of close to 100\% in the forward endcap and barrel region. 
The \icnetm shows small inefficiencies due to an overestimation of the cluster energy in the presence of beam background.
The \cgnnetm shows approximately a 2\% efficiency loss for energies below 0.3~GeV, which is a result of the chosen signal classifier threshold. 
In the majority of simulated \mumubkg{} events passing the selection, only the two signal muon \offlineclusters are present, with no additional background \offlineclusters.
This corresponds to a 2S0B event topology. 
For a fixed classifier threshold defined using the \textit{Uniform Photon Sample}, these events have a slightly higher signal efficiency than 2S1B events, resulting in a cluster finding efficiency above 97.5\% even for energies below 0.25~GeV.
In contrast, during \textit{Run A}, the background level was generally high (see \cref{tab:run_descriptions}), resulting in a larger number of \offlineclusters per event. 
In simulation, events typically contain only the two signal \offlineclusters, whereas the increased cluster multiplicity observed in data indicates additional background activity. 
Based on the simulated performance under increased cluster multiplicity, this is associated with a reduced signal efficiency of the classifier, consistent with the behavior observed in data.

\begin{figure*}[ht!]
     \centering
     \begin{subfigure}[b]{\thirdwidth\textwidth}
         \centering
         \includegraphics[width=\textwidth]{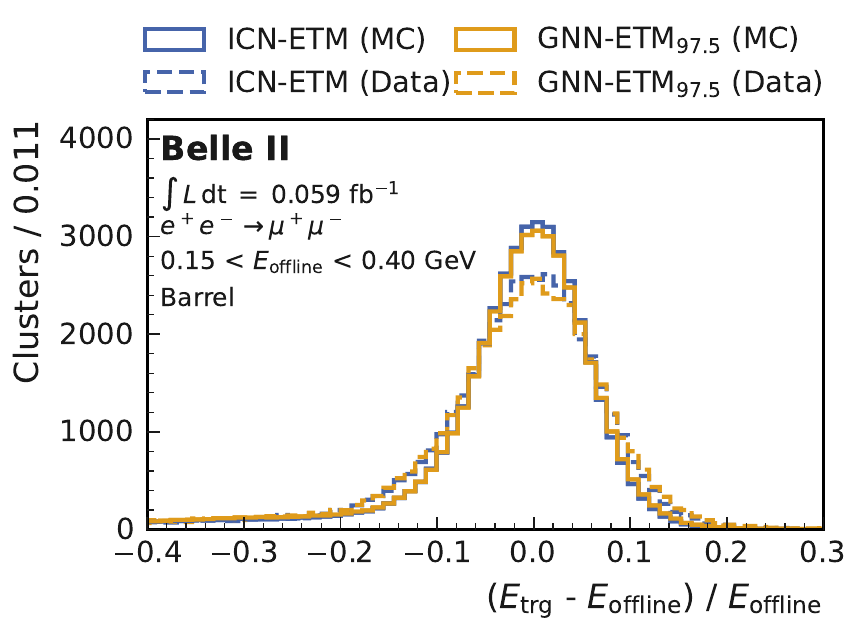}
         \caption{Energy resolution.}
         \label{fig:mumu_resolution_energy}
     \end{subfigure}\hfill
        \begin{subfigure}[b]{\thirdwidth\textwidth}
         \centering
         \includegraphics[width=\textwidth]{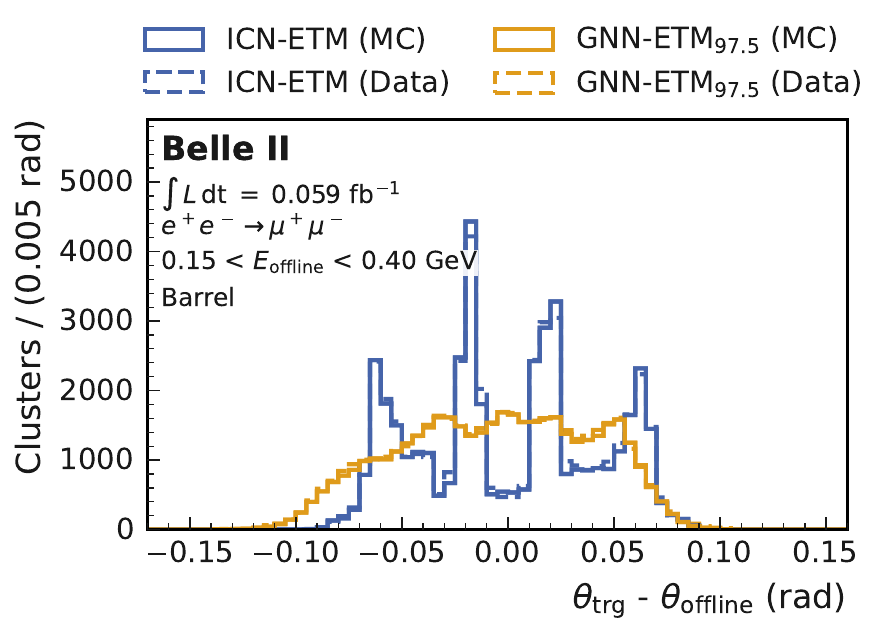}
         \caption{$\theta$ resolution.}
         \label{fig:mumu_resolution_theta}
     \end{subfigure}\hfill
        \begin{subfigure}[b]{\thirdwidth\textwidth}
         \centering
         \includegraphics[width=\textwidth]{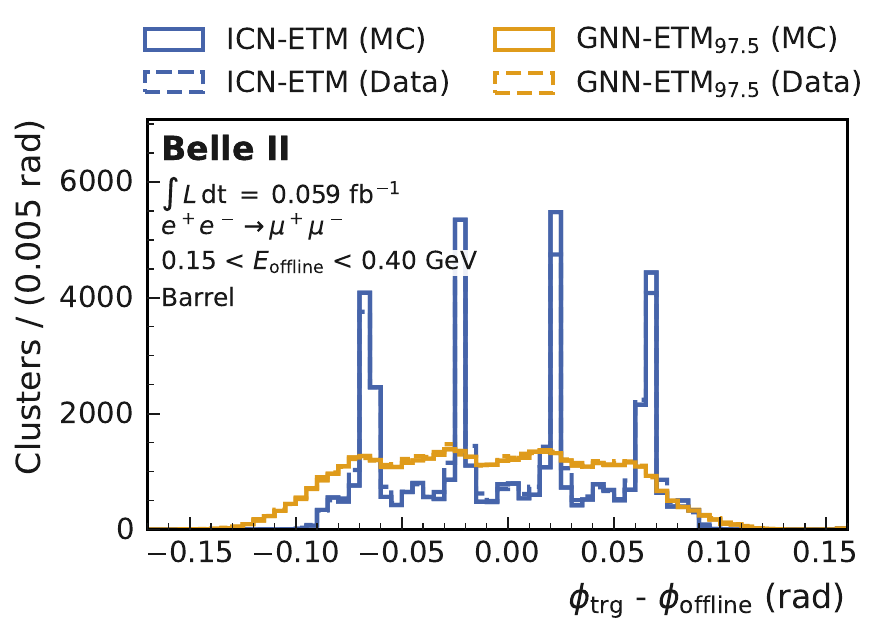}
         \caption{$\phi$ resolution.}
         \label{fig:mumu_resolution_phi}
     \end{subfigure}
\caption{Comparison of (\subref{fig:mumu_resolution_energy}) energy, (\subref{fig:mumu_resolution_theta}) polar angle $\theta$, and (\subref{fig:mumu_resolution_phi}) azimuthal angle $\phi$ resolutions for \mumubkg{} events for the \baseline~(blue) and the \cgnnetm~(orange) in the barrel for simulated events (filled lines) and for \textit{Run A} (dashed lines). Only \triggerclusters reconstructed by both algorithms and matched to an \offlinecluster selected as a muon candidate are considered. The simulated events are scaled to the luminosity of \textit{Run A}. 
}
\label{fig:mumu_resolution}
\end{figure*}

Omitting this signal classifier cut raises the cluster finding efficiency for the \gnnetm to close to 1, matching the cluster finding efficiency of \icnetm in the forward endcap and slightly surpassing it in the barrel and the backward endcap.
In \cref{fig:mumu_resolution}, the energy and angular resolution of both \cgnnetm and \icnetm are shown. 
The simulated events are scaled to the data luminosity.
Due to the rather small sample size, the resolution is not shown separately for the different \offlinecluster energies. 
The widths of all three resolution distributions are comparable for both algorithms. 
For the energy resolution, the corresponding FWHM($\eta$)/2.355 for each dataset and \trg algorithm are reported in \cref{tab:fwhm_ee_mumu}, with the \cgnnetm having a slightly worse resolution width in comparison to the \icnetm.
The \icnetm exhibits characteristic spikes in the angular distributions shown in \cref{fig:mumu_resolution_theta} and \cref{fig:mumu_resolution_phi}, which originate from the coarse TC binning.
The \cgnnetm has, in general, slightly wider angular resolution.
The angular resolution agrees for both data and simulated events for both algorithms.
The energy distribution (see \cref{fig:mumu_resolution_energy}) displays long left tails for both algorithms.
These stem from muons that have traveled through multiple crystals, where the crystals belong to separate TCs with only one TC surpassing the 100\,MeV energy threshold.
The corresponding \offlinecluster contains the entire deposited energy, while the \trg algorithms only see part of the deposited energy and therefore underestimate the full cluster energy.
The energy resolution for \textit{Run A} is slightly wider for both \icnetm and \cgnnetm than for simulated events, most likely originating from the increased beam background presence in these events.

\begin{table*}[t]
\caption{FWHM($\eta$)/2.355 calculated of the fit to the energy resolution distribution $\eta$ for \eebkg{} and \mumubkg{} events for the \icnetm and \cgnnetm for both the \textit{Run A} dataset, denoted as Data, and the simulated events, denoted as MC, for each process.}  
\centering
\renewcommand{\arraystretch}{1.1}
\begin{tabular}{ccccc}
\hline
Process & \cgnnetm (MC) & \icnetm (MC) & \cgnnetm (Data) & \icnetm (Data) \\ \hline
\eebkg{} & 0.016 $\pm$ 0.000 & 0.013 $\pm$ 0.000 & 0.022 $\pm$ 0.000 & 0.020 $\pm$ 0.000 \\
\mumubkg{} & 0.055 $\pm$ 0.000 & 0.053 $\pm$ 0.000 & 0.067 $\pm$ 0.001 & 0.064 $\pm$ 0.001 \\ \hline
\end{tabular}
\label{tab:fwhm_ee_mumu}
\end{table*}

\subsubsection{Performance on \texorpdfstring{\eebkg{}}{e+e- -> e+e-} events}

\begin{figure*}[ht!]
     \centering
     \begin{subfigure}[b]{\thirdwidth\textwidth}
         \centering
         \includegraphics[width=\textwidth]{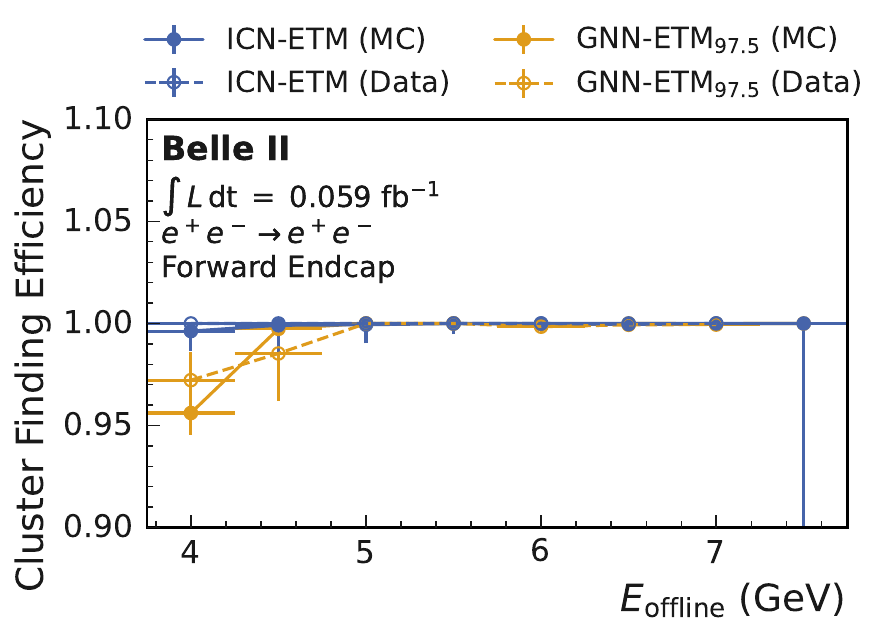}
         \caption{Forward endcap.}
         \label{fig:bhabha_performance_eff:a}
     \end{subfigure}\hfill
        \begin{subfigure}[b]{\thirdwidth\textwidth}
         \centering
         \includegraphics[width=\textwidth]{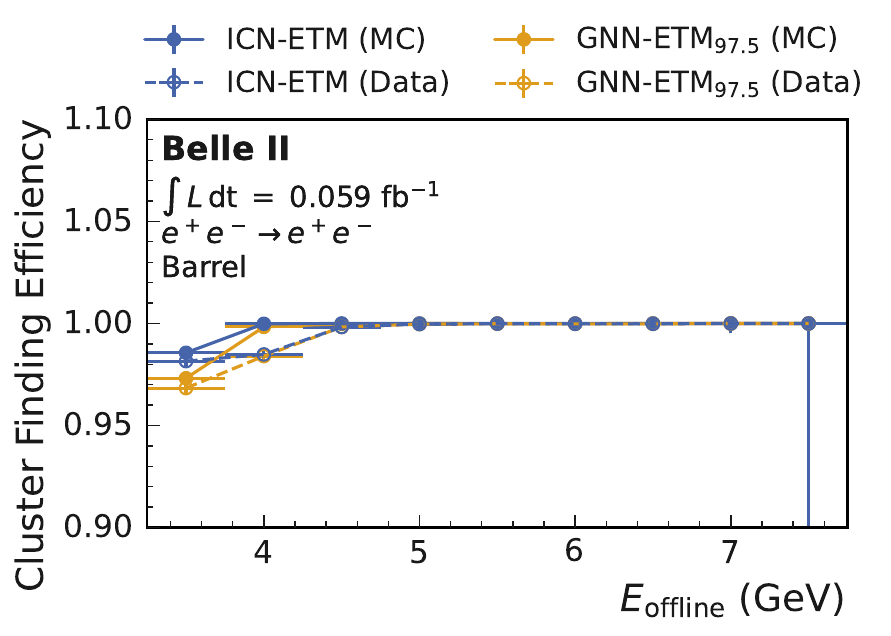}
         \caption{Barrel.}
         \label{fig:bhabha_performance_eff:b}
     \end{subfigure}\hfill
        \begin{subfigure}[b]{\thirdwidth\textwidth}
         \centering
         \includegraphics[width=\textwidth]{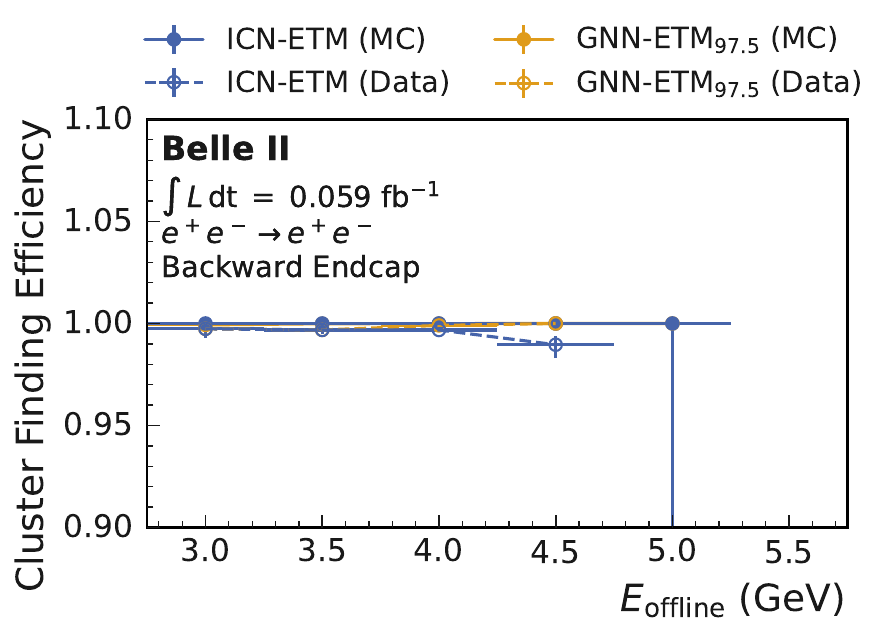}
         \caption{Backward endcap.}
         \label{fig:bhabha_performance_eff:c}
     \end{subfigure}
\caption{Cluster finding efficiency for \eebkg events as function of \offlinecluster energy $E_{\text{offline}}$ for the \baseline~(blue) and the \cgnnetm~(orange) in the (\subref{fig:bhabha_performance_eff:a}) forward endcap, (\subref{fig:bhabha_performance_eff:b}) barrel, and (\subref{fig:bhabha_performance_eff:c}) backward endcap for simulated events (filled markers) and for \textit{Run A} (unfilled markers). 
Markers are connected by solid lines to guide the eye.
The vertical error bars that show the statistical uncertainty are usually smaller than the marker size.
The horizontal error bars indicate the bin width.
The uncertainties of the two \trg algorithms are correlated since they use the same simulated events.
The rightmost bin for the backward endcap does not contain any entries.}
\label{fig:bhabha_performance_eff}
\end{figure*}

\begin{figure*}[ht!]
     \centering
     \begin{subfigure}[b]{\thirdwidth\textwidth}
         \centering
         \includegraphics[width=\textwidth]{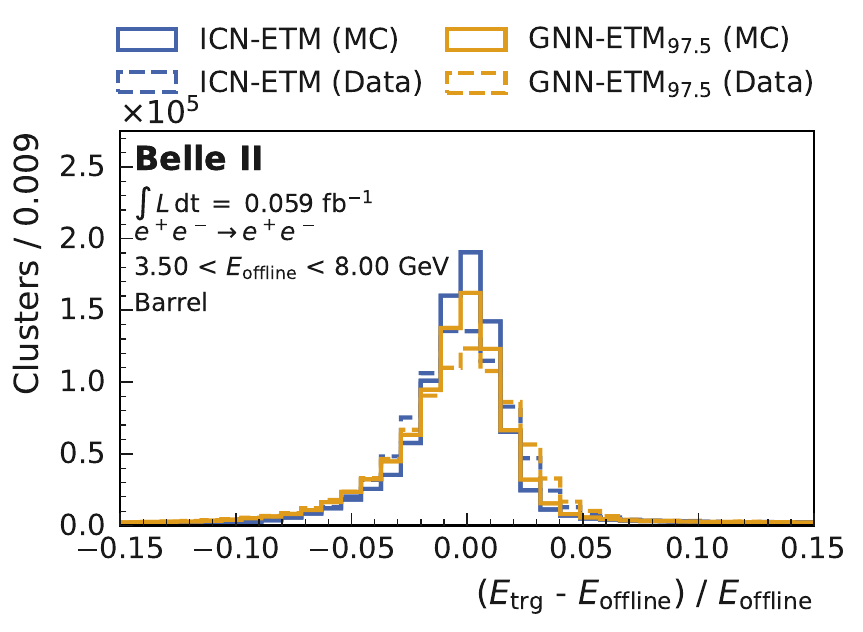}
         \caption{Energy resolution.}
         \label{fig:bhabha_resolution_energy}
     \end{subfigure}\hfill
        \begin{subfigure}[b]{\thirdwidth\textwidth}
         \centering
         \includegraphics[width=\textwidth]{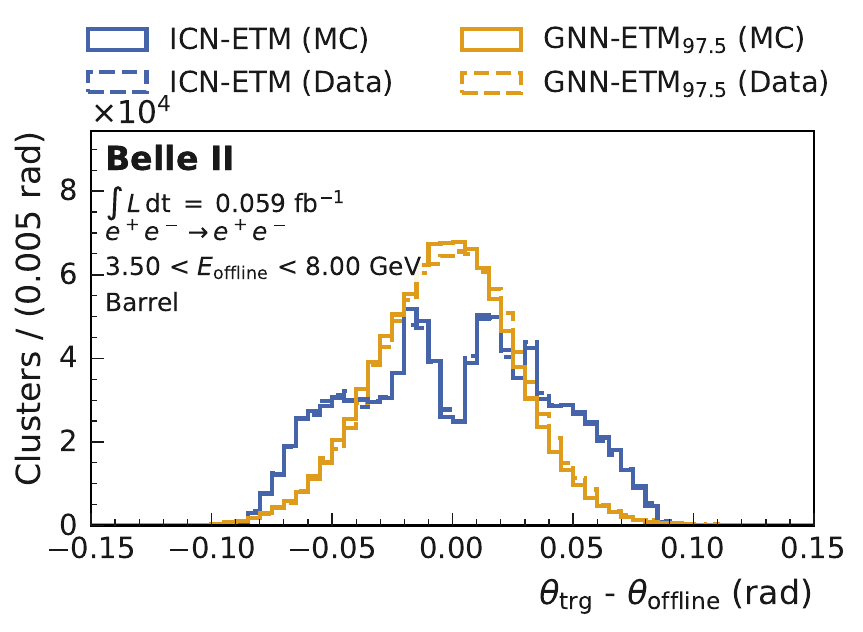}
         \caption{$\theta$ resolution.}
         \label{fig:bhabha_resolution_theta}
     \end{subfigure}\hfill
        \begin{subfigure}[b]{\thirdwidth\textwidth}
         \centering
         \includegraphics[width=\textwidth]{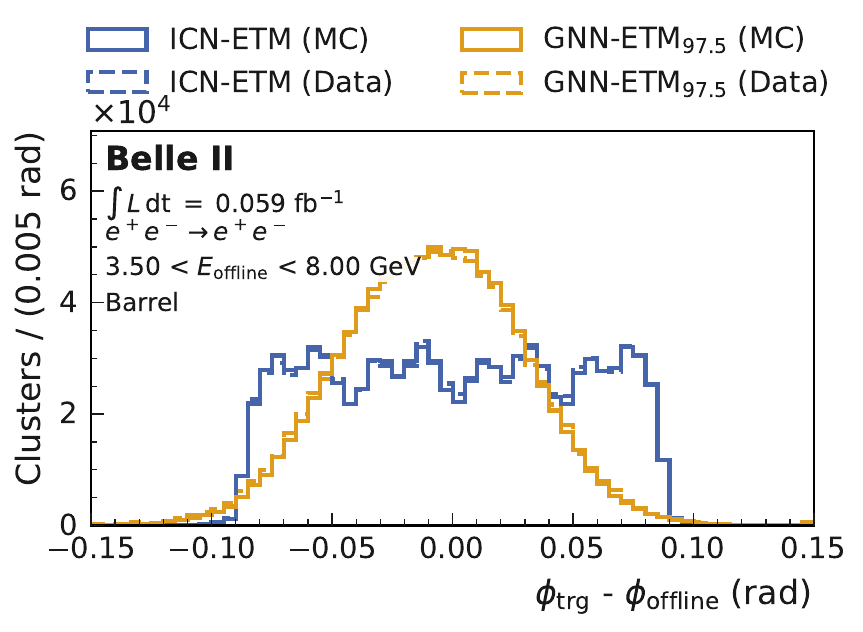}
         \caption{$\phi$ resolution.}
         \label{fig:bhabha_resolution_phi}
     \end{subfigure}
\caption{Comparison of (\subref{fig:bhabha_resolution_energy}) energy, (\subref{fig:bhabha_resolution_theta}) polar angle $\theta$, and (\subref{fig:bhabha_resolution_phi}) azimuthal angle $\phi$ resolutions for \eebkg{} for the \baseline~(blue) and the \cgnnetm~(orange) in the barrel for simulated events (filled lines) and for \textit{Run A} (dashed lines). Only \triggerclusters reconstructed by both algorithms and matched to an \offlinecluster selected as an electron/positron candidate are considered. The simulated events are scaled to the luminosity of \textit{Run A}.
}
\label{fig:bhabha_resolution}
\end{figure*}

High-energy \triggerclusters are studied using \eebkg{} events, with the selection described in~\cite{Belle-II:2024vuc}. 
The \triggerclusters are then analyzed following the same procedure as for low-energy \triggerclusters.
Only \offlineclusters with a matched electron or positron track are evaluated.
For the \cgnnetm, the same signal classifier selection as shown in \cref{fig:simulated_signal_background_efficiency}, determined on the \textit{Uniform Photon Sample} for the three regions separately, is applied. 
In \cref{fig:bhabha_performance_eff}, the cluster finding efficiency for electron/positron \offlineclusters is shown. 
The \cgnnetm and \icnetm efficiencies agree within uncertainties and reach an efficiency of 1 for higher energies. 
The signal classifier cut has no effect on the cluster finding efficiency in this energy region, since all \gnnclusters pass the signal classifier threshold.
In the majority of events, due to the event kinematics of \eebkg{} processes, the electron \offlinecluster is located in the forward region, while the positron \offlinecluster is located in the backward region.
Most electrons have energies above 4\,GeV, while most positrons have energies above 2.5\,GeV. 
Therefore, \offlineclusters with energies below these values indicate events that are not cleanly selected \eebkg{} events and are likely affected by additional processes such as bremsstrahlung. 
In such cases, the electron or positron emits a bremsstrahlung photon, which deposits part of its energy in the ECL close to the original electron or positron, thereby reducing the reconstructed energy of the corresponding \offlinecluster.
Both algorithms reconstruct a single \triggercluster, which is then matched either to the original electron or positron \offlinecluster or to the bremsstrahlung-induced \offlinecluster. 
Matching to the latter leads to a reduced cluster finding efficiency.
The energy and angular resolution for particle candidate clusters in \eebkg{} events are shown in \cref{fig:bhabha_resolution}.
The simulated events are scaled to the data luminosity.
For the energy resolution, the corresponding FWHM($\eta$)/2.355 for each dataset and \trg algorithm are reported in \cref{tab:fwhm_ee_mumu}, with the \cgnnetm having a slightly worse resolution width in comparison to the \icnetm.
The angular resolution is significantly better for the \cgnnetm. 
This improvement results from the use of position information from multiple TCs per \triggercluster, allowing the \cgnnetm to refine the overall position prediction compared with the \icnetm.
Both algorithms show the same resolutions for both the \textit{Run A} dataset and the simulated dataset.
\FloatBarrier

\FloatBarrier

\subsubsection{Beam background trigger rate}

\begin{figure}[t!]
    \centering
    \includegraphics[width=0.55\linewidth]{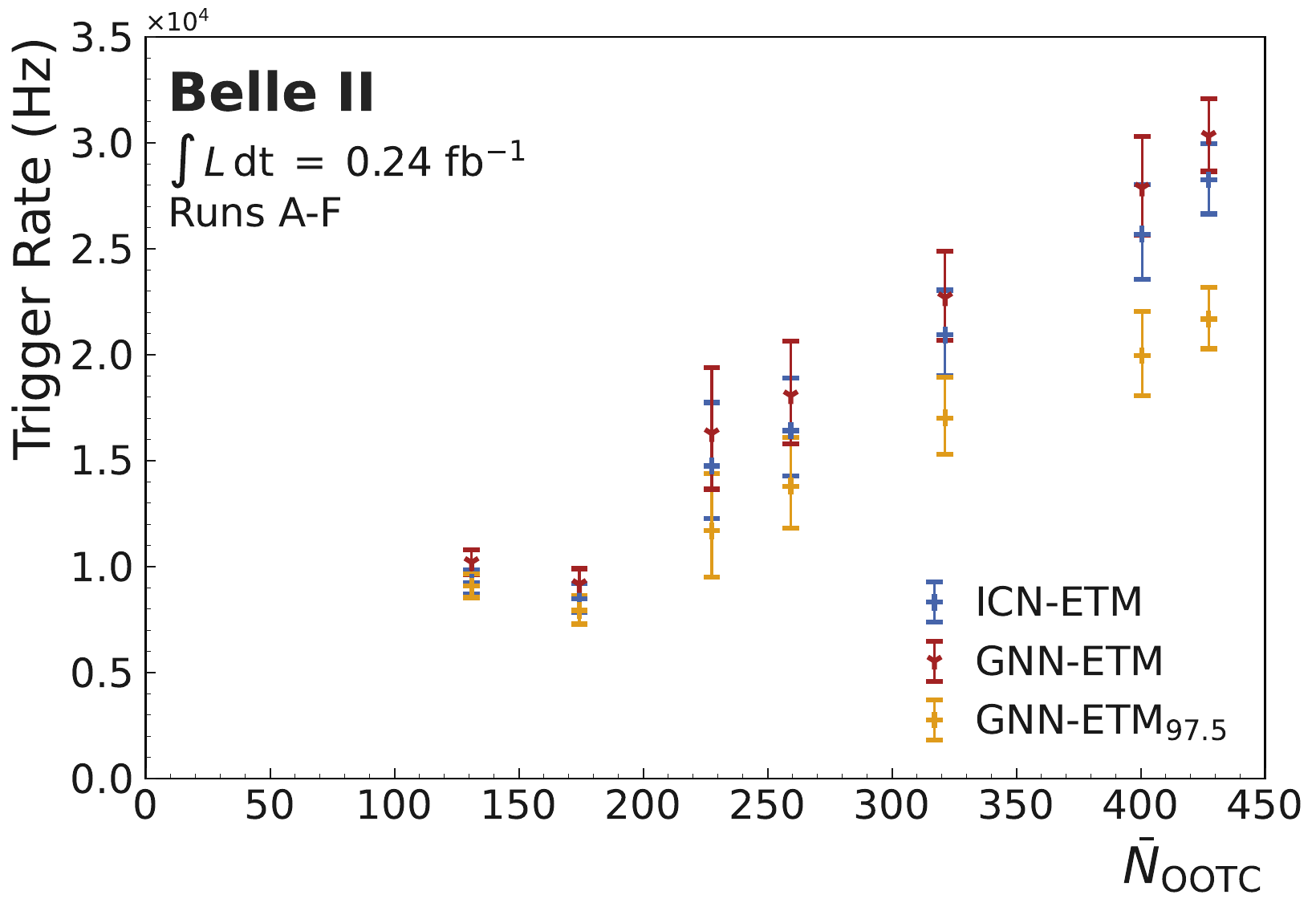}
\caption{Calculated \trg rate for the two-cluster \trg line over the average number of out-of-time crystals $\bar{N}_\mathrm{OOTC}$ for \textit{Runs A-F}~(see \cref{tab:run_descriptions}) for the \icnetm~(blue), the \gnnetm~(red), and the \cgnnetm~(orange).}
    \label{fig:c2_rate_with_signal_classifier}
\end{figure}

We evaluate the \trg rate of a high-rate \triggercluster counting \trg for the \icnetm, the \gnnetm, and the \cgnnetm.
We analyse events triggered by three random \trg lines available in \belletwo: 
the \textit{random} \trg line, which receives trigger signals in a fixed interval, 
the \textit{poisson} \trg line, which issues trigger signals at random times according to a Poisson process with a fixed average rate, 
and the \textit{background} \trg line, which issues a \trg signal five beam bunch rotations after a physics \trg signal is issued.
The cross section for physics processes of interest is small compared to the total electron positron interaction cross section. 
Consequently, events accepted by the three random \trg lines are dominated by beam related activity and background processes, including collision induced contributions such as low angle Bhabha scattering and two photon interactions, rather than signal reactions at the interaction point. 
These contributions are continuously present and represent a substantial fraction of the overall trigger rate.
We evaluate a two-cluster \trg, which issues a \trg signal when two or more \triggerclusters are present in an event, to estimate the \trg rate.
The rate is evaluated without applying vetoes against Bhabha events or events close to beam injection.
This two-cluster \trg line counts \triggerclusters with a reconstructed polar angle $\theta$ between 22.5\,$^\circ$ and 126.8\,$^\circ$.
This \trg line is currently not in active operation, as its high rate would exceed the allowed total trigger rate.
This \trg line is rather sensitive to the overall beam background level, as more energy depositions inside the ECL lead to an increase in the overall number of \triggerclusters.
We evaluate the two cluster \trg line on events selected by the three random \trg lines to estimate the trigger rate on events dominated by beam background energy depositions. 
The \gnnetm can improve the performance of the two-cluster \trg line by imposing an additional requirement on the signal classifier output for all \triggerclusters counted toward the two-cluster threshold. 
In \cref{fig:c2_rate_with_signal_classifier}, the \icnetm rate, the \gnnetm rate without a signal classifier requirement, and the \cgnnetm rate with signal classifier thresholds determined on the \textit{Uniform Photon Sample} are shown as a function of the beam background level $\bar{N}_\mathrm{OOTC}$ as defined in Eq.,\ref{eq:nootc}.
The trigger rate is calculated as the fraction of events selected by the random triggers that generate a trigger signal for a given \trg algorithm, divided by the 500\,ns minimum spacing between two trigger signals imposed by the Belle~II readout. 
Because only one trigger can be issued within this interval, the achievable average trigger rate is limited to 2\,MHz.

The \trg rates of the \icnetm and the \gnnetm without applying the signal classifier cut increase with beam background level and would imply trigger rates exceeding the hardware limit.
The \gnnetm displays a slight increase in the overall rate, which is caused by the ability of the \gnnetm to split energy depositions into multiple clusters and the different position resolution.
When applying the signal classifier of the \cgnnetm, the trigger rate still increases with beam background level, but with a significantly reduced slope. 
The rate is reduced by up to 20\% compared to the configuration without a signal classifier cut.
\FloatBarrier

\FloatBarrier
\section{Summary}
\label{sec:summary}
We have presented the implementation and detailed study of the \gnnetm, a real-time Graph Neural Network-based \trg module for the Belle II electromagnetic calorimeter.
The \gnnetm processes up to 32~sparsity-compressed \trg cells to reconstruct \triggerclusters and their properties, including a signal classifier, within the 8\,MHz throughput requirement of the \belletwo \trg system and a final latency of 3.168\,$\mu$s.
While this does not satisfy the 1.221\,$\mu$s latency constraint of the \belletwo calorimeter \trg system required to participate in the overall trigger decision, it is operated in parallel with the existing trigger system for monitoring and performance evaluation.
The model is deployed on an AMD Ultrascale XCVU190 FPGA using mixed-precision quantization, pruning, and reduced activation precision.
We have developed a complete system architecture comprising preprocessing, a GNN dataflow accelerator with GravNet and Condensation Point Clustering, and postprocessing for data encoding and readout.
The \gnnetm has been implemented on the Universal Trigger Board 4 with an integrated Belle2Link subsystem for full data readout, and its functionality has been validated using simulations on three abstraction levels, showing excellent agreement with \textit{Cosmic Data} recorded on hardware.
Performance studies using simulated and collision data recorded in December 2024 show that the \gnnetm matches the baseline ECL \trg in efficiency and energy resolution while providing significantly improved angular resolution and better cluster separation for high-energy \triggerclusters.
By employing its signal classifier, the \gnnetm rejects up to 70\% of background clusters while retaining 97.5\% of signal clusters, leading to a substantial reduction in \trg rate under increasing beam background conditions.
Running GNNs on FPGAs is an emerging research area that combines challenges from both machine learning and hardware design. 
The irregular data structures and computational demands, particularly for dynamic GNNs, are difficult to map efficiently onto FPGA architectures under tight latency and resource constraints. 
Consequently, developing real-time GNN inference on FPGAs remains an open and active topic of investigation.
To our knowledge, \gnnetm is the first GNN-based reconstruction algorithm implemented on FPGAs and operated within the readout infrastructure of a collider experiment trigger system, representing a significant step towards GNN-based real-time trigger decisions in particle physics.

\section*{Acknowledgments}
The authors would like to thank the \belletwo collaboration for useful discussions and suggestions on how to improve this work.\\
The training of the GNN-models was performed on the TOpAS GPU cluster at the Scientific Computing Center~(SCC) at the Karlsruhe Institute of Technology~(KIT).

\section*{Data Availability Statement}
This article has no associated data or the data will not be deposited.
The datasets collected, simulated, and analyzed in this study are the property of the \belletwo collaboration and are not publicly available.

\section*{Code Availability Statement}
This article has associated code in a code repository.
The instructions and code required to reproduce the training and evaluation of \gnnetm are available at \url{https://github.com/ihaide/gnnetm-software} \cite{gnnetm-software}. 
The quantized GravNet implementation is provided at \url{https://github.com/ihaide/qgravnet} \cite{qgravnet}, and the adapted \qkeras implementation at \url{https://github.com/ihaide/qkeras} \cite{qkeras-code}. 
The instructions and code for the hardware implementation are available at \url{https://github.com/marcneu/pcnhlslib/tree/gnnetm} \cite{gnnetm-hardware}.

\clearpage
\bibliographystyle{JHEP} 
\bibliography{sn-bibliography}

\clearpage

\end{document}